# Widespread quasi-steady state assumption in biological interaction modeling mischaracterizes system transitions

Pan-Jun Kim[*†]

Department of Biology, Hong Kong Baptist University, Kowloon, Hong Kong

Center for Quantitative Systems Biology, Hong Kong Baptist University, Kowloon, Hong Kong

Asia Pacific Center for Theoretical Physics, Pohang, Gyeongbuk 37673, Republic of Korea

## Abstract

From molecular, cellular, to ecological systems, the modeling of biological processes often stands on the assumption that fast components immediately reach the equilibrium at each moment (quasi-steady state) and only slow components govern the relevant system dynamics. This quasi-steady state approximation (QSSA) simplifies the modeling but discards the effects of the relaxation towards each quasi-steady state. Unclear is the QSSA's suitability around the transition point, a specific condition where the system changes to a qualitatively different state. In this regard, we here derived a theoretical framework for the near-transition dynamics of biological systems, explicitly considering the relaxation processes overlooked by the QSSA. Numerical simulations verify our predictions for cellular decision-making, metabolic oscillations, and ecological cycles. Despite the extreme slowdown near the transition point, the QSSA alone misestimates the duration of the transition from one state to another. Moreover, the QSSA erroneously predicts the transition point itself for the onset of oscillations, while the relaxation dynamics facilitates or suppresses the oscillation onset with a counterintuitive time-delay effect. Common feedback interactions between biological components are pivotal to those relaxation effects. Our study provides an analytical foundation to understand the rich transient or rhythmic dynamics of interacting biological components near the transitions.

[*]Present affiliation: C-SQ Laboratory, Kowloon, Hong Kong
[†]Contact: extutor@gmail.com

## Introduction

A variety of physical, chemical, and biological systems feature interdependent rapid to slow dynamical processes [1–9]. A common assumption for their mathematical modeling is that "fast" components equilibrate instantaneously and only "slow" components govern the relevant system dynamics [2–4, 7–14]. This assumption is based on the time-scale separation, rationalized by singular perturbation methods [5, 6, 14–18]. Depending on the context, this assumption is called by different names such as the quasi-steady state assumption (or pseudo-steady state hypothesis), slaving principle, and adiabatic approximation [6, 10–14]. Nevertheless, we will henceforth refer to any modeling with that assumption as the quasi-steady state approximation (QSSA), although this term itself has been preferred within particular areas such as chemical kinetics.

Whether deployed consciously or unconsciously, the QSSA is the routine modeling practice for a wide range of biological systems from molecular, cellular, to ecological systems [3, 4, 6–9, 12, 19–22]. The QSSA simplifies the modeling by omitting the system's relaxation process, the relatively fast intermediate process towards an equilibrium at each moment (quasi-steady state). For example, the Michaelis–Menten rate law of molecular complex formation takes the simplification that molecular complex concentration is instantaneously determined by individual component concentrations without considering an explicit relaxation process through complex association and dissociation [3, 12–14]. In the case of theoretical ecology, such relaxation-omitting simplification is exemplified by the applications of Lotka–Volterra type equations [6–8, 22].

Despite the widespread uncritical adoption of the QSSA, its suitability remains largely unclear. Numerous cellular or ecological phenomena such as signal responses, circadian oscillations, and predator–prey interactions [8, 19–21, 23, 24] exhibit actively time-varying states that may not strictly adhere to the QSSA. The deviation from the QSSA would grow as the time-scales of the fast and slow components get closer. Although this deviation can be systematically evaluated by the singular perturbation analysis [6, 14–18], the results are often case-specific without delivering an noticeable general insight. To gain more general insights into that deviation, we consider the scenario that a dynamical system enters a qualitatively new state once passing a particular condition—called the transition or bifurcation point [25, 26]. At this point, the transition is either abrupt with a finite gap between the old and new states (discontinuous), or continuous. The nearer the transition point, the extremely slower the system's pace to the final state, known as "critical slowing down" [25–29]. Over a range of dynamical systems, the transition is local and low-dimensional in the phase space, as captured by the approaches with center manifolds, reductive perturbations, and normal forms [5, 18, 25, 26]. Owing to this local and low-dimensional nature, the essence of the near-transition dynamics is not sensitive to the system's details. Hence, exploring the near-transition dynamics may help to draw the cross-system, general insights into the deviation from the QSSA. On the other hand, one may counter-argue that the QSSA by definition would well hold under the critical slowing down with the slowly-varying states and thus the near-transition dynamics would not much deviate from the QSSA itself.

In this study, we reveal that the relaxation dynamics overlooked by the QSSA plays a crucial role in the transition of biological systems, through our theory and its applications to genetic, metabolic, and ecological systems. Even for the extremely slow near-transition events, the QSSA can markedly misestimate the duration of the transition from one state to another. Moreover, the QSSA erroneously predicts the transition point itself for the onset of oscillations, while the relaxation dynamics facilitates or suppresses the oscillation onset in a counterintuitive way. These relaxation effects are attributed to common feedback interactions between biological components. Over the contexts from cellular adaptation to population cycles, our work provides a unique basis to analyze the rich transient or rhythmic dynamics of interacting biological components near the transitions.

## Results

### Theory derivation

To lay out our theory, we start with the following equation:

$$\frac{\mathrm{d}x(t)}{\mathrm{d}t} = F(x(t), y_1(t), y_2(t), \cdots, y_N(t)), \tag{1}$$

where $t$ denotes time, $x(t)$ serves as a relatively fast variable, and $y_i(t)$ ($i = 1, 2, \cdots, N$) as a relatively slow variable. Given the values of $y_1(t)$, $y_2(t)$, $\cdots$, and $y_N(t)$, let $x_{\mathrm{Q}}(t)$ be a stable fixed point of $x(t)$. In other words, $x_{\mathrm{Q}}(t)$ satisfies the following conditions:

$$F\left(x_{\mathrm{Q}}(t), y_1(t), y_2(t), \cdots, y_N(t)\right) = 0, \tag{2}$$

$$\mu(t) > 0 \text{ with } \mu(t) \equiv -\left.\frac{\partial F}{\partial x}\right|_{x = x_{\mathrm{Q}}(t)}. \tag{3}$$

In the QSSA, $x_{\mathrm{Q}}(t)$ is the quasi-steady state of $x(t)$ and $x(t) \approx x_{\mathrm{Q}}(t)$.

From Eqs. (1)–(3), the relaxation dynamics of $x(t)$ can be described by

$$\frac{\mathrm{d}x(t)}{\mathrm{d}t} \approx -\mu(t)\left(x(t) - x_{\mathrm{Q}}(t)\right). \quad (4)$$

When $\mu(t) \to \infty$, i.e., $\mu^{-1}(t) \to 0$, the QSSA becomes retrieved. Still, the QSSA may fail when the above relaxation (with the time-scale of $\mu^{-1}(t)$) is not much faster than the temporal change of $x_{\mathrm{Q}}(t)$. In the case of rhythmic $x_{\mathrm{Q}}(t)$, we indeed prove that actual $x(t)$ lags behind $x_{\mathrm{Q}}(t)$ up to the phase difference between $x(t)$ itself and $\mu^{-1}(t)x'(t)$, and owns a smaller amplitude than $x_{\mathrm{Q}}(t)$ (Supplementary Information (SI), Section A). For example, if $x_{\mathrm{Q}}(t)$ oscillates with a constant period $T$ while $\mu(t)$ is relatively static, $x(t)$ exhibits the time delay of $\sim(T/2\pi)\cdot\arctan(2\pi\mu^{-1}(t)/T)$ and the amplitude reduced by a factor of $\sim\sqrt{1+(2\pi\mu^{-1}(t)/T)^2}$ compared to $x_{\mathrm{Q}}(t)$ (SI Section A). This time delay is further estimated as $\mu^{-1}(t)$ when $\mu^{-1}(t) \ll T$, and capped by $T/4$ when $\mu^{-1}(t) \gg T$. We have recently reported such time-delay and amplitude reduction effects within the context of molecular complex formation [30], but now extend these effects to any relaxation processes governed by Eq. (4). Their mathematical forms are analogous to the sinusoidal response of an RC circuit [31].

For simplicity, we will henceforth treat $\mu(t)$ as a constant, i.e., $\mu(t) = \mu_{\mathrm{o}}$. This assumption reasonably holds for any system in the vicinity of the steady state, as explained in SI Section B. Then the asymptotic trajectory of $x(t)$ takes the following series form from Eq. (4):

$$x(t) \approx x_{\mathrm{Q}}(t) + \sum_{n=1}^{\infty}(-1)^n \mu_{\mathrm{o}}^{-n} x_{\mathrm{Q}}^{(n)}(t) = x_{\mathrm{Q}}(t) - \mu_{\mathrm{o}}^{-1}\frac{\mathrm{d}x_{\mathrm{Q}}(t)}{\mathrm{d}t} + \cdots, \quad (5)$$

where $x_{\mathrm{Q}}^{(n)}(t)$ is the $n$-th derivative of $x_{\mathrm{Q}}(t)$ and $-\mu_{\mathrm{o}}^{-1}x_{\mathrm{Q}}'(t)$ on the rightmost side indicates the time delay $\sim\mu_{\mathrm{o}}^{-1}$ of $x(t)$ compared to $x_{\mathrm{Q}}(t)$ (SI Section C). The aforementioned periodic oscillations aside, $\sim\mu_{\mathrm{o}}^{-1}$ indeed matches the time delay for a range of dynamical events (SI Section C).

One may thus expect that the transition of the system from one state to another is likely to take $\sim\mu_{\mathrm{o}}^{-1}$ longer than predicted by the QSSA. As the transition (bifurcation) point approaches, this finite difference $\sim\mu_{\mathrm{o}}^{-1}$ between the actual transition time and the QSSA would become negligible compared to the huge transition time itself with the critical slowing down, and hence the QSSA may suffice for the description of this system. However, this expectation comes only true when $x(t)$ is unilaterally affected by $x_{\mathrm{Q}}(t)$ through Eq. (4) but does not affect $x_{\mathrm{Q}}(t)$ either directly or indirectly. In biological systems, the direct or indirect feedback from $x(t)$ to $x_{\mathrm{Q}}(t)$ can naturally arise from common mutual interactions between their components, and thus $x_{\mathrm{Q}}(t)$ ever continues to be modified by the relaxation dynamics of $x(t)$. The resulting transition time may not merely differ from the QSSA by $\sim\mu_{\mathrm{o}}^{-1}$. Our analysis indeed reveals that the transition time deviation from the QSSA is not relatively negligible even near the transition point, but proportional to the transition time itself (SI Section D). For example, upon the discontinuous transition where the existing stable steady state suddenly disappears by its merging with a nearby unstable steady state (saddle-node bifurcation) [25], the transition time for the new state is predicted with small $\mu_{\mathrm{o}}^{-1}$ as follows:

$$t_{\mathrm{r}} \approx \frac{t_{\mathrm{Qr}}}{\sqrt{1-2\mu_{\mathrm{o}}^{-1} b_{\mathrm{Qm}}}}, \tag{6}$$

where $t_{\mathrm{r}}$ and $t_{\mathrm{Qr}}$ denote the transition time and its QSSA, respectively, and $b_{\mathrm{Qm}}$ is a constant determined by the feedback of $x(t)$ to $x_{\mathrm{Q}}(t)$ (SI Section D). The nature of this feedback sets the sign of $b_{\mathrm{Qm}}$: if the increase in $x(t)$ is promotive (repressive) of $x_{\mathrm{Q}}(t)$, $b_{\mathrm{Qm}} > 0$ ($b_{\mathrm{Qm}} < 0$). Owing to the above multiplicative relationship between $t_{\mathrm{Qr}}$ and $t_{\mathrm{r}}$, not an additive one, their difference $\left|t_{\mathrm{r}} - t_{\mathrm{Qr}}\right|$ is proportional to the transition time $t_{\mathrm{r}}$ itself and the QSSA can markedly underestimate (if $b_{\mathrm{Qm}} > 0$) or overestimate (if $b_{\mathrm{Qm}} < 0$) this transition time. For example, the transition time would be $\sim$30%-longer than the QSSA, according to Eq. (6) when $\mu_{\mathrm{o}}^{-1} b_{\mathrm{Qm}} = 0.2$. The predicted relationship in Eq. (6) is consistent with our numerical simulations, as we will later see for the genetic switch's response to environmental signals. The predicted mismatch between the actual transition time and QSSA around the transition point—despite the presence of critical slowing down—is caused by the near-transition ultrasensitivity that a small variation in the system's trajectory substantially perturbs its path.

The above quantitative modification aside, the relaxation dynamics can also change a qualitative outcome against the QSSA (SI Section E). Specifically, consider the continuous transition from the steady state to periodic oscillations due to a slight change in the system's condition (supercritical Hopf bifurcation) [5, 25]. According to our theory, even tiny $\mu_{\mathrm{o}}^{-1} > 0$ facilitates (suppresses) the oscillation onset and therefore advances (delays) the transition point itself compared to the QSSA, if the system satisfies

$$\left(\frac{2\pi}{T_{\mathrm{Q}}}\right)^2 > A \quad \left(\left(\frac{2\pi}{T_{\mathrm{Q}}}\right)^2 < A\right). \tag{7}$$

This prediction is confirmed by our numerical simulations of glycolytic oscillations and ecological cycles, as we will see later. In Eq. (7), $T_{\mathrm{Q}}$ and $A$ are the quantities defined at the QSSA-based transition point: $T_{\mathrm{Q}}$ denotes the QSSA-based oscillation period and $A$ is defined as

$$A \equiv \left(\frac{\partial G_1}{\partial y_1}\frac{\partial G_2}{\partial y_2} - \frac{\partial G_1}{\partial y_2}\frac{\partial G_2}{\partial y_1}\right)\Big|_{x=f(y_{1\mathrm{s}},y_{2\mathrm{s}}),y_1=y_{1\mathrm{s}},y_2=y_{2\mathrm{s}}}, \tag{8}$$

where $G_1$ and $G_2$ ($G_1 \equiv G_1(x, y_1, y_2)$ and $G_2 \equiv G_2(x, y_1, y_2)$) govern the velocity of the main oscillatory components of the system ($y_1$ and $y_2$) by $y_1' = G_1(x, y_1, y_2)$ and $y_2' = G_2(x, y_1, y_2)$ in the essentially planar phase space at the transition, $y_{1\mathrm{s}}$ and $y_{2\mathrm{s}}$ indicate the transitional steady state by $(y_1, y_2) = (y_{1\mathrm{s}}, y_{2\mathrm{s}})$, and $f(y_1, y_2)$ sets the quasi-steady state of the relaxation component $x$ in Eq. (4) as $x_\mathrm{Q} = f(y_1, y_2)$ (SI Section E).

As a result, the QSSA erroneously informs the oscillation onset, and the actual transition point becomes advanced or delayed compared to the QSSA. The delayed case contradicts a natural expectation that the relaxation dynamics would enhance the oscillation rather than suppress it [6, 32]. As we will see unexpectedly, this oscillation suppression is mainly the time-delay effect of the relaxation dynamics noted under Eq. (4), while the amplitude reduction effect noted there is relatively minor. For the physical interpretation of the condition in Eq. (7), we notice its equivalence to $\left(2\pi/T_\mathrm{Q}\right)^2 - A > 0$ or $\left(2\pi/T_\mathrm{Q}\right)^2 - A < 0$. As explained in SI Section F, $\left(2\pi/T_\mathrm{Q}\right)^2 - A$ decomposes into the several terms proportional to $\partial_x G_1$ or $\partial_x G_2$ at the transitional steady state and all these terms originate from the feedback of $x$ to $x_\mathrm{Q}$. Next, adapting the concept of classical mechanics, $\left(2\pi/T_\mathrm{Q}\right)^2$ is proportional to the centripetal acceleration of a moving object along the QSSA-based orbit in the phase space with the oscillatory state. $\left(2\pi/T_\mathrm{Q}\right)^2 - A$ then reflects the contribution of the temporal change in $x$ to this object's acceleration through the feedback of $x$ to $x_\mathrm{Q}$ (SI Section F). Hence, if $\left(2\pi/T_\mathrm{Q}\right)^2 > A$ $\left(\left(2\pi/T_\mathrm{Q}\right)^2 < A\right)$ as in Eq. (7) and thus $\left(2\pi/T_\mathrm{Q}\right)^2 - A > 0$ $\left(\left(2\pi/T_\mathrm{Q}\right)^2 - A < 0\right)$, that contribution of $x$'s change acts like the centripetal (centrifugal) force aligned with (against) the object's acceleration. Regarding the relaxation dynamics, a slight increase of $\mu_\mathrm{o}^{-1}$ from zero induces the time delay $\sim\mu_\mathrm{o}^{-1}$ in the $x$'s change, and then this repression of the change weakens its associated centripetal (centrifugal)-like force, thereby stretching (shrinking) the orbit and amplifying (suppressing) the oscillation compared to the QSSA. The full physical interpretation with its mathematical rationale is provided in SI Section F.

Thus far, we have overviewed our theory of the discrepancy between the actual and QSSA-based dynamics in biological transitions. Relaxation dynamics overlooked by the QSSA serves as the source of this discrepancy through the feedback interactions between the system's components. Next, we will test this theory over a range of biological systems.

**Genetic switch with autoregulation**

Cellular decision-making is a pivotal biological process for the adaptation to environmental changes [23, 33]. This process is exemplified by the genetic switch with positive autoregulation in Fig. 1(a) where proteins enhance their own transcription after homodimer formation and subsequent DNA promoter binding as a transcription factor (TF). This dimer–promoter interaction is facilitated by inducer molecules. The protein's dimeric cooperativity is known to confer a sigmoid TF–DNA binding curve, leading to abrupt and history-dependent transitions in the protein production [34–36].

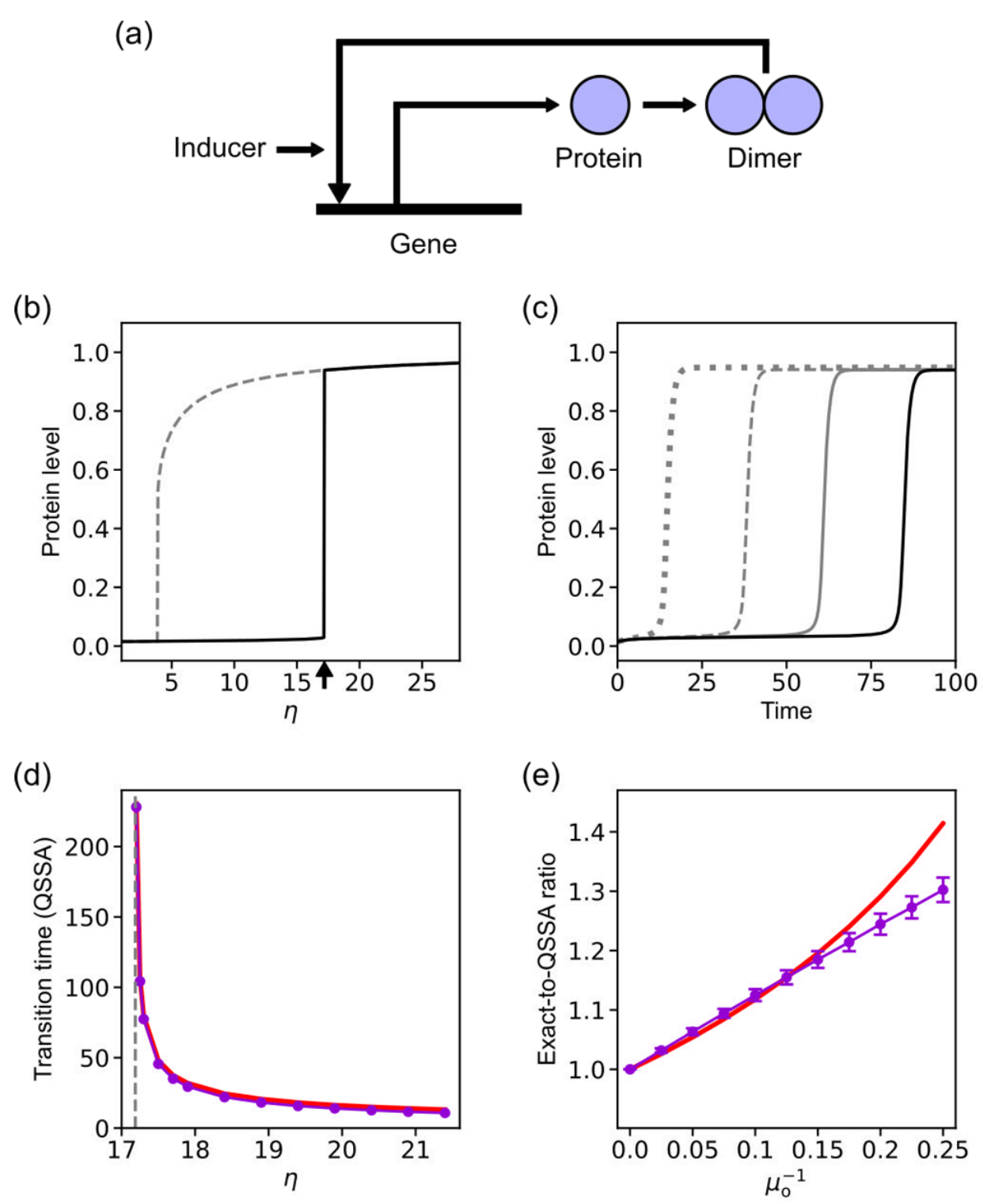

**Fig. 1. Genetic switch and induction kinetics.** (a) Protein production mechanism with positive autoregulation and induction. (b) Bifurcation diagram of the simulated protein level as a function of inducer level $\eta$ at $\sigma = 0.015$ (basal transcription rate). The same $\sigma$ is used for (c)–(e) as well, but the observation here remains valid for other $\sigma$ values (e.g., Fig. S1). The steady state is plotted as $\eta$ increases (solid line) or decreases (dashed line). In the case of increasing $\eta$, the bifurcation at $\eta = \eta_c$ is indicated by an arrow for the analytically-calculated $\eta_c$ from SI Section G. (c) Time-series of the simulated protein level upon the acute induction at time 0 from $\eta = 0$ to $\eta > \eta_c$, specifically $\eta = 20$, 17.7, 17.4, or 17.3 (left to right). Here $\mu_o^{-1} = 0.1$. (d) QSSA-based transition time from the simulated acute induction (violet dot) or its analytical estimation (red line and SI Section G) as a function of $\eta$. $\eta = \eta_c$ is marked by a vertical dashed line. (e) Ratio of the transition time to its QSSA from the simulated acute induction (violet dot and error bar) or the analytical estimation with Eq. (11) (red line) as a function of $\mu_o^{-1}$. Each error bar represents the standard deviation across $\eta$ close to $\eta_c$ (0 <

$(\eta - \eta_c)/\eta_c \leq 0.25$). $\mu_o^{-1} = 0$ corresponds to the QSSA. The analytical and simulation results are comparable over small $\mu_o^{-1}$, and even their functional forms become matched as $\mu_o^{-1}$ decreases. In (b)–(e), all the values are unitless through the scaling of the original quantities (SI Section G).

---

The following equations and the relaxation process in Eq. (4) with $\mu(t) = \mu_o$ capture the essence of this process:

$$\frac{\mathrm{d}y(t)}{\mathrm{d}t} = x(t) - y(t), \tag{9}$$

$$x_Q(t) = f(y(t)) \text{ with } f(y(t)) = (1-\sigma) \cdot \frac{\eta y^2(t)}{1+\eta y^2(t)} + \sigma. \tag{10}$$

Here time $t$, $\mu_o$, and all the other quantities were non-dimensionalized by the scaling of the original quantities (SI Section G). $x(t)$ and $y(t)$ correspond to mRNA and protein concentrations, respectively. $-y(t)$ term in Eq. (9) describes the concentration loss with the protein degradation and cell growth-associated dilution. In other words, the right-hand side of Eq. (9) presents the net protein production rate. Additionally, $f(y(t))$ in Eq. (10) provides the explicit form of the quasi-steady state of $x(t)$ (i.e., $x_Q(t)$), and the parameters $\eta$ and $\sigma$ correspond to the inducer concentration and basal transcription rate, respectively. The terms of $y^2(t)$, $(1-\sigma) \cdot \eta y^2(t)/(1+\eta y^2(t))$, and $+\sigma$ in $f(y(t))$ originate from the protein dimerization, dimer-regulated transcription, and basal transcription, respectively. $\mu_o$ in Eq. (4) reflects the relaxation rates of the typically faster processes than the protein loss, including the mRNA degradation, protein dimer dissociation, and TF–DNA dissociation. The full derivation of this model is provided in SI Section G.

As the simulated inducer level $\eta$ increases, Figs. 1(b), S1(a) show that the initially low, steady protein level undergoes an abrupt leap once $\eta > \eta_c$. This discontinuous transition at the transition point $\eta = \eta_c$, also known as a tipping point, belongs to the aforementioned saddle-node bifurcation and exists when $0 < \sigma < 1/9$ (SI Section G). Just reducing the inducer level back to the transition point does not reverse the protein level, which remains high until the inducer level becomes further reduced (Figs. 1(b), S1(a)). This history-dependent behavior, called hysteresis, indicates the coexistence of two distinct stable states of the protein level between the forward and backward transitions [35, 36].

Given these steady states, we examine the transition time—the signal response time upon the acute induction from zero to $\eta > \eta_c$. The induced protein level grows towards the new steady state, and this transition time becomes increasingly longer with critical slowing down as the inducer level $\eta$ is closer to $\eta_c$ while $\eta > \eta_c$ (Figs. 1(c),(d), S1(b),(c); the transition time here is defined as the time to reach 70% of the new steady level). Note that $x(t)$ is not only affected by $x_{\mathrm{Q}}(t)$ through Eq. (4) but also feeds back into $x_{\mathrm{Q}}(t)$ through Eqs. (9) and (10) with the positive autoregulation. This positive feedback sets the sign of $b_{\mathrm{Qm}}$ in Eq. (6) as $b_{\mathrm{Qm}} > 0$ (SI Section D). Overall, the time delay from the relaxation processes decelerates the protein production and substantially retards the protein growth due to the near-transition ultrasensitivity of Eq. (9) against the protein level. This mechanism causes the transition time to systematically deviate from the QSSA despite the critical slowing down near the transition point. Specifically, Eqs. (9) and (10) yield $b_{\mathrm{Qm}} = 1$ for Eq. (6) (SI Section G) and the following result for small $\mu_{\mathrm{o}}^{-1}$:

$$t_{\mathrm{r}} \approx \frac{t_{\mathrm{Qr}}}{\sqrt{1-2\mu_{\mathrm{o}}^{-1}}}, \tag{11}$$

where $t_{\mathrm{r}}$ and $t_{\mathrm{Qr}}$ denote the transition time and its QSSA, respectively. Owing to the above multiplicative relationship between the transition time and the QSSA, the QSSA would markedly underestimate the transition time. On the other hand, the critical slowing down itself is captured by the analytical property that $t_{\mathrm{Qr}} \propto 1/\sqrt{(\eta - \eta_{\mathrm{c}})/\eta_{\mathrm{c}}}$ (SI Section G) and thus $t_{\mathrm{r}} \propto 1/\sqrt{(\eta - \eta_{\mathrm{c}})/\eta_{\mathrm{c}}}$ from Eq. (11).

Over the ranges of small $\mu_{\mathrm{o}}^{-1}$, $\eta$ close to $\eta_{\mathrm{c}}$, and $\sigma$, the numerical simulations of the system support our predictions, including Eq. (11) (Figs. 1(d),(e), S1(c),(d); $P \leq 0.02$ and SI Section J). For example, the simulations at $\mu_{\mathrm{o}}^{-1} = 0.25$ and $\sigma = 0.015$ show that the transition time is longer than the QSSA by $30.2 \pm 2.1\%$ (avg. ± s.d.) across $\eta$, comparable with $41.4\%$ from our prediction (Fig. 1(e)). Our prediction even better agrees with the simulations in its functional form as $\mu_{\mathrm{o}}^{-1}$ decreases (Figs. 1(e), S1(d)). Taken together, our analytical and numerical studies suggest that the cellular decision-making with the genetic switch takes longer than expected by the QSSA, and this error of the QSSA is still pronounced near the tipping point under the autoregulatory feedback with the molecular relaxations.

**Glycolytic oscillation**

Glycolysis, a widely conserved metabolic process, is the step-by-step breakdown of consumed glucose to extract energy for cellular activities and maintenance. Under certain conditions, metabolite concentrations in yeast glycolysis are found to exhibit sustained oscillations, called glycolytic oscillations [24, 37, 38]. As a simple classical model of glycolytic oscillations, the Sel'kov model focuses on the phosphofructokinase reaction in Fig. 2(a) where adenosine triphosphate (ATP) donates a phosphate to fructose-6-phosphate and turns into adenosine diphosphate (ADP) [39]. This enzymatic reaction involves the inhibition by the substrate ATP and the activation by the product molecules. Adenosine monophosphate (AMP) also activates this enzyme, but its effect can be grouped to that of ADP due to the consistency of their concentration changes. The concentrations of fructose-6-phosphate and its phosphorylated product, fructose-1,6-diphosphate are treated as oscillation-irrelevant in the Sel'kov model. As a result, the model comprises the following equations:

$$\frac{\mathrm{d}y_1(t)}{\mathrm{d}t} = 1 - x(t), \tag{12}$$

$$\frac{\mathrm{d}y_2(t)}{\mathrm{d}t} = \alpha\big(x(t) - y_2(t)\big). \tag{13}$$

Here, $x(t) \equiv y_1(t)y_2^{\gamma}(t)$, and time $t$ and all the other quantities are appropriately non-dimensionalized [39]. $y_1(t)$ and $y_2(t)$ stand for ATP and ADP concentrations, respectively, and $x(t)$ for the activated and substrate-loaded enzyme concentration. $\gamma$ and $\alpha$ in $x(t)$ and Eq. (13) are positive constants. The terms of 1 and $-\alpha y_2(t)$ on the right-hand sides of Eqs. (12) and (13) describe ATP supply and ADP drainage, respectively, and the other terms $-x(t)$ and $\alpha x(t)$ describe the ATP-to-ADP conversion by the enzyme. Although $\gamma$ in $x(t)$ appears to be the number of the product molecules to activate the enzyme, its actual interpretation is the relative weakness of the substrate inhibition, with any real number of $\gamma > 1$ [39].

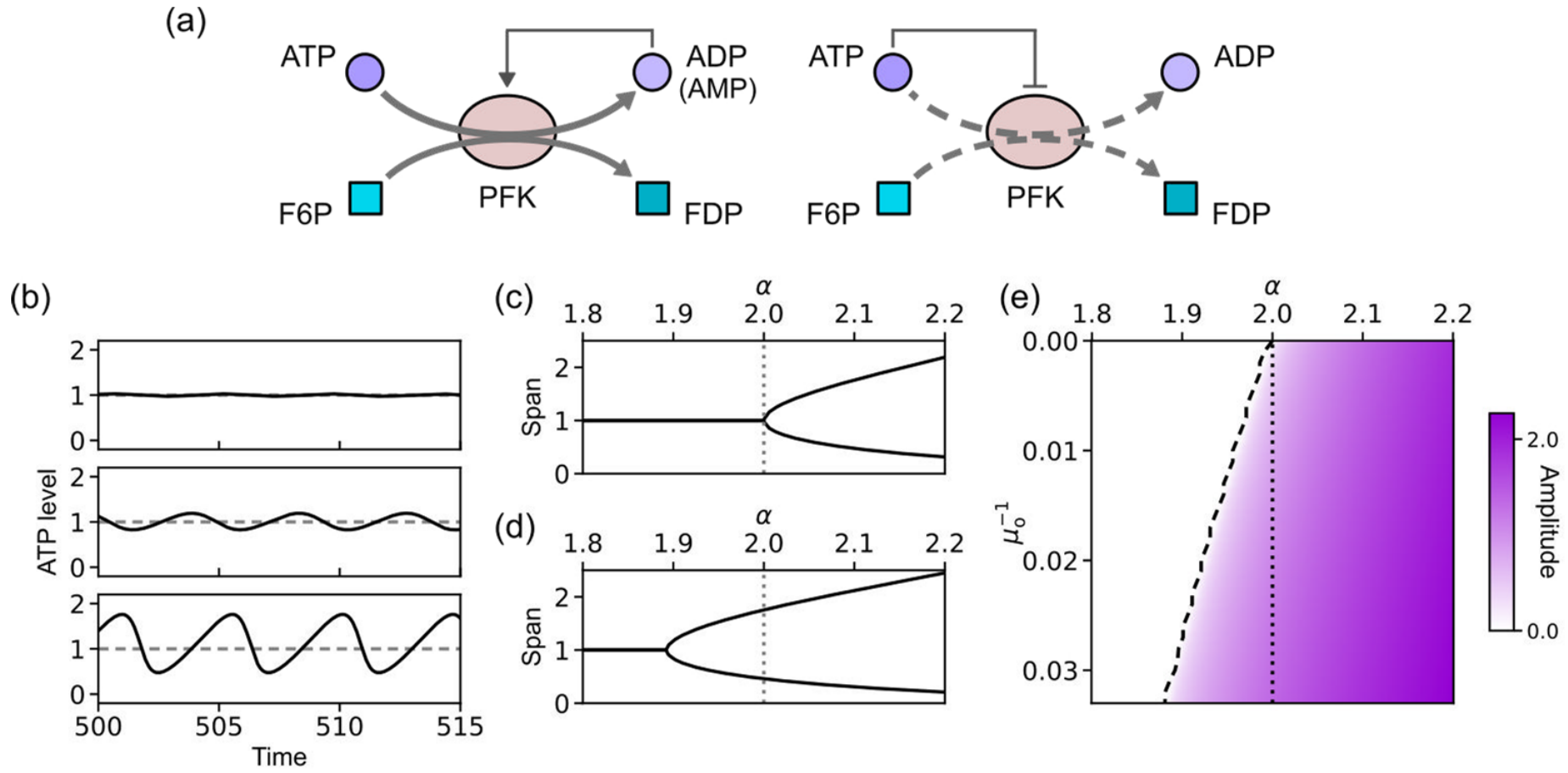

**Fig. 2. Glycolytic enzyme activity and oscillation onset.** (a) Phosphofructokinase (PFK) converts ATP and fructose-6-phosphate (F6P) into ADP and fructose-1,6-diphosphate (FDP), and involves product activation and substrate inhibition. AMP also activates PFK, but can be treated as a proxy for ADP. (b) Time-series of the ATP level from the Sel'kov model simulation (solid line) and the analytical steady state (dashed line) at $\alpha = 1.99$, 2.01, or 2.1 (top to bottom) and $\gamma = 1.5$. The same $\gamma$ is used for (c)–(e) as well, but the observation here remains valid for other $\gamma$ values (e.g., Fig. S2). (c) Bifurcation diagram of the ATP level from the Sel'kov model simulation: peak and trough levels as a function of $\alpha$ (solid line). Bifurcation at $\alpha = \alpha_{\mathrm{Qc}}$ is indicated by a vertical dotted line for the analytically-calculated

$\alpha_{\mathrm{Qc}}$ from SI Section H ($\alpha_{\mathrm{Qc}} = 1/(\gamma - 1)$). (d) Bifurcation diagram of the ATP level from the simulation of the extended Sel'kov model at $\mu_{\mathrm{o}}^{-1} = 0.03$: peak and trough levels as a function of $\alpha$ (solid line), together with $\alpha = \alpha_{\mathrm{Qc}}$ in (c) (vertical dotted line). (e) ATP level's amplitude from the simulation of the extended Sel'kov model: peak-to-trough difference at each $\alpha$ and $\mu_{\mathrm{o}}^{-1}$ (violet-colored). Bifurcation at $\alpha = \alpha_{\mathrm{c}}$ is indicated by a dashed line for numerically-calculated $\alpha_{\mathrm{c}}$ across $\mu_{\mathrm{o}}^{-1}$. For comparison, $\alpha = \alpha_{\mathrm{Qc}}$ from (c) is marked by a vertical dotted line. $\mu_{\mathrm{o}}^{-1} = 0$ corresponds to the original Sel'kov model. In (b)–(e), all the values are unitless through the scaling of the original quantities (SI Section H).

---

As $\alpha$ increases and slightly passes $\alpha = \alpha_{\mathrm{Qc}}$ ($\alpha_{\mathrm{Qc}} \equiv 1/(\gamma - 1)$), steady molecular levels $y_1 = y_2 = 1$ become unstable and start to periodically oscillate with a gradually-increased amplitude (Figs. 2(b),(c), S2(a),(b) and SI Section J). This continuous transition to the glycolytic oscillation is the aforementioned supercritical Hopf bifurcation, and the transition point $\alpha = \alpha_{\mathrm{Qc}}$ derives from the analytical calculation (SI Section H) [25, 39]. Still, the Sel'kov model relies on the QSSA of the substrate-loaded enzyme level, $x(t) = y_1(t)y_2^{\gamma}(t)$ [39].

Beyond the QSSA, we here consider the relaxation of this enzyme level, under Eq. (4) with $\mu(t) = \mu_{\mathrm{o}}$ and the following condition:

$$x_{\mathrm{Q}}(t) = f\big(y_1(t), y_2(t)\big) \text{ with } f\big(y_1(t), y_2(t)\big) = y_1(t)y_2^{\gamma}(t). \quad (14)$$

$\mu_{\mathrm{o}}$ in that relaxation corresponds to the enzyme deactivation rate and is treated as dimensionless for consistency with dimensionless time $t$. This extended model is derived in SI Section H. About Eqs. (7) and (8), we analytically obtain $\left(2\pi/T_{\mathrm{Q}}\right)^2 = \alpha_{\mathrm{Qc}}$ and $A = 0$ from Eqs. (12)–(14) and $y_{1\mathrm{s}} = y_{2\mathrm{s}} = 1$, and therefore $\left(2\pi/T_{\mathrm{Q}}\right)^2 > A$ (SI Section H). It is hence our prediction that even tiny $\mu_{\mathrm{o}}^{-1}$ would advance the oscillation transition point itself compared to the QSSA, i.e., $\alpha_{\mathrm{c}} < \alpha_{\mathrm{Qc}}$ for the actual transition point $\alpha = \alpha_{\mathrm{c}}$ when $\mu_{\mathrm{o}}^{-1} > 0$. Our numerical simulations do confirm this prediction over the range of $\gamma$, as demonstrated in Figs. 2(d),(e), S2(c),(d) (see also SI Section J). To conclude, the relaxation dynamics facilitates the glycolytic oscillation and advances its transition point compared to the QSSA, original Sel'kov model.

Next, we will dive into the system of richer relaxation effects, with both the advanced and delayed transition points together.

**Population cycle with cooperative foraging**

From bacteria to social insects and animals, the cooperation among foraging or predating individuals appears, and can also be mimicked by robots [40–43]. Given this cooperative resource foraging, we here propose the analytically-tractable model with population oscillations:

$$\frac{\mathrm{d}y_1(t)}{\mathrm{d}t} = ax(t) - by_1(t), \tag{15}$$

$$\frac{\mathrm{d}y_2(t)}{\mathrm{d}t} = 1 - x(t) - y_2(t), \tag{16}$$

where $x(t)$ follows Eq. (4) with $\mu(t) = \mu_\mathrm{o}$ and this condition:

$$x_\mathrm{Q}(t) = f\big(y_1(t), y_2(t)\big) \text{ with } f\big(y_1(t), y_2(t)\big) = y_1(t)y_2(t)\big(1 + \phi y_1(t)\big). \tag{17}$$

In the above equations, time $t$, $\mu_\mathrm{o}$, and all the other quantities were non-dimensionalized by the scaling of the original quantities (SI Section I). $y_1(t)$, $y_2(t)$, and $x(t)$ denote consumer, resource, and consumer-captured resource abundances, respectively. $a$, $b$, and $\phi$ in Eqs. (15) and (17) are positive constants. The terms of $ax(t)$, $-by_1(t)$, $1$, $-x(t)$, and $-y_2(t)$ on the right-hand sides of Eqs. (15) and (16) reflect the consumer proliferation and death/migration, and the resource supply, consumption, and drainage/decay, respectively. $\phi y_1(t)$ term in Eq. (17) reflects the consumer cooperation during the foraging, with cooperativity $\phi$. Lastly, $\mu_\mathrm{o}$ combines the resource consumption rate and other consumer–resource breakage rates (SI Section I).

We start with the QSSA by setting $x(t) = f\big(y_1(t), y_2(t)\big)$ for Eqs. (15) and (16) in the limit of $\mu_\mathrm{o}^{-1} \to 0$. As the consumer cooperativity $\phi$ increases with given $a$ and $b$ in the ranges of $4 < b$ and $b < a < b^2/4$, our numerical simulations and analytical calculations reveal the presence of the continuous transition from the steady to periodically oscillating consumer and resource levels at the following point of $\phi = \phi_\mathrm{Qc1}$ (Figs. 3(a),(b), and SI Sections I and J):

$$\phi_\mathrm{Qc1} \equiv \frac{2}{\frac{2a}{b} - 1 - \sqrt{1 - \frac{4a}{b^2}}}\left(\frac{2}{1 + \sqrt{1 - \frac{4a}{b^2}}} - 1\right), \tag{18}$$

with the transitional steady state as $(y_1, y_2) = (y_{1\mathrm{s}}, y_{2\mathrm{s}})$ below,

$$y_{1\mathrm{s}} = \frac{1}{2\phi}\left\{\frac{a}{b}\phi - 1 + \sqrt{\left(\frac{a}{b}\phi - 1\right)^2 + 4\phi\left(\frac{a}{b} - 1\right)}\right\} \text{ and } y_{2\mathrm{s}} = 1 - \frac{b}{a}y_{1\mathrm{s}}. \tag{19}$$

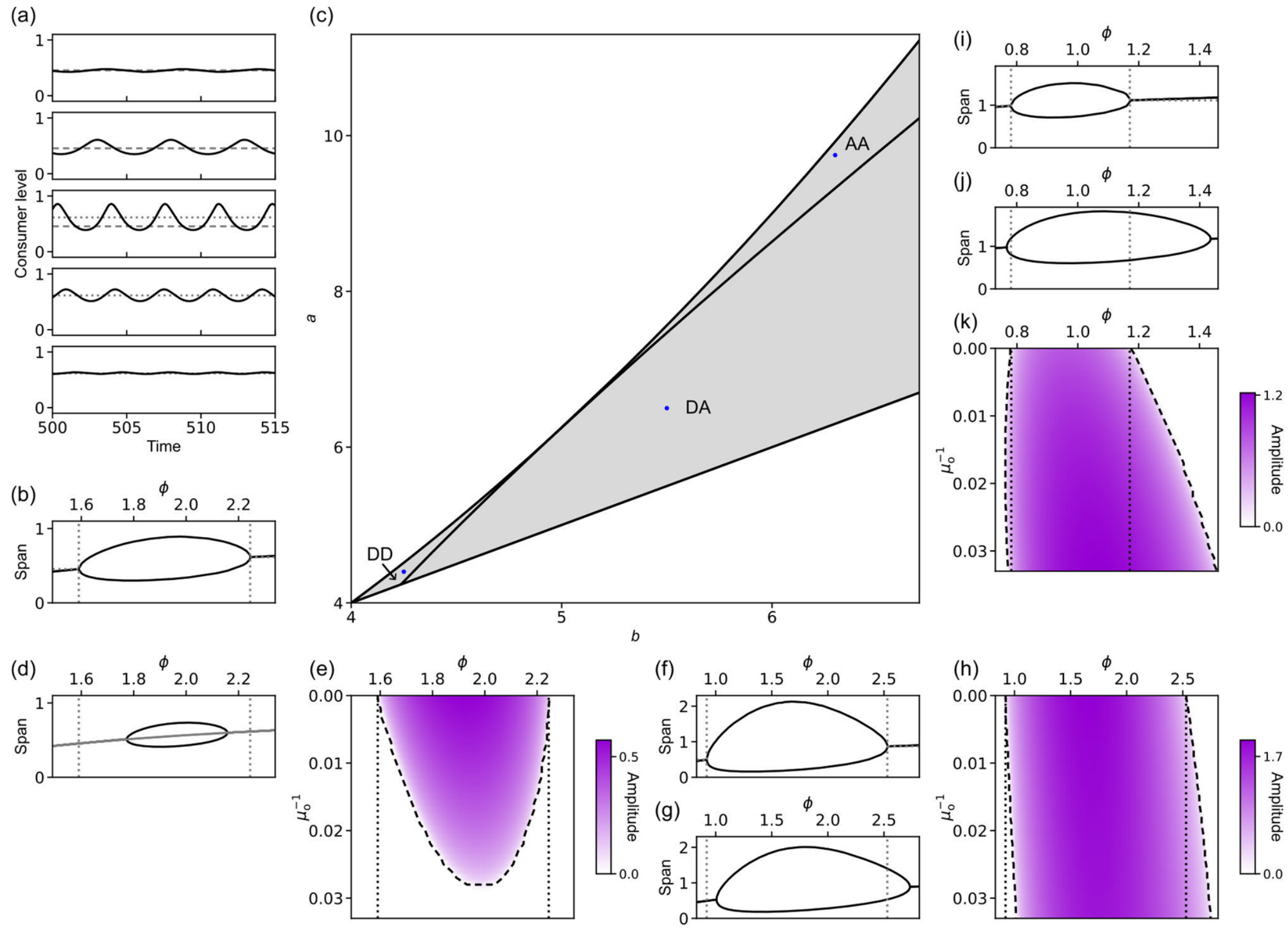

**Fig. 3. Consumer cooperativity and population oscillation.** (a) Time-series of the simulated, QSSA-based consumer level (solid line) with consumer cooperativity $\phi = 1.58, 1.63, 2.1, 2.22$, or $2.27$ (top to bottom) at $a = 4.4$ and $b = 4.25$. The analytical steady state from Eqs. (18)–(20) at $\phi = \phi_{\mathrm{Qc1}}$ (dashed line) or $\phi = \phi_{\mathrm{Qc2}}$ (dotted line) is presented for visual guidance. (b) Bifurcation diagram of the simulated, QSSA-based consumer level at $a$ and $b$ in (a): peak and trough levels as a function of $\phi$ (solid line). From Eqs. (18)–(20), the analytical bifurcation points $\phi = \phi_{\mathrm{Qc1}}, \phi_{\mathrm{Qc2}}$ are marked by the left and right vertical dotted lines, respectively, and their analytical steady states by the left and right horizontal dotted lines. (c) Analytically-predicted effect of tiny $\mu_{\mathrm{o}}^{-1} > 0$ compared to the QSSA, across $a$ and $b$ satisfying both $4 < b$ and $b < a < b^2/4$ (gray-shaded; see SI Section I). “DD”, “DA”, and “AA” areas divided by solid lines are explained in the main text. Each $a$ and $b$ example used for (a), (b), (d), and (e) (“DD”), (f)–(h) (“DA”), or (i)–(k) (“AA”) is marked by the dot within that corresponding area. (d) Bifurcation diagram of the simulated consumer level with $\mu_{\mathrm{o}}^{-1} = 0.02$ (black solid line) or $\mu_{\mathrm{o}}^{-1} = 0.03$ (gray solid line) at $a$ and $b$ in (b) (dot in the “DD” area of (c)): peak and trough consumer levels as a function of $\phi$, together with $\phi = \phi_{\mathrm{Qc1}}, \phi_{\mathrm{Qc2}}$ in (b) (left and right vertical dotted lines, respectively). The oscillation absence at $\mu_{\mathrm{o}}^{-1} = 0.03$ is caused by the continuous bifurcation shifts and collision from both sides of $\phi$ with an increase in $\mu_{\mathrm{o}}^{-1}$, as revealed by (e). (e) Amplitude of the simulated consumer level with $a$ and $b$ in (b): peak-to-trough difference at each $\phi$ and $\mu_{\mathrm{o}}^{-1}$ (violet-colored). Bifurcations at $\phi = \phi_{\mathrm{c1}}, \phi_{\mathrm{c2}}$ are indicated by a dashed line for numerically-calculated $\phi_{\mathrm{c1}}$ and $\phi_{\mathrm{c2}}$ across $\mu_{\mathrm{o}}^{-1}$. For comparison, $\phi = \phi_{\mathrm{Qc1}}, \phi_{\mathrm{Qc2}}$ from (b) are marked by vertical dotted lines. As $\mu_{\mathrm{o}}^{-1}$ increases, $\phi_{\mathrm{c1}}$ and $\phi_{\mathrm{c2}}$ become closer, eventually collide, and disappear. (f) Bifurcation diagram of the simulated, QSSA-based consumer level at $a = 6.5$ and $b = 5.5$ (dot in the “DA” area of (c)): peak and trough consumer

levels as a function of $\phi$ (solid line). Analytical bifurcation points $\phi = \phi_{\mathrm{Qc1}}, \phi_{\mathrm{Qc2}}$ are marked by the left and right vertical dotted lines, respectively, and their analytical steady states by the left and right horizontal dotted lines. (g) Bifurcation diagram of the simulated consumer level with $\mu_{\mathrm{o}}^{-1} = 0.03$ at $a$ and $b$ in (f): peak and trough levels as a function of $\phi$ (solid line), together with $\phi = \phi_{\mathrm{Qc1}}, \phi_{\mathrm{Qc2}}$ in (f) (left and right vertical dotted lines, respectively). (h) Amplitude of the simulated consumer level with $a$ and $b$ in (f): peak-to-trough difference at each $\phi$ and $\mu_{\mathrm{o}}^{-1}$ (violet-colored). Bifurcations at $\phi = \phi_{\mathrm{c1}}, \phi_{\mathrm{c2}}$ are indicated by dashed lines for numerically-calculated $\phi_{\mathrm{c1}}$ and $\phi_{\mathrm{c2}}$ across $\mu_{\mathrm{o}}^{-1}$. For comparison, $\phi = \phi_{\mathrm{Qc1}}, \phi_{\mathrm{Qc2}}$ from (f) are marked by dotted lines. Here, $\phi_{\mathrm{c2}}$ becomes more shifted than $\phi_{\mathrm{c1}}$ as $\mu_{\mathrm{o}}^{-1}$ increases. (i) Bifurcation diagram of the simulated, QSSA-based consumer level at $a = 9.75$ and $b = 6.3$ (dot in the "AA" area of (c)): peak and trough consumer levels as a function of $\phi$ (solid line). Analytical bifurcation points $\phi = \phi_{\mathrm{Qc1}}, \phi_{\mathrm{Qc2}}$ are marked by the left and right vertical dotted lines, respectively, and their analytical steady states by the left and right horizontal dotted lines. (j) Bifurcation diagram of the simulated consumer level with $\mu_{\mathrm{o}}^{-1} = 0.03$ at $a$ and $b$ in (i): peak and trough levels as a function of $\phi$ (solid line), together with $\phi = \phi_{\mathrm{Qc1}}, \phi_{\mathrm{Qc2}}$ in (i) (left and right vertical dotted lines, respectively). Here, the bifurcation shift at $\phi > \phi_{\mathrm{Qc2}}$ is far more pronounced than at $\phi < \phi_{\mathrm{Qc1}}$, as also shown in (k). (k) Amplitude of the simulated consumer level with $a$ and $b$ in (i): peak-to-trough difference at each $\phi$ and $\mu_{\mathrm{o}}^{-1}$ (violet-colored). Bifurcations at $\phi = \phi_{\mathrm{c1}}, \phi_{\mathrm{c2}}$ are indicated by dashed lines for numerically-calculated $\phi_{\mathrm{c1}}$ and $\phi_{\mathrm{c2}}$ across $\mu_{\mathrm{o}}^{-1}$. For comparison, $\phi = \phi_{\mathrm{Qc1}}, \phi_{\mathrm{Qc2}}$ from (i) are marked by dotted lines. Here, $\phi_{\mathrm{c2}}$ becomes far more shifted with $\mu_{\mathrm{o}}^{-1}$ than $\phi_{\mathrm{c1}}$ of a slight change, and this slightness is just attributed to the proximity of $a$ and $b$ to the "DA" area in (c). In (e), (h), and (k), $\mu_{\mathrm{o}}^{-1} = 0$ corresponds to the QSSA. In (a)–(k), all the values are unitless through the scaling of the original quantities (SI Section I).

---

Similar to the existing predator–prey models [6, 8], the oscillation here is maintained by the negative feedback between the consumers and resources, but its unique requirement is the acceleration of the resource exploitation by the cooperating consumers along their population growth. If $\phi$ further increases and crosses another point $\phi = \phi_{\mathrm{Qc2}}$ presented below, then the oscillation turns back into the steady state in Eq. (19) with a gradually-decreased amplitude (Figs. 3(a),(b), and SI Sections I and J):

$$\phi_{\mathrm{Qc2}} \equiv \frac{2}{\frac{2a}{b} - 1 + \sqrt{1 - \frac{4a}{b^2}}} \left( \frac{2}{1 - \sqrt{1 - \frac{4a}{b^2}}} - 1 \right). \tag{20}$$

These QSSA results suggest that the moderate range of $\phi$ ($\phi_{\mathrm{Qc1}} < \phi < \phi_{\mathrm{Qc2}}$) is required for the population oscillations. Both the transitions at $\phi = \phi_{\mathrm{Qc1}}, \phi_{\mathrm{Qc2}}$ are characterized by the supercritical Hopf bifurcation as in Fig. 3(b) (see also other panels in Fig. 3; SI Section J).

Beyond the QSSA, we next explore the relaxation effect of the consumer-captured resource level. Given the dynamics in Eqs. (4) and (15)–(17) with $\mu(t) = \mu_{\mathrm{o}}$, we applied Eqs. (7), (8), and (18)–(20), and analytically identified the following three regimes based on $\left(2\pi/T_{\mathrm{Q}}\right)^2 = (b^2 y_{1\mathrm{s}} - a)/(a - b y_{1\mathrm{s}})$ and $A = b$ at $\phi = \phi_{\mathrm{Qc1}}, \phi_{\mathrm{Qc2}}$ (SI Section I; for all these three regimes, still $4 < b$ and $b < a < b^2/4$ as mentioned before):

(*i*) "DD" area in Fig. 3(c), where $b < 5$ and $(b/4)\left\{9 - (b+3)^2/(b-1)^2\right\} < a$. In this case, we predict that tiny $\mu_{\mathrm{o}}^{-1} > 0$ would suppress the oscillation onset and delay the oscillation transition point compared to the QSSA, along the increase in $\phi$ around $\phi_{\mathrm{Qc1}}$ and the decrease in $\phi$ around $\phi_{\mathrm{Qc2}}$. In other words, $\phi_{\mathrm{c1}} > \phi_{\mathrm{Qc1}}$ and $\phi_{\mathrm{c2}} < \phi_{\mathrm{Qc2}}$ for the actual transition points $\phi = \phi_{\mathrm{c1}}, \phi_{\mathrm{c2}}$. This prediction stands on the Eq. (7)-based relation $\left(\left(2\pi/T_{\mathrm{Q}}\right)^2 < A \text{ at both } \phi = \phi_{\mathrm{c1}}, \phi_{\mathrm{c2}}\right)$ and is confirmed by the numerical simulations (Figs. 3(d),(e) and SI Section J). The larger is $\mu_{\mathrm{o}}^{-1}$, the severer are the transition delays and thus $\phi_{\mathrm{c1}}$ and $\phi_{\mathrm{c2}}$ eventually collide and disappear without any transitions to the oscillations, as demonstrated in Fig. 3(e).

(*ii*) "DA" area in Fig. 3(c), where $a < (b/4)\left\{9 - (b+3)^2/(b-1)^2\right\}$. Compared to the QSSA, we here predict that tiny $\mu_{\mathrm{o}}^{-1} > 0$ would delay the oscillation transition point along the increase in $\phi$ around $\phi_{\mathrm{Qc1}}$ as in the regime (*i*) above, but advance the transition point along the decrease in $\phi$ around $\phi_{\mathrm{Qc2}}$ by facilitating the oscillation onset. In other words, $\phi_{\mathrm{c1}} > \phi_{\mathrm{Qc1}}$ and $\phi_{\mathrm{c2}} > \phi_{\mathrm{Qc2}}$ for the actual transition points $\phi = \phi_{\mathrm{c1}}, \phi_{\mathrm{c2}}$. This prediction stands on the Eq. (7)-based relations $\left(\left(2\pi/T_{\mathrm{Q}}\right)^2 < A \text{ at } \phi = \phi_{\mathrm{Qc1}} \text{ and } \left(2\pi/T_{\mathrm{Q}}\right)^2 > A \text{ at } \phi = \phi_{\mathrm{Qc2}}\right)$ and is confirmed by the numerical simulations (Figs. 3(f)–(h) and SI Section J). In the case of Fig. 3(h), $\phi_{\mathrm{c2}}$ becomes more shifted than $\phi_{\mathrm{c1}}$ as $\mu_{\mathrm{o}}^{-1}$ increases.

(*iii*) "AA" area in Fig. 3(c), where $5 < b$ and $(b/4)\left\{9 - (b+3)^2/(b-1)^2\right\} < a$. In this case, we predict that tiny $\mu_{\mathrm{o}}^{-1} > 0$ would facilitate the oscillation onset and advance the oscillation transition point compared to the QSSA, along the increase in $\phi$ around $\phi_{\mathrm{Qc1}}$ and the decrease in $\phi$ around $\phi_{\mathrm{Qc2}}$. In other words, $\phi_{\mathrm{c1}} < \phi_{\mathrm{Qc1}}$ and $\phi_{\mathrm{c2}} > \phi_{\mathrm{Qc2}}$ for the actual transition points $\phi = \phi_{\mathrm{c1}}, \phi_{\mathrm{c2}}$. This prediction stands on the Eq. (7)-based relation $\left(\left(2\pi/T_{\mathrm{Q}}\right)^2 > A \text{ at both } \phi = \phi_{\mathrm{c1}}, \phi_{\mathrm{c2}}\right)$ and is confirmed by the numerical simulations (Figs. 3(i)–(k) and SI Section J). Yet, these transition advances are not necessarily symmetric between $\phi_{\mathrm{c1}}$ and $\phi_{\mathrm{c2}}$, as demonstrated in Fig. 3(k) where $\phi_{\mathrm{c2}}$ becomes far more shifted than $\phi_{c1}$ of a slight change (this slightness is just attributed to the proximity of $a$ and $b$ to the above regime (*ii*)).

To summarize, the QSSA erroneously informs the oscillation onset, and the actual transition point with the relaxation dynamics becomes advanced or delayed compared to the QSSA depending on $a$ and $b$. The delayed cases are particularly interesting due to the under-expected oscillation suppression by the relaxation dynamics [44, 45]. As previously noted under Eq. (4), the relaxation causes both the time delay and amplitude reduction in $x(t)$ compared to the quasi-steady state $x_{\mathrm{Q}}(t)$. One may thus attribute that oscillation suppression to this amplitude reduction. To clarify the time-delay and amplitude reduction effects, we replaced $x(t)$ by $x_{\mathrm{Q}}(t)$ but incorporating either the time delay, the amplitude reduction, or both, and then re-simulated the model (SI Section J). Surprisingly, we found that the time delay alone reproduces all the observed oscillation suppression/promotion and transition point shifts with tiny $\mu_{\mathrm{o}}^{-1}$, whereas the amplitude reduction just merely modifies them (Figs. 4(a)–(f)). As already mentioned after Eq. (7), this decisive yet nontrivial effect of the time delay can be mechanically understood by the weakening of the centripetal- or centrifugal-like force to a moving object in the phase space (SI Section F).

---

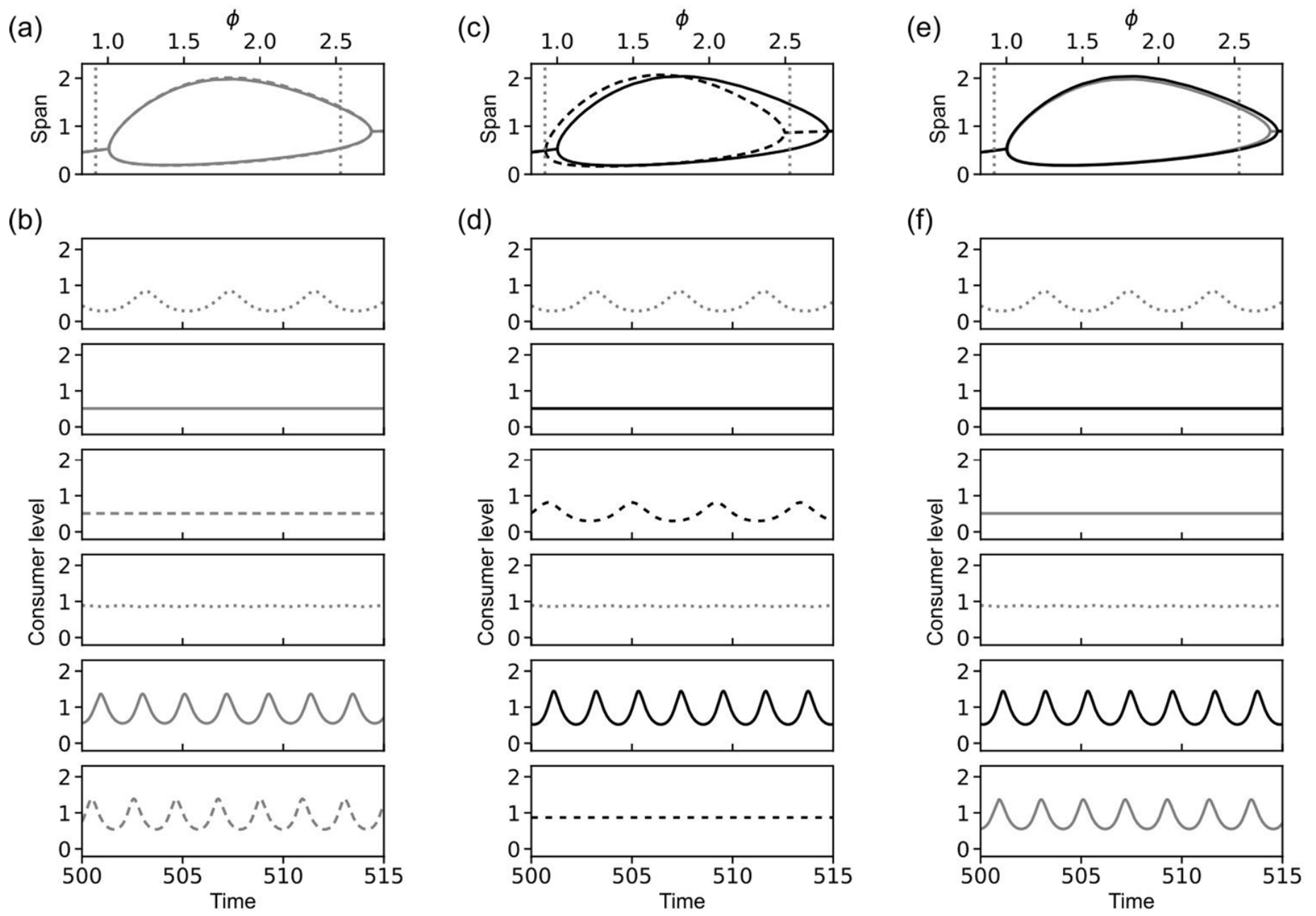

**Fig. 4. Time-delay and amplitude reduction effects of relaxation.** In (a)–(f), $a = 6.5$ and $b = 5.5$ (dot in the "DA" area of Fig. 3(c)), but the observation here remains valid for other $a$ and $b$ values. (a) Bifurcation diagram of the simulated consumer level at $\mu_{\mathrm{o}}^{-1} = 0.03$ (gray dashed line) or with the time delay and amplitude reduction in place of a relaxation process (gray solid line and SI Section J): peak

and trough consumer levels as a function of $\phi$, together with the QSSA-based bifurcation points $\phi = \phi_{\mathrm{Qc1}}, \phi_{\mathrm{Qc2}}$ (left and right vertical dotted lines, respectively). The two bifurcation diagrams overlap almost perfectly, suggesting that the time-delay and amplitude reduction effects well approximate the relaxation dynamics [30]. (b) Time-series of the simulated consumer level with the QSSA (dotted line), with the time delay and amplitude reduction instead of the relaxation process at $\mu_{\mathrm{o}}^{-1}$ in (a) (gray solid line), or with the exact relaxation process (gray dashed line). $\phi = 0.96$ (top three) or $\phi = 2.54$ (bottom three). The oscillation suppression (top three) or promotion (bottom three) by the relaxational effect is consistent with that bifurcation shift relative to the QSSA in (a). (c) Bifurcation diagram of the simulated consumer level with only the time delay (black solid line) or amplitude reduction (black dashed line) in place of the relaxation process with $\mu_{\mathrm{o}}^{-1}$ in (a) (SI Section J). The other contents are the same as (a). Clearly, the time delay reproduces every bifurcation shift in (a) relative to the QSSA, but the amplitude reduction hardly affects the bifurcation (around $\phi = \phi_{\mathrm{Qc1}}$) or just slightly delays it (along the decrease in $\phi$ around $\phi = \phi_{\mathrm{Qc2}}$) by the oscillation suppression. (d) Time-series of the simulated consumer level with only the time delay (black solid line) or amplitude reduction (black dashed line) instead of the relaxation process with $\mu_{\mathrm{o}}^{-1}$ in (a). The other contents are the same as (b), including $\phi$ at each vertical position. The oscillation suppression (top three) or promotion (bottom three) by the time delay alone is consistent with each bifurcation shift from the time delay in (c). (e) Bifurcation diagram of the simulated consumer level with only the time delay (black solid line) or both the time delay and amplitude reduction (gray solid line) in place of the relaxation process with $\mu_{\mathrm{o}}^{-1}$ in (a). The other contents are the same as (a). A mere difference between the two bifurcation diagrams (at $\phi > \phi_{\mathrm{Qc2}}$) suggests that the amplitude reduction just simply modifies the time-delay effect, in a direction consistent with (c). (f) Time-series of the simulated consumer level with only the time delay (black solid line) or both the time delay and amplitude reduction (gray solid line) instead of the relaxation process with $\mu_{\mathrm{o}}^{-1}$ in (a). The other contents are the same as (b), including $\phi$ at each vertical position. The oscillation suppression (top three) and promotion (bottom three) by either the time delay or its combination with the amplitude reduction are consistent with their close bifurcation effects in (e). In (a)–(f), all the values are unitless through the scaling of the original quantities (SI Section I).

---

## Discussion

Our theory and numerical simulations have suggested that relaxation dynamics overlooked by the QSSA quantitatively or even qualitatively alters the transitions in biological interaction systems. Despite the critical slowing down around the transition point, the QSSA may fail to reliably estimate the state-to-state transition time due to the interplay between the relaxation dynamics and near-transition ultrasensitivity. Compared to the QSSA, even the relaxation with tiny $\mu_{\mathrm{o}}^{-1}$ advances or delays the transition point itself for an oscillation onset, and our simple analytical expression in Eq. (7) correctly predicts this shift. This transition point shift is the effect of the time delay from the relaxation process, and physically interpreted as the outcome of the weakened centripetal- or centrifugal-like force in the phase space (SI Section F). The actual mechanical or electronic implementation of this time-delay effect will be an interesting future work [46, 47]. The importance of the time delay is reminiscent of our previous reports of the effective time-delay scheme (ETS) to generalize the Michaelis–Menten rate law [30, 34]. Although not actively addressed in the present study, we anticipate that much larger $\mu_{\mathrm{o}}^{-1}$ would enhance the role of the amplitude reduction from the relaxation process relative to the time delay, because the amplitude reduction and time-delay effects appear to scale as $O(\mu_{\mathrm{o}}^{-2})$ and $O(\mu_{\mathrm{o}}^{-1})$, respectively (SI Section A).

Applications of our work to other biological and even non-biological systems are warranted, as their own feedback interactions would subject the QSSA to its errors. Note that the current study has targeted the essentially low-dimensional dynamics near the transition point. The dynamics far from the transition point may also be of future interest but challenging to address, given its less generality and high-dimensional nature.

If possible, the explicit modeling of all intermediate kinetic processes is the most fundamental solution to avoid the QSSA [34, 36, 48]. However, the QSSA is practically often unavoidable given the system complexity and knowledge deficit. If one nevertheless knows a rate-limiting relaxation process, the structure of the QSSA-based model may still inform the *direction* of the output change under this relaxation process, like the transition time change with the feedback-dependent sign of $b_{\mathrm{Qm}}$ in Eq. (6). Additionally, the exact and QSSA-based models by definition have the same steady states (but not the same stability), and hence the parameters from the steady-state experiments can be adopted for both these models. Yet, to overcome the QSSA's limitation in its kinetic predictions, the "effective" parameters rather than the raw parameters may be considered for the QSSA's match with the kinetic data. However, such effective parameters would not remain constant but vary with the system's state [7, 49]; therefore, instead of using the definite parameter values, we recommend the ensemble model simulations [50] over the feasible parameter ranges with the multiplicative margins of the relaxation time-scales.

## Acknowledgments

We thank Cheol-Min Ghim, Junghun Chae, Tae-Wook Ko, and Adam G. Craig for useful discussions. This work was supported by the General Research Fund (No. 12202322) from the Research Grants Council of the Hong Kong Special Administrative Region, China.

**Supplementary Information (SI) for**

# Widespread quasi-steady state assumption in biological interaction modeling mischaracterizes system transitions

Pan-Jun Kim

Technical details of numerical simulation and statistical analysis methods are provided in Section J. Equation numbers without prefixes "A" to "I" and figure numbers without prefix "S" refer to those in the main text.

## A. Relaxation-inherent time delay and amplitude reduction

Consider a situation that $x(t)$ is rhythmic over time, with the relaxation process in Eq. (4) and $\mu(t) > 0$ from Eq. (3). At the peak or trough time of $x(t)$, $x'(t) = 0$ and therefore $x(t) = x_\mathrm{Q}(t)$. Combined with $\min_t \left(x_\mathrm{Q}(t)\right) \le x_\mathrm{Q}(t) \le \max_t \left(x_\mathrm{Q}(t)\right)$, it leads to $\min_t \left(x_\mathrm{Q}(t)\right) \le \min_t(x(t))$ and $\max_t(x(t)) \le \max_t \left(x_\mathrm{Q}(t)\right)$. Hence, $x(t)$ owns a smaller amplitude than $x_\mathrm{Q}(t)$.

To explore the further relationship between $x(t)$ and $x_\mathrm{Q}(t)$, we rewrite Eq. (4) as

$$\mu^{-1}(t)\frac{\mathrm{d}x(t)}{\mathrm{d}t} + x(t) \approx x_\mathrm{Q}(t). \quad \text{(A1)}$$

Therefore, $x_\mathrm{Q}(t)$ reaches the peak (trough) at the time between the peak (trough) times of $\mu^{-1}(t)x'(t)$ and $x(t)$. In other words, $x(t)$ exhibits a time-delayed profile compared to $x_\mathrm{Q}(t)$, with at most the peak (trough)-time difference between $x(t)$ itself and $\mu^{-1}(t)x'(t)$.

Multiplying both the left- and right-hand sides of Eq. (4) by $\exp\left(\int_{t_0}^{t} \mu(t')\,\mathrm{d}t'\right)$ (with initial time $t_0$) yields

$$\frac{\mathrm{d}}{\mathrm{d}t}\left(x(t)e^{\int_{t_0}^{t}\mu(t')\mathrm{d}t'}\right) \approx \mu(t)x_\mathrm{Q}(t)e^{\int_{t_0}^{t}\mu(t')\mathrm{d}t'}. \quad \text{(A2)}$$

The integral of Eq. (A2) from $t = t_0$ leads to

$$x(t) \approx \int_{t_0}^{t}\mu(t')x_\mathrm{Q}(t')e^{-\int_{t'}^{t}\mu(t'')\mathrm{d}t''}\mathrm{d}t' + x(t_0)e^{-\int_{t_0}^{t}\mu(t')\mathrm{d}t'}. \quad \text{(A3)}$$

For further analysis of Eq. (A3), assume that $\mu(t')$ is rather static over time $t'$ by satisfying

$$\mu(t') \approx \mu(t) \text{ for } t' \text{ in the range } t - \mu^{-1}(t) \lesssim t' \le t. \quad \text{(A4)}$$

From the above condition, $\int_{t'}^{t}\mu(t'')\mathrm{d}t'' \approx (t-t')\mu(t)$ and $\int_{t_0}^{t}\mu(t')\,\mathrm{d}t' \approx (t-t_0)\mu(t_0)$ for $t' \gtrsim t-\mu^{-1}(t)$ and $t \lesssim t_0+\mu^{-1}(t_0)$, respectively. Hence, Eq. (A3) becomes the following form for $t \gg t_0+\mu^{-1}(t_0)$:

$$x(t) \approx \mu(t)\int_{-\infty}^{t} x_{\mathrm{Q}}(t')e^{-(t-t')\mu(t)}\mathrm{d}t'. \tag{A5}$$

The right-hand side is not sensitive to the lower bound of $t'$ for the integral as long as this lower bound is $\ll t-\mu^{-1}(t)$.

In the case of periodic oscillations, it is possible to specify the analytical forms of the aforementioned amplitude reduction and time delay of $x(t)$ compared to $x_{\mathrm{Q}}(t)$. Applying the inverse Fourier transform $x_{\mathrm{Q}}(t)=\int_{-\infty}^{\infty}x_{\mathrm{Q}}^{\mathrm{f}}(\xi)e^{i2\pi\xi t}\mathrm{d}\xi$ to Eq. (A5) results in

$$\begin{aligned} x(t) &\approx \mu(t)e^{-\mu(t)t}\int_{-\infty}^{\infty}\int_{-\infty}^{t} x_{\mathrm{Q}}^{\mathrm{f}}(\xi)e^{(i2\pi\xi+\mu(t))t'}\mathrm{d}t'\mathrm{d}\xi \\ &= \int_{-\infty}^{\infty}\frac{x_{\mathrm{Q}}^{\mathrm{f}}(\xi)}{\sqrt{1+\left(2\pi\xi\mu^{-1}(t)\right)^2}}e^{i\left(2\pi\xi t-\arctan\left(2\pi\xi\mu^{-1}(t)\right)\right)}\mathrm{d}\xi. \end{aligned} \tag{A6}$$

The above form is analogous to the sinusoidal response of an RC circuit [S1]. Eq. (A6) suggests the following approximation when the system exhibits the oscillation with constant period $T$:

$$x(t) \approx \frac{x_{\mathrm{Q}}\left(t-\frac{T}{2\pi}\arctan\left(\frac{2\pi}{T}\mu^{-1}(t)\right)\right)-\langle x_{\mathrm{Q}}(t)\rangle_t}{\sqrt{1+\left(\frac{2\pi}{T}\mu^{-1}(t)\right)^2}}+\langle x_{\mathrm{Q}}(t)\rangle_t, \tag{A7}$$

where $\langle\cdots\rangle_t$ denotes a time average. Compared to $x_{\mathrm{Q}}(t)$, $x(t)$ shows the amplitude reduced by a factor of $\sim\sqrt{1+(2\pi\mu^{-1}(t)/T)^2}$ and the time delay of $\sim(T/2\pi)\cdot\arctan(2\pi\mu^{-1}(t)/T)$. The time delay here is further estimated as $\mu^{-1}(t)$ when $\mu^{-1}(t)\ll T$, and capped by $T/4$ when $\mu^{-1}(t)\gg T$. This cap of the time delay is consistent with a more general discussion under Eq. (A1). The accuracy of Eq. (A7) is numerically confirmed by our recent study in the context of molecular complex formation [S2].

## B. Relaxation rate simplification

Consider the dynamical system with a fixed point (i.e., steady state) at $\left(x(t),y_1(t),y_2(t),\cdots,y_N(t)\right)=(x_{\mathrm{s}},y_{1\mathrm{s}},y_{2\mathrm{s}},\cdots,y_{N\mathrm{s}})$. From Eq. (1), $F(x_{\mathrm{s}},y_{1\mathrm{s}},y_{2\mathrm{s}},\cdots,y_{N\mathrm{s}})=0$, and the Taylor expansion of Eq. (2) gives

$$0=-\mu_{\mathrm{o}}\left(x_{\mathrm{Q}}(t)-x_{\mathrm{s}}\right)+\sum_{i=1}^{N}(y_i(t)-y_{i\mathrm{s}})\cdot\left.\frac{\partial F}{\partial y_i}\right|_{\vec{r}=\vec{0}}+O\left(\left\|\vec{r}_{\mathrm{Q}}(t)\right\|^2\right), \tag{B1}$$

where $\mu_{\rm o} \equiv -\partial_x F|_{\vec{r}=\vec{0}}$, $\vec{r}(t) \equiv \big(x(t), y_1(t), y_2(t), \cdots, y_N(t)\big) - (x_{\rm s}, y_{1{\rm s}}, y_{2{\rm s}}, \cdots, y_{N{\rm s}})$, and $\vec{r}_{\rm Q}(t) \equiv \Big(x_{\rm Q}(t), y_1(t), y_2(t), \cdots, y_N(t)\Big) - (x_{\rm s}, y_{1{\rm s}}, y_{2{\rm s}}, \cdots, y_{N{\rm s}})$. From Eq. (B1), we obtain

$$-\frac{\partial F}{\partial x}\bigg|_{x=x_{\rm Q}(t)} = \mu_{\rm o} - \big(x_{\rm Q}(t) - x_{\rm s}\big) \cdot \frac{\partial^2 F}{\partial x^2}\bigg|_{\vec{r}=\vec{0}} - \textstyle\sum_{i=1}^{N}(y_i(t) - y_{i{\rm s}}) \cdot \frac{\partial^2 F}{\partial x\,\partial y_i}\bigg|_{\vec{r}=\vec{0}} + O\left(\left\|\vec{r}_{\rm Q}(t)\right\|^2\right)$$
$$= \mu_{\rm o} + O\left(\vec{r}_y(t)\right), \qquad \text{(B2)}$$

$$x(t) - x_{\rm Q}(t) = (x(t) - x_{\rm s}) - \mu_{\rm o}^{-1} \textstyle\sum_{i=1}^{N}(y_i(t) - y_{i{\rm s}}) \cdot \frac{\partial F}{\partial y_i}\bigg|_{\vec{r}=\vec{0}} + O\left(\left\|\vec{r}_{\rm Q}(t)\right\|^2\right)$$
$$= (x(t) - x_{\rm s}) - \mu_{\rm o}^{-1} \textstyle\sum_{i=1}^{N}(y_i(t) - y_{i{\rm s}}) \cdot \frac{\partial F}{\partial y_i}\bigg|_{\vec{r}=\vec{0}} + O\left(\left\|\vec{r}_y(t)\right\|^2\right), \qquad \text{(B3)}$$

where $\vec{r}_y(t) \equiv \big(y_1(t), y_2(t), \cdots, y_N(t)\big) - (y_{1{\rm s}}, y_{2{\rm s}}, \cdots, y_{N{\rm s}})$. These equations and the Taylor expansion of the right-hand side of Eq. (1) around $x = x_{\rm Q}(t)$ lead to

$$\frac{{\rm d}x(t)}{{\rm d}t} = \Big(x(t) - x_{\rm Q}(t)\Big) \cdot \frac{\partial F}{\partial x}\bigg|_{x=x_{\rm Q}(t)} + O\left(\Big(x(t) - x_{\rm Q}(t)\Big)^2\right)$$
$$= -\mu_{\rm o}\left\{(x(t) - x_{\rm s}) - \mu_{\rm o}^{-1} \textstyle\sum_{i=1}^{N}(y_i(t) - y_{i{\rm s}}) \cdot \frac{\partial F}{\partial y_i}\bigg|_{\vec{r}=\vec{0}}\right\} + O(\|\vec{r}(t)\|^2). \qquad \text{(B4)}$$

If the system is in the vicinity of the steady state and one can thus ignore $O(\|\vec{r}(t)\|^2)$ compared to $O\big(\vec{r}(t)\big)$, Eqs. (B3) and (B4) show that simplified Eq. (4) with the constant $\mu_{\rm o}$ in place of $\mu(t)$ would still approximate $x'(t)$.

## C. Series solution of relaxation

The direct solution of Eq. (4), i.e., Eq. (A3) and the Taylor expansion $x_{\rm Q}(t') = \sum_{n=0}^{\infty}(-1)^n(t-t')^n x_{\rm Q}^{(n)}(t)/n!$ yield the following form of $x(t)$ when $\mu(t) = \mu_{\rm o}$:

$$x(t) \approx \textstyle\sum_{n=0}^{\infty}\frac{(-1)^n}{n!}\mu_{\rm o} x_{\rm Q}^{(n)}(t)\int_{t_0}^{t}(t-t')^n e^{-\mu_{\rm o}(t-t')}{\rm d}t' + x(t_0)e^{-\mu_{\rm o}(t-t_0)}$$
$$= \textstyle\sum_{n=0}^{\infty}\frac{(-1)^n}{n!}\mu_{\rm o}^{-n} x_{\rm Q}^{(n)}(t)\int_{0}^{\mu_{\rm o}(t-t_0)}\tau^n e^{-\tau}{\rm d}\tau + x(t_0)e^{-\mu_{\rm o}(t-t_0)}$$
$$= \textstyle\sum_{n=0}^{\infty}(-1)^n\mu_{\rm o}^{-n} x_{\rm Q}^{(n)}(t)\left\{1 - e^{-\mu_{\rm o}(t-t_0)}\sum_{k=0}^{n}\frac{\mu_{\rm o}^k(t-t_0)^k}{k!}\right\} + x(t_0)e^{-\mu_{\rm o}(t-t_0)}$$
$$= \textstyle\sum_{n=0}^{\infty}(-1)^n\mu_{\rm o}^{-n} x_{\rm Q}^{(n)}(t) - e^{-\mu_{\rm o}(t-t_0)}\sum_{n=0}^{\infty}\sum_{k=0}^{n}\frac{(-1)^n\mu_{\rm o}^{-k}}{(n-k)!}(t-t_0)^{n-k}x_{\rm Q}^{(n)}(t)$$
$$+x(t_0)e^{-\mu_{\rm o}(t-t_0)}$$
$$= \textstyle\sum_{n=0}^{\infty}(-1)^n\mu_{\rm o}^{-n} x_{\rm Q}^{(n)}(t) - e^{-\mu_{\rm o}(t-t_0)}\sum_{n=0}^{\infty}(-1)^n\mu_{\rm o}^{-n}\sum_{k=0}^{\infty}\frac{(-1)^k}{k!}(t-t_0)^k x_{\rm Q}^{(n+k)}(t)$$
$$+x(t_0)e^{-\mu_{\rm o}(t-t_0)}$$
$$= \textstyle\sum_{n=0}^{\infty}(-1)^n\mu_{\rm o}^{-n} x_{\rm Q}^{(n)}(t) - e^{-\mu_{\rm o}(t-t_0)}\sum_{n=0}^{\infty}(-1)^n\mu_{\rm o}^{-n} x_{\rm Q}^{(n)}(t_0) + x(t_0)e^{-\mu_{\rm o}(t-t_0)}$$
$$= x_{\rm Q}(t) + \textstyle\sum_{n=1}^{\infty}(-1)^n\mu_{\rm o}^{-n}\left\{x_{\rm Q}^{(n)}(t) - e^{-\mu_{\rm o}(t-t_0)}x_{\rm Q}^{(n)}(t_0)\right\} + \Big(x(t_0) - x_{\rm Q}(t_0)\Big)e^{-\mu_{\rm o}(t-t_0)}. \qquad \text{(C1)}$$

The above form is reminiscent of the combined inner and outer solutions (composite solution) in singular perturbation analysis. The condition for the series convergence is

$$\mu_{\mathrm{o}}^{-1}\lim_{n\to\infty}\left|\frac{x_{\mathrm{Q}}^{(n)}(t)-e^{-\mu_{\mathrm{o}}(t-t_0)}x_{\mathrm{Q}}^{(n)}(t_0)}{x_{\mathrm{Q}}^{(n-1)}(t)-e^{-\mu_{\mathrm{o}}(t-t_0)}x_{\mathrm{Q}}^{(n-1)}(t_0)}\right|<1. \tag{C2}$$

This condition is roughly satisfied when $x_{\mathrm{Q}}(t)$ changes slower than $x(t)$ of the time-scale $\mu_{\mathrm{o}}^{-1}$. However, if the series in Eq. (C1) gives the analytically closed form of $x(t)$ as a function of $t$ for particular $x_{\mathrm{Q}}(t)$ and $x(t_0)$, the convergence condition need not be satisfied, because the resulting expression of $x'(t)$ will automatically retrieve $-\mu_{\mathrm{o}}\left(x(t)-x_{\mathrm{Q}}(t)\right)$ in Eq. (4).

As time elapses and $\exp\left(-\mu_{\mathrm{o}}(t-t_0)\right)$ vanishes with $\mu_{\mathrm{o}}>0$, the asymptotic trajectory of $x(t)$ from Eq. (C1) given $x_{\mathrm{Q}}(t)$ becomes the form in Eq. (5). For the application of Eq. (C1) to specific $x_{\mathrm{Q}}(t)$, consider the following system with Eq. (4), $\mu(t)=\mu_{\mathrm{o}}$, and new variable $y(t)$:

$$x_{\mathrm{Q}}(t)=f(y(t)), \tag{C3}$$

$$\frac{\mathrm{d}y(t)}{\mathrm{d}t}=G\big(y(t)\big). \tag{C4}$$

Let $y(t)=y_{\mathrm{s}}$ at the fixed point of the system, i.e., $G(y_{\mathrm{s}})=0$. Eq. (C4) and the linearization of $G(y)$ around that fixed point lead to

$$G(y)\approx\lambda_{\mathrm{Q}}(y-y_{\mathrm{s}}) \text{ with } \lambda_{\mathrm{Q}}\equiv\left.\frac{\mathrm{d}G}{\mathrm{d}y}\right|_{y=y_{\mathrm{s}}}, \tag{C5}$$

$$y(t)\approx y_{\mathrm{s}}+(y(t_0)-y_{\mathrm{s}})e^{\lambda_{\mathrm{Q}}(t-t_0)}. \tag{C6}$$

Combining Eqs. (C1), (C3), and (C6) under the assumption that $f^{(n)}(y)$ is relatively negligible for $n\geq 3$, we obtain

$$\begin{aligned}x_{\mathrm{Q}}(t)\approx{}& f\big(y(t_0)\big)-f'\big(y(t_0)\big)(y(t_0)-y_{\mathrm{s}})\{1-e^{\lambda_{\mathrm{Q}}(t-t_0)}\}\\ &+\tfrac{1}{2}f''\big(y(t_0)\big)(y(t_0)-y_{\mathrm{s}})^2\{1-2e^{\lambda_{\mathrm{Q}}(t-t_0)}+e^{2\lambda_{\mathrm{Q}}(t-t_0)}\},\end{aligned} \tag{C7}$$

$$x(t)\approx x_{\mathrm{Q}}(t)-\mu_{\mathrm{o}}^{-1}\lambda_{\mathrm{Q}}e^{\lambda_{\mathrm{Q}}(t-t_0)}\left[\frac{1-e^{-\mu_{\mathrm{o}}\left(1+\mu_{\mathrm{o}}^{-1}\lambda_{\mathrm{Q}}\right)(t-t_0)}}{1+\mu_{\mathrm{o}}^{-1}\lambda_{\mathrm{Q}}}(y(t_0)-y_{\mathrm{s}})f'\big(y(t_0)\big)-\left\{\frac{1-e^{-\mu_{\mathrm{o}}\left(1+\mu_{\mathrm{o}}^{-1}\lambda_{\mathrm{Q}}\right)(t-t_0)}}{1+\mu_{\mathrm{o}}^{-1}\lambda_{\mathrm{Q}}}-\frac{1-e^{-\mu_{\mathrm{o}}\left(1+2\mu_{\mathrm{o}}^{-1}\lambda_{\mathrm{Q}}\right)(t-t_0)}}{1+2\mu_{\mathrm{o}}^{-1}\lambda_{\mathrm{Q}}}e^{\lambda_{\mathrm{Q}}(t-t_0)}\right\}(y(t_0)-y_{\mathrm{s}})^2f''\big(y(t_0)\big)\right]+\left(x(t_0)-x_{\mathrm{Q}}(t_0)\right)e^{-\mu_{\mathrm{o}}(t-t_0)}. \tag{C8}$$

The above closed form of $x(t)$ does not require any series convergence, as explained under Eq. (C2). From Eqs. (C7) and (C8), the asymptotic behavior of $x(t)$ is then expressed as follows:

(*i*) If $\lambda_{\rm Q} < 0$ and $\mu_{\rm o}^{-1} < -\lambda_{\rm Q}^{-1}$, or if $\lambda_{\rm Q} > 0$,

$$x(t) \approx x_{\rm Q}\left(t - \frac{1}{\lambda_{\rm Q}}\ln\left(1 + \mu_{\rm o}^{-1}\lambda_{\rm Q}\right)\right). \quad \text{(C9)}$$

(*ii*) If $\lambda_{\rm Q} < 0$ and $\mu_{\rm o}^{-1} = -\lambda_{\rm Q}^{-1}$,

$$x(t) \approx x_{\rm Q}\left(t + \frac{1}{\lambda_{\rm Q}}\ln\left(1 - \lambda_{\rm Q}(t - t_0)\right)\right). \quad \text{(C10)}$$

(*iii*) If $\lambda_{\rm Q} < 0$ and $\mu_{\rm o}^{-1} > -\lambda_{\rm Q}^{-1}$,

$$x(t) \approx x_{\rm Q}\left(-\frac{\mu_{\rm o}}{\lambda_{\rm Q}}t + \left(1 + \frac{\mu_{\rm o}}{\lambda_{\rm Q}}\right)t_0 + \eta\right), \quad \text{(C11)}$$

where $\eta$ is a value between $\lambda_{\rm Q}^{-1}\ln\left(\mu_{\rm o}^{-1}\lambda_{\rm Q}/\left(1 + \mu_{\rm o}^{-1}\lambda_{\rm Q}\right)\right)$ and $\lambda_{\rm Q}^{-1}\ln\left(\mu_{\rm o}^{-2}\lambda_{\rm Q}^2/\{\left(1 + \mu_{\rm o}^{-1}\lambda_{\rm Q}\right)\left(1 + 2\mu_{\rm o}^{-1}\lambda_{\rm Q}\right)\}\right)$.

Among the cases (*i*)–(*iii*) above, (*i*) is the most relevant case for relatively fast relaxation processes. In this case, Eq. (C9) reduces to $x(t) \approx x_{\rm Q}(t - \mu_{\rm o}^{-1})$ for the fast relaxation with $\mu_{\rm o}^{-1} \ll |\lambda_{\rm Q}^{-1}|$. The relaxation hence causes a time delay $\sim\mu_{\rm o}^{-1}$ in $x(t)$ compared to $x_{\rm Q}(t)$.

We next consider another system, which underwent a discontinuous transition where the existing stable fixed point suddenly disappears by its merging with a nearby unstable fixed point (saddle-node bifurcation) [S3]. We focus on the case that this system still remains close to the transition point. With $G(y)$ in Eq. (C4), let $y = y_{\rm Qm}$ be the position satisfying $G'(y_{\rm Qm}) = 0$. The Taylor expansion of $G(y)$ up to $O\left((y - y_{\rm Qm})^2\right)$ results in the following normal form:

$$G(y) \approx G_{\rm Qm} + a_{\rm Qm}(y - y_{\rm Qm})^2, \quad \text{(C12)}$$

where $G_{\rm Qm} \equiv G(y_{\rm Qm})$ and $a_{\rm Qm} \equiv G''(y_{\rm Qm})/2$. For this post-transition system, we assume $a_{\rm Qm}G_{\rm Qm} > 0$, and the solution of Eqs. (C4) and (C12) is given by

$$y(t) \approx \sqrt{\frac{G_{\rm Qm}}{a_{\rm Qm}}}\tan\left(\operatorname{sgn}(G_{\rm Qm})\sqrt{a_{\rm Qm}G_{\rm Qm}}(t - t_0) + \arctan\left(\sqrt{\frac{a_{\rm Qm}}{G_{\rm Qm}}}\left(y(t_0) - y_{\rm Qm}\right)\right)\right) + y_{\rm Qm}. \quad \text{(C13)}$$

Without loss of generality, we will henceforth focus on the case of $G_{\rm Qm} > 0$, as the signs of $y(t) - y_{\rm Qm}$ and $y(t_0) - y_{\rm Qm}$ can just be inverted for $G_{\rm Qm} < 0$. For analytical simplicity, we take the following notations and approximant:

$$\tau \equiv \sqrt{a_{\mathrm{Qm}} G_{\mathrm{Qm}}}(t-t_0) + \arctan\left(\sqrt{\frac{a_{\mathrm{Qm}}}{G_{\mathrm{Qm}}}}\left(y(t_0) - y_{\mathrm{Qm}}\right)\right), \tag{C14}$$

$$\tau_0 \equiv \arctan\left(\sqrt{\frac{a_{\mathrm{Qm}}}{G_{\mathrm{Qm}}}}\left(y(t_0) - y_{\mathrm{Qm}}\right)\right) \text{ (i.e., } \tau = \tau_0 \text{ at } t = t_0\text{)}, \tag{C15}$$

$$\bar{\mu}_{\mathrm{o}} \equiv \frac{\mu_{\mathrm{o}}}{\sqrt{a_{\mathrm{Qm}} G_{\mathrm{Qm}}}}, \tag{C16}$$

$$\tan(\tau) \approx \begin{cases} \tau \text{ if } -\frac{\pi}{4} \lesssim \tau \lesssim \frac{\pi}{4}. \\ -\left(\frac{\pi}{2} + \tau\right)^{-1} \text{ if } -\frac{\pi}{2} < \tau \lesssim -\frac{\pi}{4}. \\ \left(\frac{\pi}{2} - \tau\right)^{-1} \text{ if } \frac{\pi}{4} \lesssim \tau < \frac{\pi}{2}. \end{cases} \tag{C17}$$

The above approximant for $\tan(\tau)$ is based on $\tan(\tau) \cdot \tan(2^{-1}\pi - \tau) = 1$. Regarding the calculation of $x(t)$ with Eqs. (C1), (C3), and (C13), we found that the convergence condition in Eq. (C2) is not satisfied for $|\tau| \gtrsim \pi/4$. To avoid this problem, we revisit Eqs. (A3) and (C3) with the Taylor expansions of $f(y)$ around $y = y_{\mathrm{Qm}}$ up to $O\left(\left(y - y_{\mathrm{Qm}}\right)^2\right)$ and of the exponential function, and apply Eqs. (C14)–(C17) there. The result for $-\pi/4 \lesssim \tau \lesssim \pi/4$ is

$$\begin{aligned}
x(\tau) &\approx \bar{\mu}_{\mathrm{o}} f\left(y_{\mathrm{Qm}}\right) \int_{\tau_0}^{\tau} e^{-\bar{\mu}_{\mathrm{o}}(\tau-\tau')} \mathrm{d}\tau' + \bar{\mu}_{\mathrm{o}} f'\left(y_{\mathrm{Qm}}\right) \int_{\tau_0}^{\tau} \left(y(\tau') - y_{\mathrm{Qm}}\right) e^{-\bar{\mu}_{\mathrm{o}}(\tau-\tau')} \mathrm{d}\tau' \\
&\quad + \frac{1}{2}\bar{\mu}_{\mathrm{o}} f''\left(y_{\mathrm{Qm}}\right) \int_{\tau_0}^{\tau} \left(y(\tau') - y_{\mathrm{Qm}}\right)^2 e^{-\bar{\mu}_{\mathrm{o}}(\tau-\tau')} \mathrm{d}\tau' + x(\tau_0) e^{-\bar{\mu}_{\mathrm{o}}(\tau-\tau_0)} \\
&\approx f\left(y_{\mathrm{Qm}}\right)\left\{1 - e^{-\bar{\mu}_{\mathrm{o}}(\tau-\tau_0)}\right\} + \sqrt{\frac{G_{\mathrm{Qm}}}{a_{\mathrm{Qm}}}} f'\left(y_{\mathrm{Qm}}\right) \left\{\int_{-\frac{\pi}{4}}^{\tau} \bar{\mu}_{\mathrm{o}} \tan(\tau') e^{-\bar{\mu}_{\mathrm{o}}(\tau-\tau')} \mathrm{d}\tau' + \right. \\
&\quad \left. \int_{\tau_0}^{-\frac{\pi}{4}} \bar{\mu}_{\mathrm{o}} \tan(\tau') e^{-\bar{\mu}_{\mathrm{o}}(\tau-\tau')} \mathrm{d}\tau' \right\} + \frac{1}{2}\left(\frac{G_{\mathrm{Qm}}}{a_{\mathrm{Qm}}}\right) f''\left(y_{\mathrm{Qm}}\right) \left\{\int_{-\frac{\pi}{4}}^{\tau} \bar{\mu}_{\mathrm{o}} \tan^2(\tau') e^{-\bar{\mu}_{\mathrm{o}}(\tau-\tau')} \mathrm{d}\tau' + \right. \\
&\quad \left. \int_{\tau_0}^{-\frac{\pi}{4}} \bar{\mu}_{\mathrm{o}} \tan^2(\tau') e^{-\bar{\mu}_{\mathrm{o}}(\tau-\tau')} \mathrm{d}\tau' \right\} + x(\tau_0) e^{-\bar{\mu}_{\mathrm{o}}(\tau-\tau_0)} \\
&\approx f\left(y_{\mathrm{Qm}}\right)\left\{1 - e^{-\bar{\mu}_{\mathrm{o}}(\tau-\tau_0)}\right\} + \sqrt{\frac{G_{\mathrm{Qm}}}{a_{\mathrm{Qm}}}} f'\left(y_{\mathrm{Qm}}\right) \left\{\tau - \bar{\mu}_{\mathrm{o}}^{-1} + \left(\frac{\pi}{4} + \bar{\mu}_{\mathrm{o}}^{-1}\right) e^{-\bar{\mu}_{\mathrm{o}}\left(\tau + \frac{\pi}{4}\right)} - \right. \\
&\quad \left. e^{-\bar{\mu}_{\mathrm{o}}\left(\tau + \frac{\pi}{4}\right)} \sum_{n=0}^{\infty} \frac{(-1)^n \bar{\mu}_{\mathrm{o}}^{n+1}}{n!} \int_{\frac{\pi}{2}+\tau_0}^{\frac{\pi}{4}} \frac{\left(\frac{\pi}{4} - \tau'\right)^n}{\tau'} \mathrm{d}\tau' \right\} + \frac{1}{2}\left(\frac{G_{\mathrm{Qm}}}{a_{\mathrm{Qm}}}\right) f''\left(y_{\mathrm{Qm}}\right) \left[(\tau - \bar{\mu}_{\mathrm{o}}^{-1})^2 + \right. \\
&\quad \left. \bar{\mu}_{\mathrm{o}}^{-2} - \left\{\left(\frac{\pi}{4} + \bar{\mu}_{\mathrm{o}}^{-1}\right)^2 + \bar{\mu}_{\mathrm{o}}^{-2}\right\} e^{-\bar{\mu}_{\mathrm{o}}\left(\tau + \frac{\pi}{4}\right)} + e^{-\bar{\mu}_{\mathrm{o}}\left(\tau + \frac{\pi}{4}\right)} \sum_{n=0}^{\infty} \frac{(-1)^n \bar{\mu}_{\mathrm{o}}^{n+1}}{n!} \int_{\frac{\pi}{2}+\tau_0}^{\frac{\pi}{4}} \frac{\left(\frac{\pi}{4} - \tau'\right)^n}{(\tau')^2} \mathrm{d}\tau' \right] \\
&\quad + x(\tau_0) e^{-\bar{\mu}_{\mathrm{o}}(\tau-\tau_0)} \\
&= f\left(y_{\mathrm{Qm}}\right)\left\{1 - e^{-\bar{\mu}_{\mathrm{o}}(\tau-\tau_0)}\right\} + \sqrt{\frac{G_{\mathrm{Qm}}}{a_{\mathrm{Qm}}}} f'\left(y_{\mathrm{Qm}}\right) \left[\tau - \bar{\mu}_{\mathrm{o}}^{-1} + \left(\frac{\pi}{4} + \bar{\mu}_{\mathrm{o}}^{-1}\right) e^{-\bar{\mu}_{\mathrm{o}}\left(\tau + \frac{\pi}{4}\right)} - \right. \\
&\quad \bar{\mu}_{\mathrm{o}} \left[\left(\ln \frac{\pi}{4\left(\frac{\pi}{2} + \tau_0\right)}\right) e^{-\frac{\pi}{4}\bar{\mu}_{\mathrm{o}}} + \sum_{n=1}^{\infty} \sum_{k=1}^{n} \frac{\bar{\mu}_{\mathrm{o}}^n}{k!(n-k)!k} \left\{(-1)^k \left(-\frac{\pi}{4}\right)^n - \left(-\frac{\pi}{4}\right)^{n-k} \left(\frac{\pi}{2} + \right.\right.\right.
\end{aligned}$$

$$\tau_0\Big)^k\Big\}\Big] e^{-\bar{\mu}_o\left(\tau+\frac{\pi}{4}\right)}\Big] + \frac{1}{2}\left(\frac{G_{Qm}}{a_{Qm}}\right) f''(y_{Qm}) \Big[(\tau-\bar{\mu}_o^{-1})^2 + \bar{\mu}_o^{-2} - \Big\{\left(\frac{\pi}{4}+\bar{\mu}_o^{-1}\right)^2 +$$

$$\bar{\mu}_o^{-2}\Big\} e^{-\bar{\mu}_o\left(\tau+\frac{\pi}{4}\right)} + \bar{\mu}_o\left[\frac{4}{\pi}\left\{\frac{\pi}{4\left(\frac{\pi}{2}+\tau_0\right)}-1\right\} e^{-\frac{\pi}{4}\bar{\mu}_o} + \bar{\mu}_o\left(\ln\frac{\pi}{4\left(\frac{\pi}{2}+\tau_0\right)}\right) e^{-\frac{\pi}{4}\bar{\mu}_o} +\right.$$

$$\left.\sum_{n=1}^{\infty}\sum_{k=1}^{n}\frac{(-1)^{n-k}\bar{\mu}_o^{n+1}}{(k+1)!(n-k)!k}\left\{\left(\frac{\pi}{4}\right)^n - \left(\frac{\pi}{4}\right)^{n-k}\left(\frac{\pi}{2}+\tau_0\right)^k\right\}\right] e^{-\bar{\mu}_o\left(\tau+\frac{\pi}{4}\right)}\Big] + x(\tau_0)e^{-\bar{\mu}_o(\tau-\tau_0)}. \tag{C18}$$

Likewise, the result for $\pi/4 \lesssim \tau < \pi/2$ is

$$x(\tau) \approx \bar{\mu}_o f(y_{Qm}) \int_{\tau_0}^{\tau} e^{-\bar{\mu}_o(\tau-\tau')}\mathrm{d}\tau' + \sqrt{\frac{G_{Qm}}{a_{Qm}}} f'(y_{Qm}) \Big\{\int_{\frac{\pi}{4}}^{\tau} \bar{\mu}_o \tan(\tau') e^{-\bar{\mu}_o(\tau-\tau')}\mathrm{d}\tau' +$$

$$\int_{-\frac{\pi}{4}}^{\frac{\pi}{4}} \bar{\mu}_o \tan(\tau') e^{-\bar{\mu}_o(\tau-\tau')}\mathrm{d}\tau' + \int_{\tau_0}^{-\frac{\pi}{4}} \bar{\mu}_o \tan(\tau') e^{-\bar{\mu}_o(\tau-\tau')}\mathrm{d}\tau'\Big\}$$

$$+\frac{1}{2}\left(\frac{G_{Qm}}{a_{Qm}}\right) f''(y_{Qm}) \Big\{\int_{\frac{\pi}{4}}^{\tau} \bar{\mu}_o \tan^2(\tau') e^{-\bar{\mu}_o(\tau-\tau')}\mathrm{d}\tau' + \int_{-\frac{\pi}{4}}^{\frac{\pi}{4}} \bar{\mu}_o \tan^2(\tau') e^{-\bar{\mu}_o(\tau-\tau')}\mathrm{d}\tau' +$$

$$\int_{\tau_0}^{-\frac{\pi}{4}} \bar{\mu}_o \tan^2(\tau') e^{-\bar{\mu}_o(\tau-\tau')}\mathrm{d}\tau'\Big\} + x(\tau_0)e^{-\bar{\mu}_o(\tau-\tau_0)}$$

$$\approx f(y_{Qm})\{1-e^{-\bar{\mu}_o(\tau-\tau_0)}\} + \sqrt{\frac{G_{Qm}}{a_{Qm}}} f'(y_{Qm}) \Big[\sum_{n=0}^{\infty}\frac{(-1)^n\bar{\mu}_o^{n+1}}{n!}\int_{\frac{\pi}{2}-\tau}^{\frac{\pi}{4}}\frac{\left\{\tau'-\left(\frac{\pi}{2}-\tau\right)\right\}^n}{\tau'}\mathrm{d}\tau' +$$

$$\left\{\frac{\pi}{4}-\bar{\mu}_o^{-1}+\left(\frac{\pi}{4}+\bar{\mu}_o^{-1}\right)e^{-\frac{\pi}{2}\bar{\mu}_o}\right\} e^{-\bar{\mu}_o\left(\tau-\frac{\pi}{4}\right)} - e^{-\bar{\mu}_o\left(\tau+\frac{\pi}{4}\right)}\sum_{n=0}^{\infty}\frac{(-1)^n\bar{\mu}_o^{n+1}}{n!}\int_{\frac{\pi}{2}+\tau_0}^{\frac{\pi}{4}}\frac{\left(\frac{\pi}{4}-\tau'\right)^n}{\tau'}\mathrm{d}\tau'\Big]$$

$$+\frac{1}{2}\left(\frac{G_{Qm}}{a_{Qm}}\right) f''(y_{Qm}) \Big[\sum_{n=0}^{\infty}\frac{(-1)^n\bar{\mu}_o^{n+1}}{n!}\int_{\frac{\pi}{2}-\tau}^{\frac{\pi}{4}}\frac{\left\{\tau'-\left(\frac{\pi}{2}-\tau\right)\right\}^n}{(\tau')^2}\mathrm{d}\tau' + \Big[\left(\frac{\pi}{4}-\bar{\mu}_o^{-1}\right)^2 + \bar{\mu}_o^{-2} -$$

$$\left\{\left(\frac{\pi}{4}+\bar{\mu}_o^{-1}\right)^2+\bar{\mu}_o^{-2}\right\}e^{-\frac{\pi}{2}\bar{\mu}_o}\Big] e^{-\bar{\mu}_o\left(\tau-\frac{\pi}{4}\right)} + e^{-\bar{\mu}_o\left(\tau+\frac{\pi}{4}\right)}\sum_{n=0}^{\infty}\frac{(-1)^n\bar{\mu}_o^{n+1}}{n!}\int_{\frac{\pi}{2}+\tau_0}^{\frac{\pi}{4}}\frac{\left(\frac{\pi}{4}-\tau'\right)^n}{(\tau')^2}\mathrm{d}\tau'\Big]$$

$$+x(\tau_0)e^{-\bar{\mu}_o(\tau-\tau_0)}$$

$$= f(y_{Qm})\{1-e^{-\bar{\mu}_o(\tau-\tau_0)}\} + \sqrt{\frac{G_{Qm}}{a_{Qm}}} f'(y_{Qm}) \Big[\bar{\mu}_o\Big[\left(\ln\frac{\pi}{4\left(\frac{\pi}{2}-\tau\right)}\right)e^{\bar{\mu}_o\left(\frac{\pi}{2}-\tau\right)} +$$

$$\sum_{n=1}^{\infty}\sum_{k=1}^{n}\frac{\bar{\mu}_o^n}{k!(n-k)!k}\left\{\left(-\frac{\pi}{4}\right)^k\left(\frac{\pi}{2}-\tau\right)^{n-k} - (-1)^k\left(\frac{\pi}{2}-\tau\right)^n\right\}\Big] + \Big\{\frac{\pi}{4}-\bar{\mu}_o^{-1} +$$

$$\left(\frac{\pi}{4}+\bar{\mu}_o^{-1}\right)e^{-\frac{\pi}{2}\bar{\mu}_o}\Big\} e^{-\bar{\mu}_o\left(\tau-\frac{\pi}{4}\right)} - \bar{\mu}_o\Big[\left(\ln\frac{\pi}{4\left(\frac{\pi}{2}+\tau_0\right)}\right)e^{-\frac{\pi}{4}\bar{\mu}_o} +$$

$$\sum_{n=1}^{\infty}\sum_{k=1}^{n}\frac{\bar{\mu}_o^n}{k!(n-k)!k}\left\{(-1)^k\left(-\frac{\pi}{4}\right)^n - \left(-\frac{\pi}{4}\right)^{n-k}\left(\frac{\pi}{2}+\tau_0\right)^k\right\}\Big] e^{-\bar{\mu}_o\left(\tau+\frac{\pi}{4}\right)}\Big] +$$

$$\frac{1}{2}\left(\frac{G_{Qm}}{a_{Qm}}\right) f''(y_{Qm}) \Big[\bar{\mu}_o\Big[\frac{4}{\pi}\left\{\frac{\pi}{4\left(\frac{\pi}{2}-\tau\right)}-1\right\} - \bar{\mu}_o\left(\ln\frac{\pi}{4\left(\frac{\pi}{2}-\tau\right)}\right) -$$

$$e^{-\bar{\mu}_o\left(\frac{\pi}{2}-\tau\right)}\sum_{n=1}^{\infty}\sum_{k=1}^{n}\frac{(-1)^k\bar{\mu}_o^{n+1}}{(k+1)!(n-k)!k}\left\{\left(\frac{\pi}{4}\right)^k\left(\frac{\pi}{2}-\tau\right)^{n-k} - \left(\frac{\pi}{2}-\tau\right)^n\right\}\Big] e^{\bar{\mu}_o\left(\frac{\pi}{2}-\tau\right)} +$$

$$\left[\left(\frac{\pi}{4}-\bar{\mu}_{\rm o}^{-1}\right)^2+\bar{\mu}_{\rm o}^{-2}-\left\{\left(\frac{\pi}{4}+\bar{\mu}_{\rm o}^{-1}\right)^2+\bar{\mu}_{\rm o}^{-2}\right\}e^{-\frac{\pi}{2}\bar{\mu}_{\rm o}}\right]e^{-\bar{\mu}_{\rm o}\left(\tau-\frac{\pi}{4}\right)}+\bar{\mu}_{\rm o}\left[\frac{4}{\pi}\left\{\frac{\pi}{4\left(\frac{\pi}{2}+\tau_0\right)}-1\right\}e^{-\frac{\pi}{4}\bar{\mu}_{\rm o}}+\bar{\mu}_{\rm o}\left(\ln\frac{\pi}{4\left(\frac{\pi}{2}+\tau_0\right)}\right)e^{-\frac{\pi}{4}\bar{\mu}_{\rm o}}+\sum_{n=1}^{\infty}\sum_{k=1}^{n}\frac{(-1)^{n-k}\bar{\mu}_{\rm o}^{n+1}}{(k+1)!(n-k)!k}\left\{\left(\frac{\pi}{4}\right)^n-\left(\frac{\pi}{4}\right)^{n-k}\left(\frac{\pi}{2}+\tau_0\right)^k\right\}\right]e^{-\bar{\mu}_{\rm o}\left(\tau+\frac{\pi}{4}\right)}+x(\tau_0)e^{-\bar{\mu}_{\rm o}(\tau-\tau_0)}. \quad \text{(C19)}$$

For comparison with the QSSA,

$$x_{\rm Q}(\tau)\approx f(y_{\rm Qm})+\sqrt{\frac{G_{\rm Qm}}{a_{\rm Qm}}}f'(y_{\rm Qm})\tan(\tau)+\frac{1}{2}\left(\frac{G_{\rm Qm}}{a_{\rm Qm}}\right)f''(y_{\rm Qm})\tan^2(\tau). \quad \text{(C20)}$$

If $\bar{\mu}_{\rm o}\gg 1$, e.g., near the transition (bifurcation) point with $G_{\rm Qm}\to 0^+$ from Eqs. (C16), the system exhibits the following behaviors according to Eqs. (C17)–(C20):

(*i*) When $-\pi/4\lesssim\tau\lesssim\pi/4$,

$$x(\tau)\approx x_{\rm Q}(\tau-\bar{\mu}_{\rm o}^{-1}),\ \text{or equivalently,}\ x(t)\approx x_{\rm Q}(t-\mu_{\rm o}^{-1}). \quad \text{(C21)}$$

(*ii*) When $\tau\to(\pi/2)^-$,

$$x(\tau)\approx\begin{cases}x_{\rm Q}\left(\tau-\frac{\bar{\mu}_{\rm o}^{-1}}{\ln\left(\frac{1}{\frac{\pi}{2}-\tau}\right)}\right) & \text{if } f'(y_{\rm Qm})\neq 0 \text{ and } f''(y_{\rm Qm})=0,\\ x_{\rm Q}\left(\tau-\sqrt{\bar{\mu}_{\rm o}^{-1}\left(\frac{\pi}{2}-\tau\right)}\right) & \text{if } f''(y_{\rm Qm})\neq 0,\end{cases} \quad \text{(C22)}$$

or equivalently,

$$x(t)\approx\begin{cases}x_{\rm Q}\left(t-\frac{\mu_{\rm o}^{-1}}{\ln\left(\frac{1}{\frac{\pi}{2}-\tau(t)}\right)}\right) & \text{if } f'(y_{\rm Qm})\neq 0 \text{ and } f''(y_{\rm Qm})=0.\\ x_{\rm Q}\left(t-\sqrt{\frac{\mu_{\rm o}^{-1}}{\sqrt{a_{\rm Qm}G_{\rm Qm}}}\left(\frac{\pi}{2}-\tau(t)\right)}\right) & \text{if } f''(y_{\rm Qm})\neq 0.\end{cases} \quad \text{(C23)}$$

Here, $\tau(t)$ denotes the expression of $\tau$ as a function of $t$ in Eq. (C14).

Based on the cases (*i*) and (*ii*) above, $x(t)$ is the time-shifted form of $x_{\rm Q}(t)$ with a delay $\sim\mu_{\rm o}^{-1}$ during the bottleneck period around $y(t)=y_{\rm Qm}$, but the delay becomes vanishing as time further elapses.

## D. Feedback interaction and transition time

Suppose that $x(t)$ in Eq. (4) is not only affected by $x_{\mathrm{Q}}(t)$ but also feeds back into $x_{\mathrm{Q}}(t)$. The dynamics of this system can be described by Eqs. (4) and (C3), $\mu(t) = \mu_{\mathrm{o}}$, and the following equation with a relatively slow variable $y(t)$, if the system is close to a transition point and essentially low-dimensional there:

$$\frac{\mathrm{d}y(t)}{\mathrm{d}t} = G\big(x(t), y(t)\big). \tag{D1}$$

At the fixed point $\big(x(t), y(t)\big) = (x_{\mathrm{s}}, y_{\mathrm{s}})$, it is satisfied from Eqs. (4), (C3), and (D1) that $x_{\mathrm{s}} = f(y_{\mathrm{s}})$ and $G(x_{\mathrm{s}}, y_{\mathrm{s}}) = 0$. For the linear stability analysis of this fixed point, we obtain the Jacobian matrix J and its eigenvalue $\lambda$ with the larger real part than the remaining eigenvalue:

$$\mathrm{J} = \begin{pmatrix} -\mu_{\mathrm{o}} & \mu_{\mathrm{o}}\frac{\mathrm{d}f}{\mathrm{d}y} \\ \frac{\partial G}{\partial x} & \frac{\partial G}{\partial y} \end{pmatrix}, \tag{D2}$$

$$\lambda = \begin{cases} \frac{1}{2\mu_{\mathrm{o}}^{-1}}\left\{-1 + \mu_{\mathrm{o}}^{-1}\frac{\partial G}{\partial y} + \sqrt{4\mu_{\mathrm{o}}^{-1}\lambda_{\mathrm{Q}} + \left(1 - \mu_{\mathrm{o}}^{-1}\frac{\partial G}{\partial y}\right)^2}\right\} & \text{if } 4\mu_{\mathrm{o}}^{-1}\lambda_{\mathrm{Q}} + \left(1 - \mu_{\mathrm{o}}^{-1}\frac{\partial G}{\partial y}\right)^2 \geq 0. \\ \frac{1}{2\mu_{\mathrm{o}}^{-1}}\left\{-1 + \mu_{\mathrm{o}}^{-1}\frac{\partial G}{\partial y} \pm i\sqrt{-4\mu_{\mathrm{o}}^{-1}\lambda_{\mathrm{Q}} - \left(1 - \mu_{\mathrm{o}}^{-1}\frac{\partial G}{\partial y}\right)^2}\right\} & \text{if } 4\mu_{\mathrm{o}}^{-1}\lambda_{\mathrm{Q}} + \left(1 - \mu_{\mathrm{o}}^{-1}\frac{\partial G}{\partial y}\right)^2 < 0. \end{cases} \tag{D3}$$

The derivatives in Eqs. (D2) and (D3) are taken at the fixed point, and for simplicity we will use these notations until Eq. (D6) below. $\lambda_{\mathrm{Q}}$ in Eq. (D3) is the QSSA version of $\lambda$, with this expression:

$$\lambda_{\mathrm{Q}} = \frac{\partial G}{\partial x}\frac{\mathrm{d}f}{\mathrm{d}y} + \frac{\partial G}{\partial y}. \tag{D4}$$

According to Eq. (D3), $\lambda \to \lambda_{\mathrm{Q}}$ for $\mu_{\mathrm{o}}^{-1} \to 0$, and we express $\lambda$ in powers of $\mu_{\mathrm{o}}^{-1}$ as

$$\lambda = \left(1 - \mu_{\mathrm{o}}^{-1}\frac{\partial G}{\partial x}\frac{\mathrm{d}f}{\mathrm{d}y}\right)\lambda_{\mathrm{Q}} + O(\mu_{\mathrm{o}}^{-2}). \tag{D5}$$

If the system moves from one state to another and this event is rate-limited by the repulsion (attraction) of the fixed point of the former (latter) state, the transition time $t_{\mathrm{r}}$ and its QSSA $t_{\mathrm{Qr}}$ follow the relations $t_{\mathrm{r}} \propto \lambda^{-1}$, $t_{\mathrm{Qr}} \propto \lambda_{\mathrm{Q}}^{-1}$, and $\lambda t_{\mathrm{r}} \sim \lambda_{\mathrm{Q}} t_{\mathrm{Qr}}$. These relations and Eq. (D5) lead to

$$t_{\mathrm{r}} \approx \frac{t_{\mathrm{Qr}}}{1 - \mu_{\mathrm{o}}^{-1}\frac{\partial G}{\partial x}\frac{\mathrm{d}f}{\mathrm{d}y}}. \tag{D6}$$

If $\partial_x G \cdot f'$ above takes a finite non-zero value at the transition point with $\lambda_{\mathrm{Q}} = 0$ (or equivalently, $t_{\mathrm{Qr}} \to \infty$), we can apply that value to the near-transition description of Eq. (D6). In this case, the multiplicative relationship between $t_{\mathrm{Qr}}$ and $t_{\mathrm{r}}$ renders their difference $|t_{\mathrm{r}} - t_{\mathrm{Qr}}|$ proportional to the transition time $t_{\mathrm{r}}$ itself. Hence, if the increase in $x(t)$ is promotive (repressive) of $x_{\mathrm{Q}}(t)$ so that $\partial_x G \cdot f' > 0$ ($\partial_x G \cdot f' < 0$), the QSSA can severely underestimate (overestimate) the transition time according to Eq. (D6). The stronger the feedback of $x(t)$ to $x_{\mathrm{Q}}(t)$ (i.e., the higher $|\partial_x G \cdot f'|$), the larger this QSSA's error. A similar result also arises in the absence of the fixed point, as will be detailed next.

We now consider another system, which underwent a saddle-node bifurcation with the disappearance of the fixed points and then stays near the bifurcation point. This system can be described by Eqs. (4), (C3), and (D1) with $\mu(t) = \mu_{\mathrm{o}}$. Hereafter $x_*(y(t))$ will refer to the asymptotic trajectory of $x(t)$ in the phase space. Around the saddle-node remnant, we adopt Eq. (5) up to $O(\mu_{\mathrm{o}}^{-1})$ for small $\mu_{\mathrm{o}}^{-1}$ and obtain

$$x_*(y) \approx f(y) - \mu_{\mathrm{o}}^{-1} G(f(y), y) \frac{\mathrm{d}f(y)}{\mathrm{d}y}, \tag{D7}$$

$$\frac{\mathrm{d}y}{\mathrm{d}t} = G(x_*(y), y) \approx G(f(y), y) \left(1 - \mu_{\mathrm{o}}^{-1} \frac{\mathrm{d}f(y)}{\mathrm{d}y} \frac{\partial G(x,y)}{\partial x}\bigg|_{x=f(y)}\right). \tag{D8}$$

Let $y$ change most slowly at $y = y_{\mathrm{m}}$ as its bottleneck position. Based on Eq. (D8), $y_{\mathrm{m}}$ shall satisfy

$$0 \approx \frac{\mathrm{d}}{\mathrm{d}y}\left\{G(f(y), y) \left(1 - \mu_{\mathrm{o}}^{-1} \frac{\mathrm{d}f(y)}{\mathrm{d}y} \frac{\partial G(x,y)}{\partial x}\bigg|_{x=f(y)}\right)\right\}\bigg|_{y=y_{\mathrm{m}}}, \tag{D9}$$

and therefore,

$$0 \approx \lambda_{\mathrm{m}} - \mu_{\mathrm{o}}^{-1}\{2a_{\mathrm{m}} G_{\mathrm{m}}(1 + u_{\mathrm{m}}) + \lambda_{\mathrm{m}}(\lambda_{\mathrm{m}} + b_{\mathrm{m}})\}, \tag{D10}$$

$$G_{\mathrm{m}} \equiv G(f(y_{\mathrm{m}}), y_{\mathrm{m}}), \tag{D11}$$

$$\lambda_{\mathrm{m}} \equiv \frac{\mathrm{d}G(f(y),y)}{\mathrm{d}y}\bigg|_{y=y_{\mathrm{m}}}, \tag{D12}$$

$$a_{\mathrm{m}} \equiv \frac{1}{2} \frac{\mathrm{d}^2 G(f(y),y)}{\mathrm{d}y^2}\bigg|_{y=y_{\mathrm{m}}}, \tag{D13}$$

$$b_{\mathrm{m}} \equiv -\frac{\partial G(x,y)}{\partial y}\bigg|_{x=f(y_{\mathrm{m}}),y=y_{\mathrm{m}}}, \tag{D14}$$

$$u_{\mathrm{m}} \equiv -\frac{1}{2a_{\mathrm{m}}}\left(\frac{\partial^2 G(x,y)}{\partial x\,\partial y} \frac{\mathrm{d}f(y)}{\mathrm{d}y} + \frac{\partial^2 G(x,y)}{\partial y^2}\right)\bigg|_{x=f(y_{\mathrm{m}}),y=y_{\mathrm{m}}}. \tag{D15}$$

From Eq. (D10),

$$\lambda_{\mathrm{m}} \approx 2\mu_{\mathrm{o}}^{-1} a_{\mathrm{m}} G_{\mathrm{m}}(1+u_{\mathrm{m}}). \tag{D16}$$

Let $y_{\mathrm{Qm}}$ be the QSSA version of $y_{\mathrm{m}}$. $y_{\mathrm{Qm}}$ is then the solution of

$$0 = \lambda_{\mathrm{Qm}} \equiv \left.\frac{\mathrm{d}G(f(y),y)}{\mathrm{d}y}\right|_{y=y_{\mathrm{Qm}}}. \tag{D17}$$

Similarly, we define $G_{\mathrm{Qm}}$, $a_{\mathrm{Qm}}$, $b_{\mathrm{Qm}}$, and $u_{\mathrm{Qm}}$. From Eq. (D16), we obtain

$$y_{\mathrm{m}} - y_{\mathrm{Qm}} \approx \mu_{\mathrm{o}}^{-1} G_{\mathrm{Qm}}(1+u_{\mathrm{Qm}}). \tag{D18}$$

The Taylor expansion of Eq. (D8) up to $O((\mathrm{y}-y_{\mathrm{m}})^2)$ then results in the following normal form around $y = y_{\mathrm{m}}$:

$$\frac{\mathrm{d}y}{\mathrm{d}t} \approx G_{\mathrm{Qm}}(1-\mu_{\mathrm{o}}^{-1} b_{\mathrm{Qm}}) + a_{\mathrm{Qm}}(1-\mu_{\mathrm{o}}^{-1} c_{\mathrm{Qm}})(\mathrm{y}-y_{\mathrm{m}})^2, \tag{D19}$$

where $c_{\mathrm{Qm}}$ is given by

$$c_{\mathrm{Qm}} \equiv b_{\mathrm{Qm}} - \frac{G_{\mathrm{Qm}}}{2a_{\mathrm{Qm}}}\left\{u_{\mathrm{Qm}}\frac{\mathrm{d}^3G(f(y),y)}{\mathrm{d}y^3} + \frac{\partial^3 G(x,y)}{\partial y^3} + 2\frac{\partial^3 G(x,y)}{\partial \mathrm{x}\,\partial y^2}\frac{\mathrm{d}f(y)}{\mathrm{d}y} + \frac{\partial^3 G(x,y)}{\partial x^2\,\partial y}\left(\frac{\mathrm{d}f(y)}{\mathrm{d}y}\right)^2 + \left.\frac{\partial^2 G(x,y)}{\partial x\,\partial y}\frac{\mathrm{d}^2 f(y)}{\mathrm{d}y^2}\right\}\right|_{x=f(y_{\mathrm{Qm}}),y=y_{\mathrm{Qm}}}. \tag{D20}$$

About this post-bifurcation system with Eq. (D19), we assume $a_{\mathrm{Qm}}G_{\mathrm{Qm}} > 0$.

Note that $G_{\mathrm{Qm}} \to 0$ as the system nears the bifurcation point. Away from the bottleneck position $y = y_{\mathrm{m}}$, Eq. (D19) would still hold if

$$o((\mathrm{y}-y_{\mathrm{m}})^{-1}) = \frac{\mathrm{d}^3G(x_*(y),y)}{\mathrm{d}y^3} = \left\{\frac{\partial^3G(x,y)}{\partial x^3}\left(\frac{\mathrm{d}x_*(y)}{\mathrm{d}y}\right)^3 + \frac{\partial^3G(x,y)}{\partial y^3} + 3\frac{\partial^3G(x,y)}{\partial x^2\,\partial y}\left(\frac{\mathrm{d}x_*(y)}{\mathrm{d}y}\right)^2 + 3\frac{\partial^3G(x,y)}{\partial x\,\partial y^2}\frac{\mathrm{d}x_*(y)}{\mathrm{d}y} + 3\frac{\partial^2G(x,y)}{\partial x^2}\frac{\mathrm{d}^2x_*(y)}{\mathrm{d}y^2}\frac{\mathrm{d}x_*(y)}{\mathrm{d}y} + 3\frac{\partial^2G(x,y)}{\partial x\,\partial y}\frac{\mathrm{d}^2x_*(y)}{\mathrm{d}y^2} + \left.\frac{\partial G(x,y)}{\partial x}\frac{\mathrm{d}^3x_*(y)}{\mathrm{d}y^3}\right\}\right|_{x=x_*(y)}, \tag{D21}$$

where $x_*(y)$ is the aforementioned asymptotic trajectory. If we consider $f(y)$ with relatively negligible $f^{(n)}(y)$ for $n \geq 3$, $x_*(y)$ becomes bounded by $O((\mathrm{y}-y_{\mathrm{m}})^2)$ due to Eq. (A3) with the Taylor expansion of $f(y)$ around $y = y_{\mathrm{m}}$. As a result, the example to meet Eq. (D21) is

$$G(x,y) = O(x) + x\cdot o(y^0) + o(y^2). \tag{D22}$$

In such cases to satisfy Eq. (D21), one can extend the use of Eq. (D19) beyond the vicinity of $y = y_{\mathrm{m}}$, whether Eqs. (D7) and (D8) hold or not based on their convergence like the discussion under Eq. (C17). This need to meet Eq. (D21) will nevertheless diminish as $G_{\mathrm{Qm}} \to 0$ because most of the system's time will be spent near $y = y_{\mathrm{m}}$.

Owing to Eq. (D19), the transition time $t_{\mathrm{r}}$ for a certain state far from the bottleneck position $y = y_{\mathrm{m}}$ is estimated as

$$t_{\mathrm{r}} \approx \int_{-\infty}^{\infty} \frac{\mathrm{d}y}{|G_{\mathrm{Qm}}(1-\mu_{\mathrm{o}}^{-1}b_{\mathrm{Qm}})+a_{\mathrm{Qm}}(1-\mu_{\mathrm{o}}^{-1}c_{\mathrm{Qm}})(y-y_{\mathrm{m}})^2|} \approx \frac{\pi}{\sqrt{a_{\mathrm{Qm}}G_{\mathrm{Qm}}\{1-\mu_{\mathrm{o}}^{-1}(b_{\mathrm{Qm}}+c_{\mathrm{Qm}})\}}}. \quad \text{(D23)}$$

For a simpler interpretation of Eq. (D23), we redefine $y_{\mathrm{Qm}}$ (in the calculation of $G_{\mathrm{Qm}}$), $a_{\mathrm{Qm}}$, and $b_{\mathrm{Qm}}$ as those at the bifurcation point, provided that $a_{\mathrm{Qm}}$ and $b_{\mathrm{Qm}}$ take finite non-zero values there. The relative errors of the new $G_{\mathrm{Qm}}$, $a_{\mathrm{Qm}}$, and $b_{\mathrm{Qm}}$ against their original values do disappear as the bifurcation point is closer, owing to Eq. (D17). In this case towards the bifurcation, $c_{\mathrm{Qm}} \to b_{\mathrm{Qm}}$ with $G_{\mathrm{Qm}} \to 0$ and

$$t_{\mathrm{r}} \approx \frac{\pi}{\sqrt{a_{\mathrm{Qm}}G_{\mathrm{Qm}}(1-2\mu_{\mathrm{o}}^{-1}b_{\mathrm{Qm}})}}. \quad \text{(D24)}$$

At $\mu_{\mathrm{o}}^{-1} = 0$, the QSSA ($t_{\mathrm{Qr}}$) of $t_{\mathrm{r}}$ is given by

$$t_{\mathrm{Qr}} \approx \frac{\pi}{\sqrt{a_{\mathrm{Qm}}G_{\mathrm{Qm}}}}. \quad \text{(D25)}$$

Combining Eqs. (D24) and (D25) reveals the relation between $t_{\mathrm{Qr}}$ and $t_{\mathrm{r}}$ in Eq. (6). Because $b_{\mathrm{Qm}} = -\partial_y G|_{x=f(y_{\mathrm{Qm}}),y=y_{\mathrm{Qm}}} = f'(y_{\mathrm{Qm}}) \cdot \partial_x G|_{x=f(y_{\mathrm{Qm}}),y=y_{\mathrm{Qm}}}$ from Eq. (D17), it is obvious that $b_{\mathrm{Qm}} > 0$ ($b_{\mathrm{Qm}} < 0$) for Eq. (6) if the increase in $x(t)$ is promotive (repressive) of $x_{\mathrm{Q}}(t)$. The multiplicative relationship between $t_{\mathrm{Qr}}$ and $t_{\mathrm{r}}$ in Eq. (6) renders the difference $|t_{\mathrm{r}} - t_{\mathrm{Qr}}|$ proportional to the transition time $t_{\mathrm{r}}$ itself, and therefore the QSSA can noticeably underestimate (overestimate) the transition time in the presence of the promotive (repressive) action of $x(t)$ on $x_{\mathrm{Q}}(t)$. The stronger this feedback effect (i.e., the higher $|b_{\mathrm{Qm}}|$), the larger the QSSA's error. Interestingly, the structure of Eq. (6) matches that of Eq. (D6) when $|\mu_{\mathrm{o}}^{-1}b_{\mathrm{Qm}}| \ll 1$, despite their different contexts.

Of note, the use of $\mu(t) = \mu_{\mathrm{o}}$ here for Eq. (4) stands on the following observation related to Section B: as $G_{\mathrm{Qm}} \to 0$, Eq. (D18) shows $y_{\mathrm{m}} \to y_{\mathrm{Qm}}$ and the system thus mostly resides near $y = y_{\mathrm{Qm}}$. At the bifurcation, this position coincides with its fixed point, i.e., saddle-node fixed point. Therefore, as $G_{\mathrm{Qm}} \to 0$, $\mu(t)$ becomes essentially constant with $\mu_{\mathrm{o}} = -\partial_x F|_{x=f(y_{\mathrm{Qm}}),y=y_{\mathrm{Qm}}}$ calculated at the bifurcation, according to Section B.

### E. Oscillation transition point

Consider the dynamical system with a continuous transition from the steady state to periodic oscillations (supercritical Hopf bifurcation) [S3, S4] when based on the QSSA. The dynamics of this system can be described by Eq. (4), $\mu(t) = \mu_\mathrm{o}$, and the following equations if the system is close to the bifurcation (transition) point and essentially low-dimensional there:

$$\frac{\mathrm{d}y_1(t)}{\mathrm{d}t} = G_1\big(x(t), y_1(t), y_2(t)\big), \tag{E1}$$

$$\frac{\mathrm{d}y_2(t)}{\mathrm{d}t} = G_2(x(t), y_1(t), y_2(t)), \tag{E2}$$

$$x_\mathrm{Q}(t) = f(y_1(t), y_2(t)). \tag{E3}$$

Here, $y_1(t)$ and $y_2(t)$ are slower variables than $x(t)$ of the time-scale $\mu_\mathrm{o}^{-1}$. $x(t)$ is not only affected by $x_\mathrm{Q}(t)$ through Eq. (4) but also feeds back into $x_\mathrm{Q}(t)$ through Eqs. (E1)–(E3).

Of importance to the supercritical Hopf bifurcation is a spiral fixed point, i.e., the steady state with an oscillatory potential [S3]. This fixed point $\big(x(t), y_1(t), y_2(t)\big) = (x_\mathrm{s}, y_{1\mathrm{s}}, y_{2\mathrm{s}})$ satisfies the conditions from Eqs. (4) and (E1)–(E3) that $x_\mathrm{s} = f(y_{1\mathrm{s}}, y_{2\mathrm{s}})$, $G_1(x_\mathrm{s}, y_{1\mathrm{s}}, y_{2\mathrm{s}}) = 0$, and $G_2(x_\mathrm{s}, y_{1\mathrm{s}}, y_{2\mathrm{s}}) = 0$. For the linear stability analysis of the fixed point, we obtain the Jacobian matrix J and calculate its eigenvalues $\lambda$ and $\lambda^*$ with the larger real part than the remaining eigenvalue:

$$\mathrm{J} = \begin{pmatrix} -\mu_\mathrm{o} & \mu_\mathrm{o}\frac{\partial f}{\partial y_1} & \mu_\mathrm{o}\frac{\partial f}{\partial y_2} \\ \frac{\partial G_1}{\partial x} & \frac{\partial G_1}{\partial y_1} & \frac{\partial G_1}{\partial y_2} \\ \frac{\partial G_2}{\partial x} & \frac{\partial G_2}{\partial y_1} & \frac{\partial G_2}{\partial y_2} \end{pmatrix}, \tag{E4}$$

$$\begin{aligned} 0 = \lambda^3 &+ \left(\mu_\mathrm{o} - \frac{\partial G_1}{\partial y_1} - \frac{\partial G_2}{\partial y_2}\right)\lambda^2 + \left\{\frac{\partial G_1}{\partial y_1}\frac{\partial G_2}{\partial y_2} - \frac{\partial G_1}{\partial y_2}\frac{\partial G_2}{\partial y_1} - \mu_\mathrm{o}\left(\frac{\partial G_1}{\partial y_1} + \frac{\partial G_2}{\partial y_2} + \frac{\partial G_1}{\partial x}\frac{\partial f}{\partial y_1} + \frac{\partial G_2}{\partial x}\frac{\partial f}{\partial y_2}\right)\right\}\lambda \\ &+ \mu_\mathrm{o}\left\{\left(\frac{\partial G_1}{\partial y_1} + \frac{\partial G_1}{\partial x}\frac{\partial f}{\partial y_1}\right)\left(\frac{\partial G_2}{\partial y_2} + \frac{\partial G_2}{\partial x}\frac{\partial f}{\partial y_2}\right) - \left(\frac{\partial G_1}{\partial y_2} + \frac{\partial G_1}{\partial x}\frac{\partial f}{\partial y_2}\right)\left(\frac{\partial G_2}{\partial y_1} + \frac{\partial G_2}{\partial x}\frac{\partial f}{\partial y_1}\right)\right\}. \end{aligned} \tag{E5}$$

The derivatives in Eqs. (E4) and (E5) are taken at the fixed point and we will keep using these notations for the sake of simplicity.

The QSSA gives the same fixed point but different stability, with the eigenvalues $\lambda_\mathrm{Q}$ and $\lambda_\mathrm{Q}^*$ as the solutions of Eq. (E5) at $\mu_\mathrm{o}^{-1} \to 0$:

$$\lambda_\mathrm{Q} = \mathrm{Re}\big(\lambda_\mathrm{Q}\big) + i\mathrm{Im}\big(\lambda_\mathrm{Q}\big), \tag{E6}$$

$$\mathrm{Re}\big(\lambda_\mathrm{Q}\big) = \frac{1}{2}\left(\frac{\partial G_1}{\partial y_1} + \frac{\partial G_2}{\partial y_2} + \frac{\partial G_1}{\partial x}\frac{\partial f}{\partial y_1} + \frac{\partial G_2}{\partial x}\frac{\partial f}{\partial y_2}\right), \tag{E7}$$

$$\mathrm{Im}(\lambda_{\mathrm{Q}}) = \pm\sqrt{D_{\mathrm{Q}}} \text{ with } D_{\mathrm{Q}} \equiv \left(\frac{\partial G_1}{\partial y_1} + \frac{\partial G_1}{\partial x}\frac{\partial f}{\partial y_1}\right)\left(\frac{\partial G_2}{\partial y_2} + \frac{\partial G_2}{\partial x}\frac{\partial f}{\partial y_2}\right) - \left(\frac{\partial G_1}{\partial y_2} + \frac{\partial G_1}{\partial x}\frac{\partial f}{\partial y_2}\right)\left(\frac{\partial G_2}{\partial y_1} + \frac{\partial G_2}{\partial x}\frac{\partial f}{\partial y_1}\right) - \mathrm{Re}^2(\lambda_{\mathrm{Q}}). \quad \text{(E8)}$$

For the QSSA-based spiral, $D_{\mathrm{Q}}$ in Eq. (E8) should satisfy

$$D_{\mathrm{Q}} > 0. \quad \text{(E9)}$$

Under this condition, $|\mathrm{Im}(\lambda_{\mathrm{Q}})|$ is the angular velocity of this spiral, i.e., $|\mathrm{Im}(\lambda_{\mathrm{Q}})| = 2\pi/T_{\mathrm{Q}}$ with the period $T_{\mathrm{Q}}$ at its center. If $\mathrm{Re}(\lambda_{\mathrm{Q}}) < 0$, the spiral is a stable fixed point. As $\mathrm{Re}(\lambda_{\mathrm{Q}})$ increases and crosses zero, the bifurcation occurs, and the spiral develops into a sustained oscillation (limit cycle) at $\mathrm{Re}(\lambda_{\mathrm{Q}}) > 0$. The nearer the bifurcation with $\mathrm{Re}(\lambda_{\mathrm{Q}}) \to 0^+$, the smaller the oscillation amplitude and the more the oscillation period matches $T_{\mathrm{Q}}$ [S3].

From Eqs. (E5)–(E8), we express $\lambda$ in powers of $\mu_{\mathrm{o}}^{-1}$ as

$$\lambda = \mathrm{Re}(\lambda) + i\mathrm{Im}(\lambda), \quad \text{(E10)}$$

$$\mathrm{Re}(\lambda) = \mathrm{Re}(\lambda_{\mathrm{Q}}) + \frac{\mu_{\mathrm{o}}^{-1}}{2}\left(\mathrm{Im}^2(\lambda_{\mathrm{Q}}) - A - 2B\mathrm{Re}(\lambda_{\mathrm{Q}}) - 3\mathrm{Re}^2(\lambda_{\mathrm{Q}})\right) + O(\mu_{\mathrm{o}}^{-2}), \quad \text{(E11)}$$

$$\mathrm{Im}(\lambda) = \mathrm{Im}(\lambda_{\mathrm{Q}}) + \frac{\mu_{\mathrm{o}}^{-1}}{2}\left\{\frac{\mathrm{Re}(\lambda_{\mathrm{Q}})}{\mathrm{Im}(\lambda_{\mathrm{Q}})}\left(A + B\mathrm{Re}(\lambda_{\mathrm{Q}}) + \mathrm{Re}^2(\lambda_{\mathrm{Q}})\right) - \mathrm{Im}(\lambda_{\mathrm{Q}})\left(B + 3\mathrm{Re}(\lambda_{\mathrm{Q}})\right)\right\} + O(\mu_{\mathrm{o}}^{-2}), \quad \text{(E12)}$$

$$A \equiv \frac{\partial G_1}{\partial y_1}\frac{\partial G_2}{\partial y_2} - \frac{\partial G_1}{\partial y_2}\frac{\partial G_2}{\partial y_1}, \quad \text{(E13)}$$

$$B \equiv -\frac{\partial G_1}{\partial y_1} - \frac{\partial G_2}{\partial y_2}. \quad \text{(E14)}$$

When $\mu_{\mathrm{o}}^{-1}$ is small enough for negligible $O(\mu_{\mathrm{o}}^{-2})$ in Eqs. (E11) and (E12), $O\left(\mu_{\mathrm{o}}^{-1}\mathrm{Re}(\lambda_{\mathrm{Q}})\right)$ can also be ignored near the bifurcation because $|\mathrm{Re}(\lambda_{\mathrm{Q}})|$ is small there and vanishes at the bifurcation. We thus obtain

$$\mathrm{Re}(\lambda) \approx \mathrm{Re}(\lambda_{\mathrm{Q}}) + \frac{\mu_{\mathrm{o}}^{-1}}{2}\left(\mathrm{Im}^2(\lambda_{\mathrm{Q}}) - A\right), \quad \text{(E15)}$$

$$\mathrm{Im}(\lambda) \approx \left(1 - \frac{1}{2}\mu_{\mathrm{o}}^{-1}B\right)\mathrm{Im}(\lambda_{\mathrm{Q}}). \quad \text{(E16)}$$

When $\mathrm{Im}^2(\lambda_{\mathrm{Q}}) - A > 0$ ($\mathrm{Im}^2(\lambda_{\mathrm{Q}}) - A < 0$) at the QSSA-based bifurcation, $\mathrm{Re}(\lambda)$ in Eq. (E15) with tiny $\mu_{\mathrm{o}}^{-1} > 0$ crosses zero at $\mathrm{Re}(\lambda_{\mathrm{Q}}) < 0$ ($\mathrm{Re}(\lambda_{\mathrm{Q}}) > 0$)—that is, the small $\mu_{\mathrm{o}}^{-1} > 0$ facilitates (suppresses) the oscillation onset and therefore advances (delays) the bifurcation point itself compared to the QSSA. This result proves the bifurcation point shift associated with Eq. (7) as $A$ in Eq. (E13) matches that in Eqs. (7) and (8), at the QSSA-based bifurcation.

## F. Physical explanation of Eq. (7)

According to Eqs. (E8) and (E13) with the discussion under Eq. (E9), $(2\pi/T_{\mathrm{Q}})^2 - A$ from Eq. (7) equals $(\partial_x G_1 \cdot \partial_{y_1} f)\,\partial_{y_2} G_2 + (\partial_x G_2 \cdot \partial_{y_2} f)\,\partial_{y_1} G_1 - (\partial_x G_1 \cdot \partial_{y_1} G_2 \cdot \partial_{y_2} f) - (\partial_x G_2 \cdot \partial_{y_2} G_1 \cdot \partial_{y_1} f)$ at the fixed point for the bifurcation (transition), and the parenthesized terms here come from the feedback of $x(t)$ to $x_{\mathrm{Q}}(t)$ in Eq. (E3). In other words, the feedback is required for the bifurcation point shift associated with Eq. (7).

To gain further insights into Eq. (7), we apply Eq. (5) to Eqs. (E1)–(E3) and obtain the following equations up to $O(\mu_{\mathrm{o}}^{-1})$ for small $\mu_{\mathrm{o}}^{-1}$:

$$\frac{\mathrm{d}y_1}{\mathrm{d}t} \approx G_1(f(y_1,y_2),y_1,y_2) - \mu_{\mathrm{o}}^{-1}\left(\frac{\partial f(y_1,y_2)}{\partial y_1}\frac{\mathrm{d}y_1}{\mathrm{d}t} + \frac{\partial f(y_1,y_2)}{\partial y_2}\frac{\mathrm{d}y_2}{\mathrm{d}t}\right)\cdot \left.\frac{\partial G_1(x,y_1,y_2)}{\partial x}\right|_{x=f(y_1,y_2)}, \quad \text{(F1)}$$

$$\frac{\mathrm{d}y_2}{\mathrm{d}t} \approx G_2(f(y_1,y_2),y_1,y_2) - \mu_{\mathrm{o}}^{-1}\left(\frac{\partial f(y_1,y_2)}{\partial y_1}\frac{\mathrm{d}y_1}{\mathrm{d}t} + \frac{\partial f(y_1,y_2)}{\partial y_2}\frac{\mathrm{d}y_2}{\mathrm{d}t}\right)\cdot \left.\frac{\partial G_2(x,y_1,y_2)}{\partial x}\right|_{x=f(y_1,y_2)}. \quad \text{(F2)}$$

For the fixed point $(x(t),y_1(t),y_2(t)) = (x_{\mathrm{s}},y_{1\mathrm{s}},y_{2\mathrm{s}})$ defined under Eq. (E3), we introduce $\Delta y_1 \equiv y_1 - y_{1\mathrm{s}}$ and $\Delta y_2 \equiv y_2 - y_{2\mathrm{s}}$. For small $|\Delta y_1|$ and $|\Delta y_2|$ near the QSSA-based bifurcation, Eqs. (F1) and (F2) with the linearization of $G_1$ and $G_2$ yield the following equation up to $O(\mu_{\mathrm{o}}^{-1})$:

$$\left\{\mathrm{I} + \mu_{\mathrm{o}}^{-1}\begin{pmatrix} \frac{\partial G_1}{\partial x}\frac{\partial f}{\partial y_1} & \frac{\partial G_1}{\partial x}\frac{\partial f}{\partial y_2} \\ \frac{\partial G_2}{\partial x}\frac{\partial f}{\partial y_1} & \frac{\partial G_2}{\partial x}\frac{\partial f}{\partial y_2}\end{pmatrix}\right\}\begin{pmatrix}\frac{\mathrm{d}(\Delta y_1)}{\mathrm{d}t} \\ \frac{\mathrm{d}(\Delta y_2)}{\mathrm{d}t}\end{pmatrix} \approx \begin{pmatrix} \frac{\partial G_1}{\partial x}\frac{\partial f}{\partial y_1} + \frac{\partial G_1}{\partial y_1} & \frac{\partial G_1}{\partial x}\frac{\partial f}{\partial y_2} + \frac{\partial G_1}{\partial y_2} \\ \frac{\partial G_2}{\partial x}\frac{\partial f}{\partial y_1} + \frac{\partial G_2}{\partial y_1} & \frac{\partial G_2}{\partial x}\frac{\partial f}{\partial y_2} + \frac{\partial G_2}{\partial y_2}\end{pmatrix}\begin{pmatrix}\Delta y_1 \\ \Delta y_2\end{pmatrix}, \quad \text{(F3)}$$

where $\mathrm{I}$ is an identify matrix and the partial derivatives are taken at the fixed point. In other words,

$$\begin{pmatrix}\frac{\mathrm{d}(\Delta y_1)}{\mathrm{d}t} \\ \frac{\mathrm{d}(\Delta y_2)}{\mathrm{d}t}\end{pmatrix} \approx \left\{\mathrm{I} - \mu_{\mathrm{o}}^{-1}\begin{pmatrix} \frac{\partial G_1}{\partial x}\frac{\partial f}{\partial y_1} & \frac{\partial G_1}{\partial x}\frac{\partial f}{\partial y_2} \\ \frac{\partial G_2}{\partial x}\frac{\partial f}{\partial y_1} & \frac{\partial G_2}{\partial x}\frac{\partial f}{\partial y_2}\end{pmatrix}\right\}\begin{pmatrix} \frac{\partial G_1}{\partial x}\frac{\partial f}{\partial y_1} + \frac{\partial G_1}{\partial y_1} & \frac{\partial G_1}{\partial x}\frac{\partial f}{\partial y_2} + \frac{\partial G_1}{\partial y_2} \\ \frac{\partial G_2}{\partial x}\frac{\partial f}{\partial y_1} + \frac{\partial G_2}{\partial y_1} & \frac{\partial G_2}{\partial x}\frac{\partial f}{\partial y_2} + \frac{\partial G_2}{\partial y_2}\end{pmatrix}\begin{pmatrix}\Delta y_1 \\ \Delta y_2\end{pmatrix}. \quad \text{(F4)}$$

Based on the above equation, we now perform the "thought experiment" that the QSSA-based limit cycle near the bifurcation point becomes perturbed by the sudden slight increase of $\mu_{\mathrm{o}}^{-1}$ from zero. The resulting velocity after the transient period, combining Eqs. (E7), (E8), (E13), and (F4) with $\mathrm{Re}(\lambda_{\mathrm{Q}}) \approx 0$, is given by

$$\begin{pmatrix}\frac{\mathrm{d}(\Delta y_1)}{\mathrm{d}t} \\ \frac{\mathrm{d}(\Delta y_2)}{\mathrm{d}t}\end{pmatrix} \approx \begin{pmatrix}\frac{\mathrm{d}(\Delta y_1)}{\mathrm{d}t} \\ \frac{\mathrm{d}(\Delta y_2)}{\mathrm{d}t}\end{pmatrix}_{\mathrm{Q}} - \mu_{\mathrm{o}}^{-1}\left\{(\Delta y_1)\left(\frac{\partial G_2}{\partial y_1}\frac{\partial f}{\partial y_2} - \frac{\partial G_2}{\partial y_2}\frac{\partial f}{\partial y_1}\right) + (\Delta y_2)\left(\frac{\partial G_1}{\partial y_2}\frac{\partial f}{\partial y_1} - \frac{\partial G_1}{\partial y_1}\frac{\partial f}{\partial y_2}\right)\right\}\begin{pmatrix}\frac{\partial G_1}{\partial x} \\ \frac{\partial G_2}{\partial x}\end{pmatrix}$$

$$= \begin{pmatrix} \frac{\mathrm{d}(\Delta y_1)}{\mathrm{d}t} \\ \frac{\mathrm{d}(\Delta y_2)}{\mathrm{d}t} \end{pmatrix}_{\mathrm{Q}} + \mu_{\mathrm{o}}^{-1}\left(\mathrm{Im}^2(\lambda_{\mathrm{Q}}) - A\right)\begin{pmatrix} \Delta y_1 \\ \Delta y_2 \end{pmatrix}_{\mathrm{p}}, \tag{F5}$$

where $((\Delta y_1)', (\Delta y_2)')_{\mathrm{Q}}$ denotes the QSSA-based velocity at the position $(\Delta y_1, \Delta y_2)$, and $(\Delta y_1, \Delta y_2)_{\mathrm{p}}$ denotes the component of the vector $(\Delta y_1, \Delta y_2)$ on the basis vector $(\partial_x G_1, \partial_x G_2)$ while the other basis is $\left((\partial_{y_2} G_1 \cdot \partial_{y_1} f) - (\partial_{y_1} G_1 \cdot \partial_{y_2} f), (\partial_{y_2} G_2 \cdot \partial_{y_1} f) - (\partial_{y_1} G_2 \cdot \partial_{y_2} f)\right)$. Accordingly, the switch-on of $\mu_{\mathrm{o}}^{-1}$ in the near-bifurcation condition modifies the velocity by $\mu_{\mathrm{o}}^{-1}\left(\mathrm{Im}^2(\lambda_{\mathrm{Q}}) - A\right) \cdot (\Delta y_1, \Delta y_2)_{\mathrm{p}}$. Hence, if $\mathrm{Im}^2(\lambda_{\mathrm{Q}}) > A$ ($\mathrm{Im}^2(\lambda_{\mathrm{Q}}) < A$), this perturbation kicks the trajectory out (in) and thus extends (shrinks) its motion around the fixed point, magnifying (diminishing) the oscillation in agreement with the Eq. (7)-related bifurcation point shift.

When $(2\pi/T_{\mathrm{Q}})^2 < A$ in Eq. (7), worthy of note with the above discussion and Eq. (A7) is that a small $\mu_{\mathrm{o}}^{-1}$-based, short time-delay in $x$ relative to $x_{\mathrm{Q}}$ suffices to diminish the near-bifurcation oscillation, even without the amplitude reduction in $x$ relative to $x_{\mathrm{Q}}$. The reason is that the time delay and amplitude reduction take the magnitudes of $O(\mu_{\mathrm{o}}^{-1})$ and $O(\mu_{\mathrm{o}}^{-2})$ in Eq. (A7), respectively. We numerically confirmed this counterintuitive effect—short delay-induced oscillation suppression—by applying Eq. (A7) but with only either the time-delay or amplitude reduction effect to the simulation of the ecological cycles with cooperative foraging as in Figs. 4(a)–(f) and Section J. Likewise, when $(2\pi/T_{\mathrm{Q}})^2 > A$ in Eq. (7), the time-delay effect suffices for the oscillation promotion as we numerically confirmed in the same way, but this result is readily expected given the well-known delay's effect for oscillation promotion [S5, S6].

Combining all these analyses in view of Eq. (F5), we physically interpret the bifurcation point shift associated with Eq. (7): the acceleration of a moving object along the QSSA-based near-bifurcation limit cycle in the planar phase space of $y_1$ and $y_2$ is $\sim r_{\mathrm{Q}} \cdot \mathrm{Im}^2(\lambda_{\mathrm{Q}})$, like a centripetal force. Here, $r_{\mathrm{Q}}$ is the characteristic radius of this limit cycle and $\mathrm{Im}(\lambda_{\mathrm{Q}})$ is equal to the angular velocity. The QSSA-based temporal change in $x$ contributes $\sim r_{\mathrm{Q}} \cdot \left(\mathrm{Im}^2(\lambda_{\mathrm{Q}}) - A\right)$ to this acceleration, as obtained from Eq. (E8) by leaving only its terms with $\partial_x G_1$ and $\partial_x G_2$ at $\mathrm{Re}(\lambda_{\mathrm{Q}}) \approx 0$. In other words, if $\mathrm{Im}^2(\lambda_{\mathrm{Q}}) > A$ ($\mathrm{Im}^2(\lambda_{\mathrm{Q}}) < A$), this contribution acts like the centripetal (centrifugal) force. As already described, a slight increase of $\mu_{\mathrm{o}}^{-1}$ from zero induces a short time-delay to the change in $x$ and thus retards this change. This repression of the $x$'s change relieves its associated centripetal (centrifugal)-like force if $\mathrm{Im}^2(\lambda_{\mathrm{Q}}) > A$ ($\mathrm{Im}^2(\lambda_{\mathrm{Q}}) < A$), and hence the orbit starts to stretch (shrink) with a magnified (diminishing) oscillation. Due to that delay in the $x$'s change, the deviation of the velocity of $(y_1, y_2)$ from the QSSA after the transient period becomes proportional to the delay $\sim\mu_{\rm o}^{-1}$, and returns Eq. (F5). All this story is the physical basis of Eq. (F5) and the bifurcation point shift associated with Eq. (7). A key message is that, if the QSSA-based temporal change in $x$ tends to curve (straighten) the motion of $(y_1, y_2)$ like the centripetal (centrifugal) force, small yet non-zero $\mu_{\rm o}^{-1}$ would amplify (suppress) the oscillation.

## G. Genetic switch analysis

Consider a genetic switch with positive autoregulation in Fig. 1(a) where proteins enhance their own transcription after homodimer formation and subsequent DNA-promoter binding as a TF. This dimer–promoter interaction is facilitated by inducer molecules.

The following equations describe the molecular processes of this system [S7]:

$$\frac{\mathrm{d}C(t)}{\mathrm{d}t} = \frac{k_{\rm a}}{2}\big(P(t) - 2C(t)\big)^2 - (k_{\rm d} + r_1)\big(C(t) - C_{\rm TF}(t)\big) - k_{\rm dlt}C(t)$$
$$\approx \frac{k_{\rm a}}{2}P^2(t) - (k_{\rm d} + r_1 + k_{\rm dlt})C(t), \tag{G1}$$

$$\frac{\mathrm{d}C_{\rm TF}(t)}{\mathrm{d}t} = k_{\rm TFa}C(t)\left(\frac{1}{V} - C_{\rm TF}(t)\right) - (k_{\rm TFd} + k_{\rm dlt})C_{\rm TF}(t), \tag{G2}$$

$$\frac{\mathrm{d}M(t)}{\mathrm{d}t} = \frac{s_0}{V} + (s_1 - s_0)C_{\rm TF}(t) - (r_2 + k_{\rm dlt})M(t), \tag{G3}$$

$$\frac{\mathrm{d}P(t)}{\mathrm{d}t} = s_2M(t) - r_1\big(P(t) - 2C_{\rm TF}(t)\big) - k_{\rm dlt}P(t)$$
$$\approx s_2M(t) - (r_1 + k_{\rm dlt})P(t). \tag{G4}$$

Here, $t$ is time and $M(t)$, $P(t)$, $C(t)$, and $C_{\rm TF}(t)$ are the total mRNA, protein, protein dimer, and promoter-binding dimer concentrations, respectively. $s_0$, $s_1$, and $s_2$ denote the basal and maximal transcription rates ($s_0 < s_1$) and translation efficiency, respectively. $r_1$, $r_2$, and $k_{\rm dlt}$ denote the protein and mRNA degradation rates and cell growth-associated dilution rate, respectively. $k_{\rm a}$, $k_{\rm d}$, $k_{\rm TFa}$, and $k_{\rm TFd}$ denote the dimer association and dissociation rates and the dimer–promoter association and dissociation rates, respectively. $k_{\rm TFa}$ is proportional to the inducer level. $V$ denotes the cellular volume. Eq. (G2) is rigorously derived from the chemical master equation for stochastic promoter binding and unbinding events [S2, S7]. We here assume that the promoter-binding dimers are neither dissociated to monomers nor degraded. The approximations in Eqs. (G1) and (G4) hold under the condition $C_{\rm TF}(t) \ll C(t) \ll P(t)/2$.

Eqs. (G1)–(G4) are non-dimensionalized as

$$\frac{\mathrm{d}z(t)}{\mathrm{d}t} \approx -\mu_{\rm d}(z(t) - y^2(t)), \tag{G5}$$

$$\frac{\mathrm{d}z_{\mathrm{TF}}(t)}{\mathrm{d}t} = -\hat{\mu}_{\mathrm{TFd}}\big(1+\eta z(t)\big)\left(z_{\mathrm{TF}}(t) - \frac{\eta z(t)}{1+\eta z(t)}\right), \tag{G6}$$

$$\frac{\mathrm{d}x(t)}{\mathrm{d}t} = -\mu_{\mathrm{r}}[x(t) - \{(1-\sigma)z_{\mathrm{TF}}(t) + \sigma\}], \tag{G7}$$

$$\frac{\mathrm{d}y(t)}{\mathrm{d}t} \approx x(t) - y(t), \tag{G8}$$

where we redefine $t$ by multiplying itself with $(r_1 + k_{\mathrm{dlt}})$ and define the other quantities as $x(t) \equiv (r_2 + k_{\mathrm{dlt}})s_1^{-1}VM(t)$, $y(t) \equiv (r_1 + k_{\mathrm{dlt}})(r_2 + k_{\mathrm{dlt}})(s_1 s_2)^{-1}VP(t)$, $z(t) \equiv 2(k_{\mathrm{d}} + r_1 + k_{\mathrm{dlt}})\{(r_1 + k_{\mathrm{dlt}})(r_2 + k_{\mathrm{dlt}})(s_1 s_2)^{-1}V\}^2 k_{\mathrm{a}}^{-1}C(t)$, $z_{\mathrm{TF}}(t) \equiv VC_{\mathrm{TF}}(t)$, $\eta \equiv k_{\mathrm{TFa}}k_{\mathrm{a}}(s_1 s_2)^2[2(k_{\mathrm{TFd}} + k_{\mathrm{dlt}})(k_{\mathrm{d}} + r_1 + k_{\mathrm{dlt}})\{V(r_1 + k_{\mathrm{dlt}})(r_2 + k_{\mathrm{dlt}})\}^2]^{-1}$, $\sigma \equiv s_0 s_1^{-1} < 1$, $\mu_{\mathrm{r}} \equiv (r_1 + k_{\mathrm{dlt}})^{-1}(r_2 + k_{\mathrm{dlt}})$, $\mu_{\mathrm{d}} \equiv (r_1 + k_{\mathrm{dlt}})^{-1}(k_{\mathrm{d}} + r_1 + k_{\mathrm{dlt}})$, and $\hat{\mu}_{\mathrm{TFd}} \equiv (r_1 + k_{\mathrm{dlt}})^{-1}(k_{\mathrm{TFd}} + k_{\mathrm{dlt}})$.

We notice that $y(t)$ is the dimensionless, total protein concentration with typically slower changes than the other variables such as the mRNA concentration $x(t)$. As described in the main text, we will focus on the saddle-node bifurcation of this system. As $\eta$ increases and crosses the bifurcation point $\eta = \eta_{\mathrm{c}}$, the low protein level at the fixed point undergoes an abrupt leap. Let $y(t) = y_{\mathrm{Qm}}$ be that low protein level at the fixed point of the bifurcation (saddle-node fixed point). About the rate-limiting dynamics near the bifurcation point with $\eta \to \eta_{\mathrm{c}}^{+}$, we estimate $\hat{\mu}_{\mathrm{TFd}}\big(1+\eta z(t)\big)$ in Eq. (G6) as $\hat{\mu}_{\mathrm{TFd}}\big(1+\eta_{\mathrm{c}} y_{\mathrm{Qm}}^2\big)$ from Eq. (G5), according to the last part of Section D. Applying this estimation and Eq. (5) to Eqs. (G5)–(G7) suggests the asymptotic trajectory of $x(t)$ as

$$x(t) \approx f\big(y(t)\big) - \left(\mu_{\mathrm{r}}^{-1} + \mu_{\mathrm{d}}^{-1} + \frac{\hat{\mu}_{\mathrm{TFd}}^{-1}}{1+\eta_{\mathrm{c}} y_{\mathrm{Qm}}^2}\right)\frac{\mathrm{d}f\big(y(t)\big)}{\mathrm{d}t} + \cdots, \tag{G9}$$

where the higher-order terms of $\mu_{\mathrm{r}}^{-1}$, $\mu_{\mathrm{d}}^{-1}$, and $\hat{\mu}_{\mathrm{TFd}}^{-1}\big(1+\eta_{\mathrm{c}} y_{\mathrm{Qm}}^2\big)^{-1}$ are omitted and $f\big(y(t)\big)$ is given by

$$f\big(y(t)\big) \equiv (1-\sigma)\cdot\frac{\eta y^2(t)}{1+\eta y^2(t)} + \sigma. \tag{G10}$$

Regarding Eq. (G9), it is easy to satisfy Eq. (C2) at $y(t) \approx y_{\mathrm{Qm}}$ due to the extreme slowdown of $y(t)$ there. $f\big(y(t)\big)$ in Eq. (G10) is interpreted as the lumping of the mRNA synthesis processes including the protein dimerization, dimer-regulated transcription, and basal transcription. Comparing Eqs. (G9) and (G10) to Eq. (5) suggests that $x'(t)$ is approximated by Eqs. (4) and (10) with $\mu(t) = \mu_{\mathrm{o}}$ and

$$\mu_{\mathrm{o}} \approx \left(\mu_{\mathrm{r}}^{-1} + \mu_{\mathrm{d}}^{-1} + \frac{\hat{\mu}_{\mathrm{TFd}}^{-1}}{1+\eta_{\mathrm{c}} y_{\mathrm{Qm}}^2}\right)^{-1}. \tag{G11}$$

In other words, the effective $\mu_{\mathrm{o}}$ is one-third of the harmonic mean of $\mu_{\mathrm{r}}$, $\mu_{\mathrm{d}}$, and $\hat{\mu}_{\mathrm{TFd}}\left(1+\eta_{\mathrm{c}} y_{\mathrm{Qm}}^{2}\right)$, which are the dimensionless relaxation rates of the typically faster processes than the protein loss by $-y(t)$ in Eq. (G8). Naturally, $\mu_{\mathrm{o}}$ here is dimensionless as well. Together, Eqs. (G8), (G10), and (G11) suggest Eqs. (4), (9), and (10) with $\mu(t)=\mu_{\mathrm{o}}$, i.e., the genetic-switch model in the main text. This reduced model captures the bifurcation of the original model with Eqs. (G1)–(G4) [S7] and is still analytically tractable—thus the target of our analysis.

From Eqs. (9) and (D1),

$$G\big(x(t), y(t)\big)=x(t)-y(t). \tag{G12}$$

Eqs. (10) and (G12) satisfy Eq. (D22) and its associated discussion. From Eqs. (D17) and $G\big(f(y_{\mathrm{Qm}}), y_{\mathrm{Qm}}\big)=0$ at the bifurcation point $\eta=\eta_{\mathrm{c}}$, we obtain

$$\eta_{\mathrm{c}}=\frac{3}{1-\frac{1}{16}(1-9\sigma)\left(\sqrt{1-9\sigma}+3\sqrt{1-\sigma}\right)^{2}}, \tag{G13}$$

$$y_{\mathrm{Qm}}=\frac{1}{3}\left(1-\sqrt{1-\frac{3}{\eta_{\mathrm{c}}}}\right)=\frac{1}{3}\left\{1-\frac{1}{4}\left(\sqrt{1-9\sigma}+3\sqrt{1-\sigma}\right)\sqrt{1-9\sigma}\right\}. \tag{G14}$$

Apparently, these solutions are feasible when $0<\sigma \leq 1/9$. However, $\sigma=1/9$ lacks distinctly low and high protein states across all $\eta \geq 0$. Therefore, the bifurcation only exists when $0<\sigma<1/9$.

To calculate the low-to-high protein transition time for $\eta>\eta_{\mathrm{c}}$, *we* combine Eq. (D17), the notes under Eqs. (D17) and (D23) (in relation to Eqs. (D11), (D13), and (D14)), and the linearization of $G_{\mathrm{Qm}}$ by $\eta-\eta_{\mathrm{c}}$ ($G_{\mathrm{Qm}}=0$ at $\eta=\eta_{\mathrm{c}}$) and then obtain

$$G_{\mathrm{Qm}} \approx \frac{(1-\sigma) y_{\mathrm{Qm}}^{2}}{\left(1+\eta_{\mathrm{c}} y_{\mathrm{Qm}}^{2}\right)^{2}}\left(\eta-\eta_{\mathrm{c}}\right)=\left(\frac{y_{\mathrm{Qm}}}{2}\right)\left(\frac{\eta-\eta_{\mathrm{c}}}{\eta_{\mathrm{c}}}\right), \tag{G15}$$

$$a_{\mathrm{Qm}}=\left(\frac{2}{y_{\mathrm{Qm}}}\right)\left(\frac{1-y_{\mathrm{Qm}}}{1-\sigma}-\frac{3}{4}\right), \tag{G16}$$

$$b_{\mathrm{Qm}}=1. \tag{G17}$$

Plugging Eqs. (D15)–(D17) in Eqs. (D24) and (D25) leads to the transition time $t_{\mathrm{r}}$ in Eq. (11) with its QSSA $t_{\mathrm{Qr}}$ as

$$t_{\mathrm{Qr}} \approx \frac{\pi}{\sqrt{\left(\frac{1-y_{\mathrm{Qm}}}{1-\sigma}-\frac{3}{4}\right)\left(\frac{\eta-\eta_{\mathrm{c}}}{\eta_{\mathrm{c}}}\right)}}, \tag{G18}$$

where $\eta_{\mathrm{c}}$ and $y_{\mathrm{Qm}}$ are given by Eqs. (G13) and (G14). Critical slowing down near the bifurcation is captured by Eqs. (11) and (G18) that $t_{\mathrm{Qr}} \propto 1/\sqrt{(\eta - \eta_{\mathrm{c}})/\eta_{\mathrm{c}}}$ and thus $t_{\mathrm{r}} \propto 1/\sqrt{(\eta - \eta_{\mathrm{c}})/\eta_{\mathrm{c}}}$.

## H. Glycolytic oscillation analysis

As a simple classical model of glycolytic oscillations, the Sel'kov model focuses on the phosphofructokinase reaction in Fig. 2(a) where ATP donates a phosphate to fructose-6-phosphate and turns into ADP. This enzymatic reaction involves substrate inhibition and product activation. AMP also activates this enzyme, but its effect can be grouped to that of ADP due to the consistency of their concentration changes. The model derivation starts from the following equations [S8]:

$$\frac{\mathrm{d}s_1(t)}{\mathrm{d}t} = v - k_{+1}s_1(t)x_1(t) + k_{-1}x_2(t), \tag{H1}$$

$$\frac{\mathrm{d}s_2(t)}{\mathrm{d}t} = k_{+2}x_2(t) - k_{+3}s_2^{\gamma}(t)e_1(t) + k_{-3}x_1(t) - k_2 s_2(t), \tag{H2}$$

$$\frac{\mathrm{d}x_1(t)}{\mathrm{d}t} = -k_{+1}s_1(t)x_1(t) + (k_{-1} + k_{+2})x_2(t) + k_{+3}s_2^{\gamma}(t)e_1(t) - k_{-3}x_1(t), \tag{H3}$$

$$\frac{\mathrm{d}x_2(t)}{\mathrm{d}t} = k_{+1}s_1(t)x_1(t) - (k_{-1} + k_{+2})x_2(t), \tag{H4}$$

$$\frac{\mathrm{d}e_1(t)}{\mathrm{d}t} = -k_{+3}s_2^{\gamma}(t)e_1(t) + k_{-3}x_1(t). \tag{H5}$$

Here, $t$ is time and $s_1(t)$, $s_2(t)$, $e_1(t)$, $x_1(t)$, and $x_2(t)$ are the concentrations of ATP, ADP, and the free, product-activated, and substrate-loaded enzymes, respectively. $v$, $k_{+3}$, $k_{-3}$, $k_{+1}$, $k_{-1}$, $k_{+2}$, and $k_2$ denote the rates of ATP supply, enzyme activation and deactivation, substrate loading and release, substrate-to-product conversion, and ADP drainage, respectively. Although $\gamma$ appears to be the number of the product molecules to activate the enzyme, its known stoichiometry is of the first order and the actual interpretation of $\gamma$ is the relative weakness of the substrate inhibition, with any real number of $\gamma > 1$ [S8].

The total enzyme concentration $e_0 = e_1(t) + x_1(t) + x_2(t)$ remains constant over time and thus $e_1(t)$ in Eqs. (H2) and (H3) is replaced by $e_0 - x_1(t) - x_2(t)$. We now assume that $k_2 \lesssim k_{-3} \ll k_{+2} + k_{-1}$, $v \ll \min\left\{e_0 k_{+2}, (e_0 k_{+3})^{\frac{1}{1-\gamma}} k_2^{\frac{\gamma}{\gamma-1}}, k_2 (k_{-3} k_{+3}^{-1})^{\frac{1}{\gamma}}\right\}$, and $e_0 \lesssim k_{-3}(k_{+2} + k_{-1})(k_{+1}k_{+2})^{-1}$. Roughly, these conditions mean the following statements: $x_2(t)$ is the fastest variable and $x_2(t) \sim k_{+1}(k_{+2} + k_{-1})^{-1} s_1(t) x_1(t)$; $x_1(t)$ is slower, but still a faster variable than $s_1(t)$ and $s_2(t)$; $k_{-3}k_{+1}^{-1}(1+k_{-1}k_{+2}^{-1}) \ll s_1(t)$, $s_2^{\gamma}(t) \ll k_{-3}k_{+3}^{-1}$, and $s_1(t)s_2^{\gamma}(t) \ll k_{-3}(k_{+2}+k_{-1})(k_{+1}k_{+3})^{-1}$. As a result, Eqs. (H1)–(H3) reduce to

$$\frac{\mathrm{d}s_1(t)}{\mathrm{d}t} \approx v - k_{+2}x_2(t), \tag{H6}$$

$$\frac{\mathrm{d}s_2(t)}{\mathrm{d}t} \approx k_{+2}x_2(t) - k_2 s_2(t), \tag{H7}$$

$$\frac{\mathrm{d}x_1(t)}{\mathrm{d}t} \approx k_{+3}e_0 s_2^{\gamma}(t) - k_{-3}x_1(t). \tag{H8}$$

Eqs. (H4) and (H6)–(H8) can be non-dimensionalized as

$$\frac{\mathrm{d}y_1(t)}{\mathrm{d}t} \approx 1 - x(t), \tag{H9}$$

$$\frac{\mathrm{d}y_2(t)}{\mathrm{d}t} \approx \alpha\big(x(t) - y_2(t)\big), \tag{H10}$$

$$\frac{\mathrm{d}z(t)}{\mathrm{d}t} \approx -\mu_1\left(z(t) - y_2^{\gamma}(t)\right), \tag{H11}$$

$$\frac{\mathrm{d}x(t)}{\mathrm{d}t} = -\mu_2\big(x(t) - y_1(t)z(t)\big), \tag{H12}$$

where we redefine $t$ by multiplying itself with $e_0(vk_2^{-1})^{\gamma}k_{+1}k_{+2}k_{+3}k_{-3}^{-1}(k_{+2}+k_{-1})^{-1}$ and define the other quantities as $x(t) \equiv v^{-1}k_{+2}x_2(t)$, $y_1(t) \equiv e_0 v^{\gamma-1}k_2^{-\gamma}k_{+1}k_{+2}k_{+3}k_{-3}^{-1}(k_{+2}+k_{-1})^{-1}s_1(t)$, $y_2(t) \equiv v^{-1}k_2 s_2(t)$, $z(t) \equiv e_0^{-1}(v^{-1}k_2)^{\gamma}k_{-3}k_{+3}^{-1}x_1(t)$, $\alpha \equiv e_0^{-1}(v^{-1}k_2)^{\gamma}k_2 k_{-3}(k_{+2}+k_{-1})(k_{+1}k_{+2}k_{+3})^{-1}$, $\mu_1 \equiv e_0^{-1}(v^{-1}k_2)^{\gamma}k_{-3}^2(k_{+2}+k_{-1})(k_{+1}k_{+2}k_{+3})^{-1}$, and $\mu_2 \equiv e_0^{-1}(v^{-1}k_2)^{\gamma}k_{-3}(k_{+2}+k_{-1})^2(k_{+1}k_{+2}k_{+3})^{-1}$.

As mentioned earlier, $s_1(t)$ and $s_2(t)$, and thus $y_1(t)$ and $y_2(t)$ are slower variables than the others. The QSSA of the faster, $x(t)$ and $z(t)$ in Eqs. (H9)–(H12) retrieves the Sel'kov model in Eqs. (12) and (13) with $x(t) = y_1(t)y_2^{\gamma}(t)$. This model was reported to exhibit a supercritical Hopf bifurcation [S9].

At the fixed point $\big(y_1(t), y_2(t)\big) = (y_{1\mathrm{s}}, y_{2\mathrm{s}})$, it is straightforward that $y_{1\mathrm{s}} = y_{2\mathrm{s}} = 1$ with the following linear stability expressed by the notations in Eqs. (E7) and (E8):

$$\mathrm{Re}\big(\lambda_{\mathrm{Q}}\big) = \frac{1}{2}\{(\gamma-1)\alpha - 1\}, \tag{H13}$$

$$D_{\mathrm{Q}} = \alpha - \mathrm{Re}^2\big(\lambda_{\mathrm{Q}}\big). \tag{H14}$$

Because $\gamma > 1$, increasing $\alpha$ away from zero changes $\mathrm{Re}\big(\lambda_{\mathrm{Q}}\big)$ from negative to positive values, with zero at $\alpha = \alpha_{\mathrm{Qc}}$ where $\alpha_{\mathrm{Qc}} \equiv 1/(\gamma-1)$. At this bifurcation point $\alpha = \alpha_{\mathrm{Qc}}$, $D_{\mathrm{Q}}$ in Eq. (H14) satisfies Eq. (E9) and thus indicates the onset of the limit cycle for the oscillatory state, consistent with the known Hopf bifurcation of the Sel'kov model [S8, S9].

To consider relaxation dynamics, we apply Eq. (5) to Eqs. (H11) and (H12). Therefore,

$$\begin{aligned} x(t) &\approx f\big(y_1(t), y_2(t)\big) - \mu_1^{-1} y_1(t) \frac{\mathrm{d}y_2^{\gamma}(t)}{\mathrm{d}t} - \mu_2^{-1} \frac{\mathrm{d}}{\mathrm{d}t} f\big(y_1(t), y_2(t)\big) + \cdots \\ &\approx f\big(y_1(t), y_2(t)\big) - \left(\frac{\mu_1^{-1}\gamma\alpha}{\gamma\alpha - 1} + \mu_2^{-1}\right) \frac{\mathrm{d}}{\mathrm{d}t} f\big(y_1(t), y_2(t)\big) + \cdots, \end{aligned} \tag{H15}$$

where the higher-order terms of $\mu_1^{-1}$ and $\mu_2^{-1}$ are omitted and $f\big(y_1(t), y_2(t)\big)$ is given by

$$f\big(y_1(t), y_2(t)\big) \equiv y_1(t) y_2^{\gamma}(t). \tag{H16}$$

The last part of Eq. (H15) is based on the observation that $y_1(t)\left(y_2^{\gamma}(t)\right)' \sim \gamma\alpha(\gamma\alpha - 1)^{-1} f'\big(y_1(t), y_2(t)\big)$ from the linearization of the Sel'kov model around the fixed point. The near-bifurcation case with $\alpha \sim \alpha_{\mathrm{Qc}}$ satisfies $\gamma\alpha > 1$ and thus $\gamma\alpha(\gamma\alpha - 1)^{-1} > 0$. Comparing Eqs. (H15) and (H16) to Eq. (5) suggests that $x'(t)$ is approximated by Eq. (4) with $\mu(t) = \mu_{\mathrm{o}}$, and

$$\mu_{\mathrm{o}} \approx \left(\frac{\mu_1^{-1}\gamma\alpha}{\gamma\alpha - 1} + \mu_2^{-1}\right)^{-1}. \tag{H17}$$

$\mu_{\mathrm{o}}$ itself is dimensionless and the reversal of the time scaling dimensionalizes it as $\mu_{\mathrm{o}} \propto k_{-3}$, i.e., the enzyme deactivation rate. Together, Eqs. (H9), (H10), (H16), and (H17) provide the extended version of the Sel'kov model, which now comprises Eqs. (4) and (12)–(14) with $\mu(t) = \mu_{\mathrm{o}}$. Our numerical simulations show the presence of the supercritical Hopf bifurcation, whether based on the original QSSA model or the current extended model with $\mu_{\mathrm{o}}^{-1} > 0$ (Figs. 2(b)–(d), S2(a)–(c) and Section J).

From Eqs. (12), (13), (E1), and (E2),

$$G_1\big(x(t), y_1(t), y_2(t)\big) = 1 - x(t), \tag{H18}$$

$$G_2\big(x(t), y_1(t), y_2(t)\big) = \alpha\big(x(t) - y_2(t)\big). \tag{H19}$$

At the fixed point $\big(x(t), y_1(t), y_2(t)\big) = (x_{\mathrm{s}}, y_{1\mathrm{s}}, y_{2\mathrm{s}})$, clearly $y_{1\mathrm{s}} = y_{2\mathrm{s}} = 1$ and $x_{\mathrm{s}} = f(y_{1\mathrm{s}}, y_{2\mathrm{s}}) = 1$. About the bifurcation point shift compared to the QSSA in light of Eq. (7), we apply Eqs. (H18) and (H19) to Eq. (E13) at the fixed point, then consider Eq. (H14), the discussion under Eq. (E9), and $\mathrm{Re}\big(\lambda_{\mathrm{Q}}\big) = 0$ at the QSSA-based bifurcation $\alpha = \alpha_{\mathrm{Qc}}$, and finally obtain $A = 0$ and $\big(2\pi/T_{\mathrm{Q}}\big)^2 = \alpha_{\mathrm{Qc}}$. Obviously, $\big(2\pi/T_{\mathrm{Q}}\big)^2 > A$ for Eq. (7). Therefore, even small $\mu_{\mathrm{o}}^{-1} > 0$ would advance the bifurcation point compared to the QSSA, i.e., $\alpha_{\mathrm{c}} < \alpha_{\mathrm{Qc}}$ for the actual bifurcation point $\alpha = \alpha_{\mathrm{c}}$ with $\mu_{\mathrm{o}}^{-1} > 0$.

## I. Population cycle analysis

To describe the population oscillation with the resource foraging activity of cooperating individuals, we create the following equations:

$$\frac{\mathrm{d}C(t)}{\mathrm{d}t} = (k_1 r_1 - r_2)wX(t) - r_3\big(C(t) - wX(t)\big) = (k_1 r_1 - r_2 + r_3)wX(t) - r_3 C(t), \quad \text{(I1)}$$

$$\frac{\mathrm{d}R(t)}{\mathrm{d}t} = v - (r_1 + r_4)X(t) - (r_4 + r_5)\big(R(t) - X(t)\big)$$
$$= v - (r_1 - r_5)X(t) - (r_4 + r_5)R(t), \quad \text{(I2)}$$

$$\frac{\mathrm{d}X(t)}{\mathrm{d}t} = k_2\big\{1 + k_3\big(C(t) - wX(t)\big)\big\}\big(C(t) - wX(t)\big)\big(R(t) - X(t)\big) - (r_1 + r_2 + r_4 + r_6)X(t)$$
$$\approx k_2(1 + k_3 C(t))C(t)R(t) - (r_1 + r_2 + r_4 + r_6)X(t). \quad \text{(I3)}$$

Here, $t$ is time and $C(t)$, $R(t)$, and $X(t)$ denote the total consumer, resource, and consumer-captured resource abundances, respectively. $w$, $k_2$, and $k_3$ denote the foraging consumer-to-resource ratio, foraging rate coefficient, and consumer cooperativity, respectively. $k_1$, $r_3$, and $r_2$ denote the consumer reproduction yield, free consumer death/migration rate, and forager death rate, respectively. $v$, $r_1$, $r_4$, $r_5$, and $r_6$ are the resource supply, consumption, decay, drainage, and release rates, respectively. The approximation in Eq. (I3) holds under condition $X(t) \ll \min\big(w^{-1}C(t), R(t)\big)$.

Eqs. (I1)–(I3) are non-dimensionalized as

$$\frac{\mathrm{d}y_1(t)}{\mathrm{d}t} = ax(t) - by_1(t), \quad \text{(I4)}$$

$$\frac{\mathrm{d}y_2(t)}{\mathrm{d}t} = 1 - x(t) - y_2(t), \quad \text{(I5)}$$

$$\frac{\mathrm{d}x(t)}{\mathrm{d}t} \approx -\mu_{\mathrm{o}}\{x(t) - y_1(t)y_2(t)(1 + \phi y_1(t))\}, \quad \text{(I6)}$$

where we redefine $t$ by multiplying itself with $(r_4 + r_5)$, and under the condition $\max\{k_1^{-1}(r_2 - r_3), r_5\} < r_1$, we define the other quantities as $x(t) \equiv v^{-1}(r_1 - r_5)X(t)$, $y_1(t) \equiv k_2(r_1 - r_5)(r_4 + r_5)^{-1}(r_1 + r_2 + r_4 + r_6)^{-1}C(t)$, $y_2(t) \equiv v^{-1}(r_4 + r_5)R(t)$, $a \equiv wvk_2(k_1 r_1 - r_2 + r_3)(r_4 + r_5)^{-2}(r_1 + r_2 + r_4 + r_6)^{-1}$, $b \equiv r_3(r_4 + r_5)^{-1}$, $\phi \equiv k_3 k_2^{-1}(r_4 + r_5)(r_1 + r_2 + r_4 + r_6)(r_1 - r_5)^{-1}$, and $\mu_{\mathrm{o}} \equiv (r_1 + r_2 + r_4 + r_6)(r_4 + r_5)^{-1}$. In other words, $\mu_{\mathrm{o}}$ combines the resource consumption rate and other consumer–resource breakage rates. From Eqs. (4), (E3), and (I6),

$$f\big(y_1(t), y_2(t)\big) = y_1(t)y_2(t)(1 + \phi y_1(t)). \quad \text{(I7)}$$

The above equations then offer the model of Eqs. (4) and (15)–(17) with $\mu(t) = \mu_{\mathrm{o}}$ in the main text.

From Eqs. (E1), (E2), (I4), and (I5),

$$G_1\big(x(t), y_1(t), y_2(t)\big) = ax(t) - by_1(t), \tag{I8}$$

$$G_2(x(t), y_1(t), y_2(t)) = 1 - x(t) - y_2(t). \tag{I9}$$

As will be explained soon, we here target the following ranges of $a$ and $b$:

$$4 < b \text{ and } b < a < \tfrac{1}{4}b^2. \tag{I10}$$

The fixed point $\big(x(t), y_1(t), y_2(t)\big) = (x_{\mathrm{s}}, y_{1\mathrm{s}}, y_{2\mathrm{s}})$ satisfies $x_{\mathrm{s}} = f(y_{1\mathrm{s}}, y_{2\mathrm{s}})$, $G_1(x_{\mathrm{s}}, y_{1\mathrm{s}}, y_{2\mathrm{s}}) = 0$, and $G_2(x_{\mathrm{s}}, y_{1\mathrm{s}}, y_{2\mathrm{s}}) = 0$, and the QSSA provides the same fixed point. One fixed point to satisfy these conditions is given by Eq. (19) in the main text and the other fixed point by $y_{1\mathrm{s}} = 0$ and $y_{2\mathrm{s}} = 1$. According to the QSSA-based linear stability analysis under Eq. (I10), the former fixed point can experience the Hopf bifurcation as analyzed below, while the latter is neither a spiral nor stable. We will focus on the former fixed point of the Hopf-bifurcation relevance.

From Eqs. (19), (E7), (E8), (E13), and (I7)–(I10),

$$\mathrm{Re}\big(\lambda_{\mathrm{Q}}\big) = \frac{1}{2}\Big\{b(1 + y_{1\mathrm{s}}) - \frac{by_{1\mathrm{s}}}{a - by_{1\mathrm{s}}} - a - 1\Big\}, \tag{I11}$$

$$D_{\mathrm{Q}} = \frac{b^2 y_{1\mathrm{s}} - a}{a - by_{1\mathrm{s}}} + 1 - \big\{\mathrm{Re}\big(\lambda_Q\big) + 1\big\}^2, \tag{I12}$$

$$A = b. \tag{I13}$$

The ranges of $a$ and $b$ in Eq. (I10) allow $\mathrm{Re}\big(\lambda_{\mathrm{Q}}\big)$ to span negative to positive values. From Eqs. (I11) and (I12), $D_{\mathrm{Q}}$ readily satisfies Eq. (E9) at $\mathrm{Re}\big(\lambda_{\mathrm{Q}}\big) = 0$. In other words, $\mathrm{Re}\big(\lambda_{\mathrm{Q}}\big) = 0$ corresponds to the QSSA-based Hopf bifurcation. According to Eqs. (19) and (I11), this bifurcation appears at $\phi = \phi_{\mathrm{Qc1}}$ or $\phi_{\mathrm{Qc2}}$ in Eqs. (18) and (20). Specifically, as $\phi$ increases and crosses the point $\phi = \phi_{\mathrm{Qc1}}$ ($\phi_{\mathrm{Qc2}}$) with given $a$ and $b$, the system transitions from the steady (oscillatory) to oscillatory (steady) state. Our numerical simulations confirm it as the supercritical Hopf bifurcation, which is also observed for $\mu_{\mathrm{o}}^{-1} > 0$ beyond the QSSA (Figs. 3(a),(b),(d),(f),(g),(i),(j) and Section J).

About the bifurcation point shift with $\mu_{\mathrm{o}}^{-1} > 0$ relative to the QSSA in view of Eq. (7), we apply Eqs. (18)–(20) to Eq. (I12) at $\mathrm{Re}\big(\lambda_Q\big) = 0$, follow the discussion under Eq. (E9), and compare the results with Eq. (I13) for Eq. (7). Eq. (7) then leads to the predictions in the main text and Fig. 3(c) that tiny $\mu_{\mathrm{o}}^{-1} > 0$ would advance or delay the bifurcation point depending on $a$ and $b$, as verified by the numerical simulations in Figs. 3(b),(d)–(k) (Section J).

## J. Numerical simulation and analysis

Numerical simulations and analyses were performed with Python 3.11.9 and Julia 1.11.3. Ordinary differential equations were solved by LSODA (scipy.integrate.solve_ivp) in the Python library SciPy v1.14.1 with shorter time steps than the system's relevant time-scales, and further time-step reductions did not alter the simulation results. Delay differential equations (DDEs) were solved by DDEProblem and Tsit5 in the Julia package DifferentialEquations v7.15.0 with the relative and absolute tolerances of $10^{-9}$.

***Genetic switch with autoregulation***

To test the significance of the similarity between the exact-to-QSSA transition time ratio ($t_{\mathrm{r}}/t_{\mathrm{Qr}}$) from the numerical simulation and its analytical prediction $f_{t_{\mathrm{r}}/t_{\mathrm{Qr}}}(\mu_{\mathrm{o}}^{-1}) \equiv (1-2\mu_{\mathrm{o}}^{-1})^{-1/2}$ from Eq. (11), we first calculated their difference as $\Delta_{t_{\mathrm{r}}/t_{\mathrm{Qr}}}(\mu_{\mathrm{o}}^{-1},\sigma) \equiv \left|\langle t_{\mathrm{r}}/t_{\mathrm{Qr}}\rangle_{\eta} - f_{t_{\mathrm{r}}/t_{\mathrm{Qr}}}(\mu_{\mathrm{o}}^{-1})\right|$ for each $\mu_{\mathrm{o}}^{-1}$ and $\sigma$, where $\langle\cdots\rangle_{\eta}$ denotes the average over $\eta$ close to $\eta_{\mathrm{c}}$ ($0 < (\eta-\eta_{\mathrm{c}})/\eta_{\mathrm{c}} \le 0.25$). We then averaged $\Delta_{t_{\mathrm{r}}/t_{\mathrm{Qr}}}(\mu_{\mathrm{o}}^{-1},\sigma)$ for the first- and second-half ranges of small $\mu_{\mathrm{o}}^{-1}$ ($0 \le \mu_{\mathrm{o}}^{-1} \le 0.25$; $\mu_{\mathrm{o}}^{-1} = 0$ for the QSSA), and refer to these first and second averages as $\Delta_{t_{\mathrm{r}}/t_{\mathrm{Qr}},1}(\sigma)$ and $\Delta_{t_{\mathrm{r}}/t_{\mathrm{Qr}},2}(\sigma)$, respectively. For comparison, we introduced a function $f_{t_{\mathrm{r}}/t_{\mathrm{Qr}}}^{\mathrm{null}}(\mu_{\mathrm{o}}^{-1}|q_1,q_2) \equiv (1-q_2\mu_{\mathrm{o}}^{-1})^{-q_1}$ as the null analytical expression of $t_{\mathrm{r}}/t_{\mathrm{Qr}}$. Like $f_{t_{\mathrm{r}}/t_{\mathrm{Qr}}}(\mu_{\mathrm{o}}^{-1})$ above, $f_{t_{\mathrm{r}}/t_{\mathrm{Qr}}}^{\mathrm{null}}(\mu_{\mathrm{o}}^{-1}|q_1,q_2)$ satisfies $f_{t_{\mathrm{r}}/t_{\mathrm{Qr}}}^{\mathrm{null}}(0|q_1,q_2) = 1$ and monotonically increases with $\mu_{\mathrm{o}}^{-1} \ge 0$ for positive $q_1$ and $q_2$, ensuring the conservative nature of our statistical test. Using $f_{t_{\mathrm{r}}/t_{\mathrm{Qr}}}^{\mathrm{null}}(\mu_{\mathrm{o}}^{-1}|q_1,q_2)$ instead of $f_{t_{\mathrm{r}}/t_{\mathrm{Qr}}}(\mu_{\mathrm{o}}^{-1})$, we defined $\Delta_{t_{\mathrm{r}}/t_{\mathrm{Qr}},1}^{\mathrm{null}}(\sigma|q_1,q_2)$ and $\Delta_{t_{\mathrm{r}}/t_{\mathrm{Qr}},2}^{\mathrm{null}}(\sigma|q_1,q_2)$ in the same way as $\Delta_{t_{\mathrm{r}}/t_{\mathrm{Qr}},1}(\sigma)$ and $\Delta_{t_{\mathrm{r}}/t_{\mathrm{Qr}},2}(\sigma)$. Through the brute-force search of $q_1$ and $q_2$ ($0 < q_1 \le 2$ and $0 < q_2 < 4$), we identified the probability that $f_{t_{\mathrm{r}}/t_{\mathrm{Qr}}}^{\mathrm{null}}(\mu_{\mathrm{o}}^{-1}|q_1,q_2)$ matches the overall pattern of the simulated $t_{\mathrm{r}}/t_{\mathrm{Qr}}$ at least as much as $f_{t_{\mathrm{r}}/t_{\mathrm{Qr}}}(\mu_{\mathrm{o}}^{-1})$ does, with the conditions $\Delta_{t_{\mathrm{r}}/t_{\mathrm{Qr}},1}^{\mathrm{null}}(\sigma|q_1,q_2) \le \Delta_{t_{\mathrm{r}}/t_{\mathrm{Qr}},1}(\sigma)$ and $\Delta_{t_{\mathrm{r}}/t_{\mathrm{Qr}},2}^{\mathrm{null}}(\sigma|q_1,q_2) \le \Delta_{t_{\mathrm{r}}/t_{\mathrm{Qr}},2}(\sigma)$. This probability now serves as the *P* value of the accuracy of the analytical prediction with Eq. (11) for $t_{\mathrm{r}}/t_{\mathrm{Qr}}$, and only a one-tailed test is relevant here. We finally obtained $0.01 \le P \le 0.02$ across $\sigma$ with the saddle-node bifurcation, i.e., $0 < \sigma < 1/9$ from Section G.

In a similar fashion, we computed the *P* value of the accuracy of the analytical prediction Eq. (G18) for $t_{\mathrm{Qr}}$ itself, with the result of $0.001 \le P \le 0.01$ across $\sigma$. Specifically, we calculated $\Delta_{t_{\mathrm{Qr}}}(\eta,\sigma) \equiv \left|t_{\mathrm{Qr}} - f_{t_{\mathrm{Qr}}}(\eta,\sigma)\right|$ for each $\eta$ and $\sigma$, with $t_{\mathrm{Qr}}$ from the simulation and $f_{t_{\mathrm{Qr}}}(\eta,\sigma) \equiv \pi\left[\{(1-y_{\mathrm{Qm}})(1-\sigma)^{-1} - (3/4)\}\{(\eta-\eta_{\mathrm{c}})/\eta_{\mathrm{c}}\}\right]^{-1/2}$ from Eq. (G18). We then averaged $\Delta_{t_{\mathrm{Qr}}}(\eta,\sigma)$ for the first- and second-half ranges of $\eta$ close to $\eta_{\mathrm{c}}$ ($0 < (\eta-\eta_{\mathrm{c}})/\eta_{\mathrm{c}} \leq 0.25$), and refer to these first and second averages as $\Delta_{t_{\mathrm{Qr}},1}(\sigma)$ and $\Delta_{t_{\mathrm{Qr}},2}(\sigma)$, respectively. For comparison, we took a function $f_{t_{\mathrm{Qr}}}^{\mathrm{null}}(\eta,\sigma|q_1,q_2) \equiv q_2\{(\eta-\eta_{\mathrm{c}})/\eta_{\mathrm{c}}\}^{-q_1}$ as the null analytical expression of $t_{\mathrm{Qr}}$. Like $f_{t_{\mathrm{Qr}}}(\eta,\sigma)$ above, $f_{t_{\mathrm{Qr}}}^{\mathrm{null}}(\eta,\sigma|q_1,q_2)$ diverges as $\eta \to \eta_{\mathrm{c}}^{+}$ and monotonically decreases with $\eta > \eta_{\mathrm{c}}$ for positive $q_1$ and $q_2$, ensuring the conservative nature of our statistical test. Using $f_{t_{\mathrm{Qr}}}^{\mathrm{null}}(\eta,\sigma|q_1,q_2)$ instead of $f_{t_{\mathrm{Qr}}}(\eta,\sigma)$, we defined $\Delta_{t_{\mathrm{Qr}},1}^{\mathrm{null}}(\sigma|q_1,q_2)$ and $\Delta_{t_{\mathrm{Qr}},2}^{\mathrm{null}}(\sigma|q_1,q_2)$ in the same way as $\Delta_{t_{\mathrm{Qr}},1}(\sigma)$ and $\Delta_{t_{\mathrm{Qr}},2}(\sigma)$. Through the brute-force search of $q_1$ and $q_2$ ($0 < q_1 \leq 2$ and $0 < q_2 \leq 20$), we identified the probability that $f_{t_{\mathrm{Qr}}}^{\mathrm{null}}(\eta,\sigma|q_1,q_2)$ matches the overall pattern of the simulated $t_{\mathrm{Qr}}$ at least as much as $f_{t_{\mathrm{Qr}}}(\eta,\sigma)$ does, with the conditions $\Delta_{t_{\mathrm{Qr}},1}^{\mathrm{null}}(\sigma|q_1,q_2) \leq \Delta_{t_{\mathrm{Qr}},1}(\sigma)$ and $\Delta_{t_{\mathrm{Qr}},2}^{\mathrm{null}}(\sigma|q_1,q_2) \leq \Delta_{t_{\mathrm{Qr}},2}(\sigma)$. This probability now serves as the above-mentioned *P* value and only a one-tailed test is relevant here.

***Glycolytic oscillation***

For each parameter set, we examined the presence of sustained oscillations (limit cycles) in the numerical solutions of $x(t)$, $y_1(t)$, and $y_2(t)$ up to $t - t_0 = 12{,}000$ or $20{,}000$, and quantified the peaks and troughs by averaging their last five consecutive values. The individual peaks and troughs themselves were identified by scipy.signal.find_peaks in SciPy v1.14.1. On the other hand, the derivation of the simulated model (Eqs. (4) and (12)–(14) with $\mu(t) = \mu_{\mathrm{o}}$) involves the partial, linear reduction of Eqs. (H1)–(H3) to Eqs. (H6)–(H8) (Section H). Therefore, the initial conditions for the simulation were chosen near the fixed point, ensuring the physically sensible, non-negative solutions.

***Population cycle with cooperative foraging***

For each parameter set, we examined the presence of sustained oscillations (limit cycles) in the numerical solutions of $x(t)$, $y_1(t)$, and $y_2(t)$ up to $t - t_0 = 12{,}000$ or $20{,}000$, and quantified the peaks and troughs by averaging their last five consecutive values. The individual peaks and troughs themselves were identified by scipy.signal.find_peaks in SciPy v1.14.1. On the other hand, the derivation of the simulated model (Eqs. (4) and (15)–(17) with $\mu(t) = \mu_{\mathrm{o}}$) involves the linear reduction of $X(t)$ in Eq. (I3) (Section I). Therefore, the initial conditions for the simulation were chosen near the fixed point, ensuring the physically sensible, non-negative solutions.

Eq. (A7) captures the relaxation's time-delay and amplitude reduction effects on $x(t)$. To examine their contributions to the bifurcation point shifts associated with Eq. (7), we analytically expressed $T$ and $\langle x_{\mathrm{Q}}(t)\rangle$ in Eq. (A7) using the following near-bifurcation relations: (*i*) $2\pi/T \approx |\mathrm{Im}(\lambda)|$ up to $O(\mu_{\mathrm{o}}^{-1})$ from Eq. (E12) and Eqs. (19), (E14), (I8), (I9), and (I11)–(I13), and (*ii*) $\langle x_{\mathrm{Q}}(t)\rangle_t \approx f(y_{1\mathrm{s}}, y_{2\mathrm{s}})$ from Eqs. (17) and (19). We then combined this modified Eq. (A7) with Eqs. (15)–(17) and $\mu(t) = \mu_{\mathrm{o}}$, attaining the DDE-based model with both the time-delay and amplitude reduction effects. Likewise, the model with only the time-delay (amplitude reduction) effect was constructed by ignoring the Eq. (A7)'s term $\sqrt{1 + (2\pi\mu_{\mathrm{o}}^{-1}/T)^2}$ ($(T/2\pi) \cdot \arctan(2\pi\mu_{\mathrm{o}}^{-1}/T)$) with its assignment of negligible $\mu_{\mathrm{o}}^{-1}$ (= $10^{-6}$). These three DDE-based models were numerically solved as described above, and their results were examined for the bifurcation point shifts. To aid this analysis, we identified the peaks and troughs of the solutions by findmaxima and findminima in the Julia package Peaks v0.6.1 and averaged their last five consecutive values for the peak or trough quantification.

## SI Figures

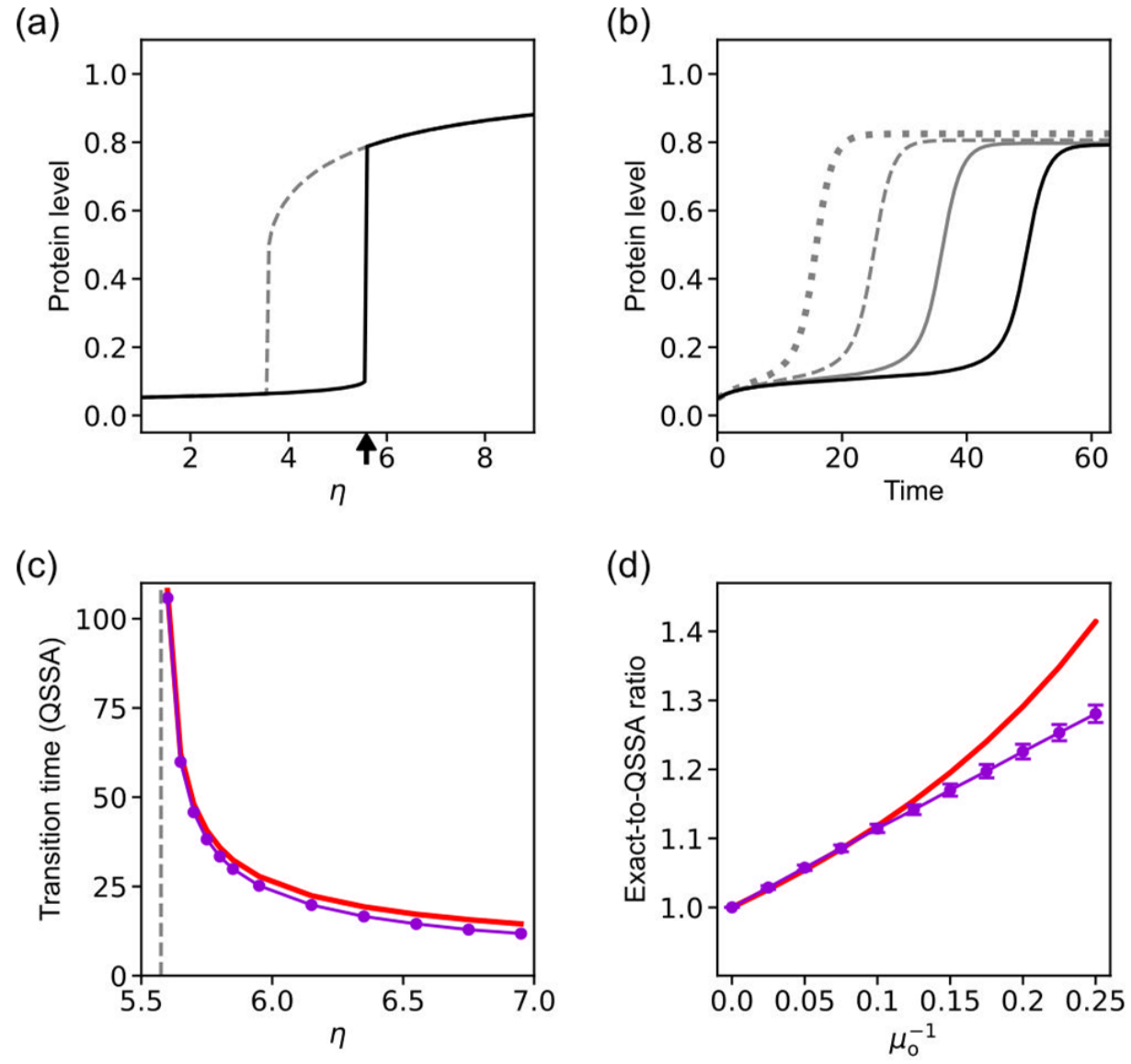

**Fig. S1. Genetic switch and induction kinetics.** (a)–(d) are analogous to Figs. 1(b)–(e), but based on different $\sigma$ (basal transcription rate). (a) Bifurcation diagram of the simulated protein level as a function of inducer level $\eta$ at $\sigma = 0.05$. The same $\sigma$ is used for (b)–(d) as well. The steady state is plotted as $\eta$ increases (solid line) or decreases (dashed line). In the case of increasing $\eta$, the bifurcation at $\eta = \eta_c$ is indicated by an arrow for the analytically-calculated $\eta_c$ from Section G. (b) Time-series of the simulated protein level upon the acute induction at time 0 from $\eta = 0$ to $\eta > \eta_c$, specifically $\eta = 6.5$, 6, 5.8, or 5.7 (left to right). Here $\mu_o^{-1} = 0.1$. (c) QSSA-based transition time from the simulated acute induction (violet dot) or its analytical estimation (red line and Section G) as a function of $\eta$. $\eta = \eta_c$ is marked by a vertical dashed line. (d) Ratio of the transition time to its QSSA from the simulated acute induction (violet dot and error bar) or the analytical estimation with Eq. (11) (red line) as a function of $\mu_o^{-1}$. Each error bar represents the standard deviation across $\eta$ close to $\eta_c$ $(0 < (\eta - \eta_c)/\eta_c \leq 0.25)$. $\mu_o^{-1} = 0$ corresponds to the QSSA. The analytical and simulation results are comparable over small $\mu_o^{-1}$, and even their functional forms become matched as $\mu_o^{-1}$ decreases. In (a)–(d), all the values are unitless through the scaling of the original quantities (Section G).

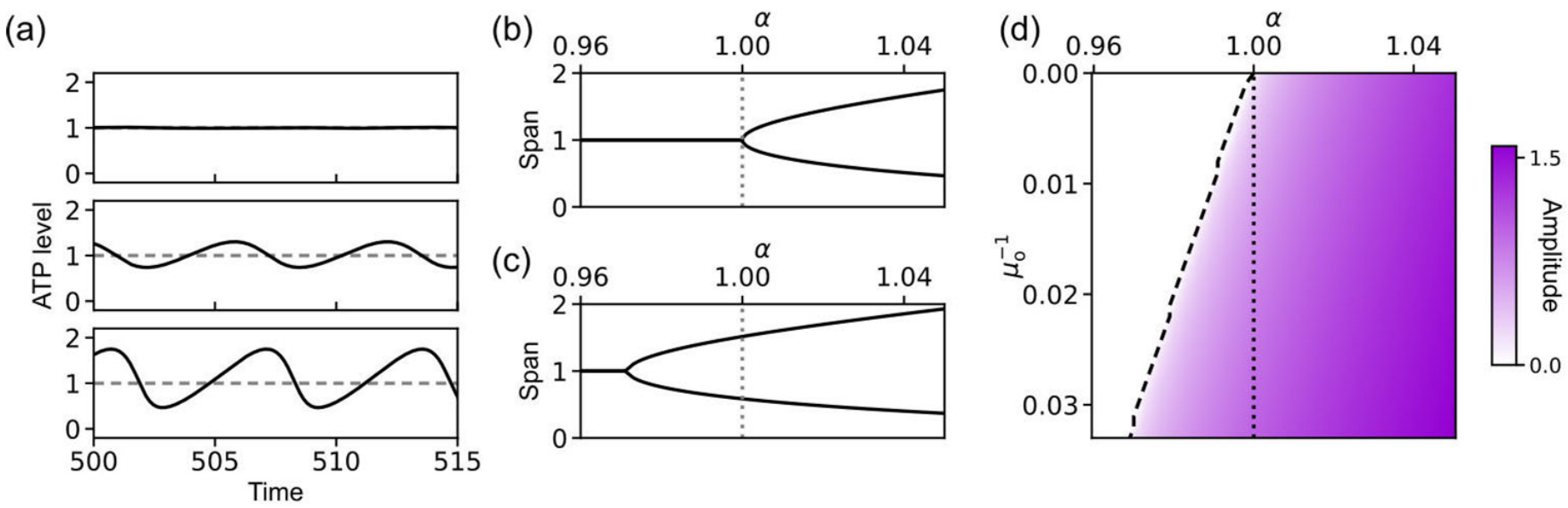

**Fig. S2. Glycolytic enzyme activity and oscillation onset.** (a)–(d) are analogous to Figs. 2(b)–(e), but based on different $\gamma$. (a) Time-series of the ATP level from the Sel'kov model simulation (solid line) and the analytical steady state (dashed line) at $\alpha = 0.99$, $1.01$, or $1.05$ (top to bottom) and $\gamma = 2$. The same $\gamma$ is used for (b)–(d) as well. (b) Bifurcation diagram of the ATP level from the Sel'kov model simulation: peak and trough levels as a function of $\alpha$ (solid line). Bifurcation at $\alpha = \alpha_{\mathrm{Qc}}$ is indicated by a vertical dotted line for the analytically-calculated $\alpha_{\mathrm{Qc}}$ from Section H ($\alpha_{\mathrm{Qc}} = 1/(\gamma - 1)$). (c) Bifurcation diagram of the ATP level from the simulation of the extended Sel'kov model at $\mu_{\mathrm{o}}^{-1} = 0.03$: peak and trough levels as a function of $\alpha$ (solid line), together with $\alpha = \alpha_{\mathrm{Qc}}$ in (b) (vertical dotted line). (d) ATP level's amplitude from the simulation of the extended Sel'kov model: peak-to-trough difference at each $\alpha$ and $\mu_{\mathrm{o}}^{-1}$ (violet-colored). Bifurcation at $\alpha = \alpha_{\mathrm{c}}$ is indicated by a dashed line for numerically-calculated $\alpha_{\mathrm{c}}$ across $\mu_{\mathrm{o}}^{-1}$. For comparison, $\alpha = \alpha_{\mathrm{Qc}}$ from (b) is marked by a vertical dotted line. $\mu_{\mathrm{o}}^{-1} = 0$ corresponds to the original Sel'kov model. In (a)–(d), all the values are unitless through the scaling of the original quantities (Section H).

## SI References